\documentclass[aps,pre,twocolumn,floatfix,groupedaddress]{revtex4}
\usepackage{amsfonts}
\usepackage{amsmath}
\usepackage{graphicx}
\usepackage{times}
\usepackage{amssymb}
\usepackage{verbatim}
\renewcommand{\vec}{\mathbf}
\begin{document}

\title{Meteor generated plasma columns in E-region ionosphere: fields and diffusion}

\author{Y. S. Dimant}
\email{dimant@bu.edu}%
\author{M. M. Oppenheim}
\email{meerso@bu.edu}%
\affiliation {Boston University}
\date{\today}

\begin{abstract}
A meteoroid penetrating the Earth's atmosphere leaves behind a trail
of dense plasma in the E-region ionosphere, a region where electrons
are strongly magnetized while ions are demagnetized due to their
frequent collisions with neutrals. While radar measurements of
meteor trail evolution have been collected and used to infer meteor
and atmospheric properties since the 1950s, no accurate quantitative
model of trail fields and diffusion exists.  This paper describes a
theory and simulations of trail plasma physics which applies to the
majority of small meteors.  Unlike earlier papers, our theory
assumes a significant angle between the geomagnetic field and the
plasma trail and includes the important interaction between the
trail and the background ionospheric plasma. This study provides
quantitative knowledge of the spatial distribution and dynamics of
the plasma density and electric field. This should enable meteor and
atmospheric researchers to more accurately interpret radar
observations of specular and non-specular meteor echoes.
\end{abstract}


\maketitle

\section{Introduction\label{introduction}}

Meteoroids impacting the Earth's upper atmosphere have a number of
important consequences by: depositing material, most notably metals
and dust; damaging spacecraft; creating layers of material which
radars and ionosondes detect; modifying the plasma density and
conductivity of the lower ionosphere; leaving plasma columns which
can be used for meteor burst communication and to monitor
atmospheric conditions in the lower thermosphere. While large
meteoroids generate spectacular optical displays, particles much
smaller than a sand grain comprise the majority of all meteoroids
and represent the major source of all extra-terrestrial material
deposited in the Earth's atmosphere
\cite{Bronshten:Physics83,Ceplecha:Meteor98}. Such small meteoroids
can rarely be observed by a naked eye or even by a sensitive optical
technique, but radars can easily detect them. Such observations
become possible because meteoroids frequently enter the Earth's
atmosphere with a sufficient speed (11--72\ km/s) and energy to
cause the formation of a dense plasma visible to radars. The front
edge of this plasma can often be observed by high-power
large-aperture (HPLA) radars, while smaller, lower-power radars can
detect the residual trail as either specular or non-specular echoes.

\begin{figure}
\includegraphics[width=0.47\textwidth]{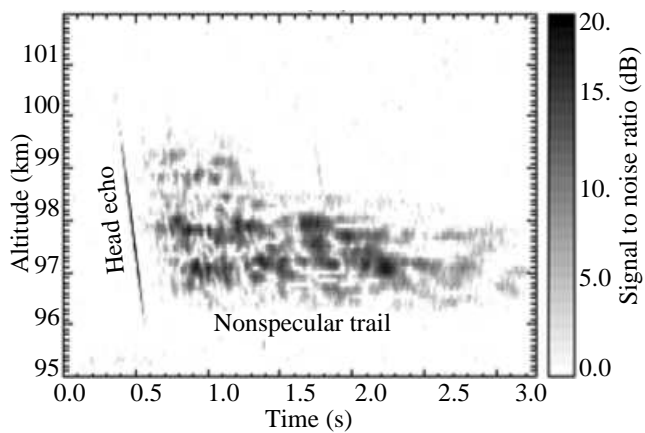}
\caption{\label{dimantFig1}Non-specular radar echo (Jicamarca Radio
Observatory near the magnetic equator, July 12, 2005, 3:43~AM local
time).}
\end{figure}

Specular echoes, usually observed by small-aperture and HF-VHF
radars, originate from parts of the plasma trail where the
wavevector of the backscattered wave is nearly perpendicular to the
axis of a slowly diffusing, quasi-cylindrical plasma trail (for
review, see \cite{Ceplecha:Meteor98}).  Non-specular meteor echoes
observed by HPLA VHF or UHF radars like the one shown in
Fig.~\ref{dimantFig1} originate from trails where the specular
condition is not necessarily satisfied but where the radar typically
points close to perpendicular to the geomagnetic field
\cite{Chapin:Plasma94,Oppenheim:Electrodynamics00,Close:Analysis00,Zhou:Implications01}.
Non-specular echoes appear to result from small-scale electron
density irregularities caused by plasma turbulence and measured when
the radar wavevector lies parallel to the irregularity wavevector
but with a wavenumber twice the irregularity wavenumber, satisfying
the Bragg condition. A polarization electric field, resulting from
the plasma trail ambipolar diffusion, drives field-aligned
instabilities which generate these irregularities
\cite{Oppenheim:Plasma02a,Dyrud:Interpretation02}.

Modeling specular echoes requires knowledge of the spatial and
temporal distribution of the plasma trail density, while modeling
non-specular trails also requires knowledge of the evolution and
structure of the polarization electric field which drives the
instabilities. In a series of papers
\cite{OppEndtDyrGrl00:Electrodyn,Dyrud:anomalous01,Dyrud:Interpretation02,%
Oppenheim:Plasma03b,Dyrud2005JASTP}, Oppenheim, Dyrud and others
have simulated the development of plasma instabilities in the meteor
trail under conditions when the axis of an axially-symmetric
cylindrical plasma trail was perfectly aligned with the geomagnetic
field [or, in a two-dimensional (2D) case, making the trail a slab
instead of a cylinder]. This is, however, a degenerate case because
most of meteor trails are oriented at a significant angle with
respect to the magnetic field. In the general case, electrons have
anisotropic responses to the electric field, which significantly
complicates theoretical description of the dense plasma trail
evolution.

A number of studies have looked at the evolution of a dense plasma
column in a collisional magnetized plasma. The earliest simulations
\cite{KaiserPickWatk69:Ambipolar,PickWindle70:Diffusion,LyatskaiaKlimov88:Diffusion}
had restricted box sizes which could not properly describe the
actual ionospheric situation. Furthermore, numerical simulations
alone cannot provide the parameter dependence of the plasma density
and polarization electric field, which is of importance for modeling
plasma instabilities. At the same time, analytical theory, even an
approximate one, could provide such dependencies and would be an
indispensable tool for the accurate interpretation of radar
observations, meteoric and ionospheric diagnostics.

The analytical theory by Jones \cite{JonesDiffusion1991} represents
a significant step forward. He proposed a 2D self-similar solution
(SSS) of the meteor trail evolution, starting from an initial line
density. Using a combination of a Gaussian spatial distribution of
the plasma density with a parabolic distribution of the electric
potential, Jones developed a mathematical scheme, which describes
the initial evolution and structure of the trail plasma density
reasonably well. However, it improperly describes the spatial
structure of the polarization electric field because the assumed
plasma density structure requires that the field goes to infinity
with distance from the trail axis. As we show in this paper, that
theory also fails to predict significant deviations from the
self-similar diffusion at a later stage of the meteor trail
evolution. The reason for this is that the SSS does not account for
interaction with the background ionospheric plasma.

In the earlier stage of plasma trail diffusion, the background
plasma density is usually small compared to the plasma density
within the trail. However, this low-density plasma plays a crucial
role for carrying electric currents originating from the trail
during its ambipolar diffusion. This current significantly affects
the evolution and structure of the trail density in the later stage.
Further, the electrodynamic interaction of the meteor trail with the
background ionosphere provides a natural restriction for the
polarization electric field which drives plasma instabilities. The
research presented in this paper includes the interaction of the
trail with the background ionosphere.

Note that in a recent theoretical paper on meteor trail diffusion,
Robson \cite{Robson2001:Dispersion} (see also
\cite{ElfordElford2001}) attempted to revise the previous theories.
Robson's approach, however, seems invalid because, in addition to
the quasineutral assumption,
Robson assumed that the electron and ion fluxes along the magnetic
field are equal. In the general case, the latter assumption causes
the resulting electric field to have a significant non-zero curl.
Estimates show, however, that the contribution of the induction
electric field to the trail dynamics in the E region is negligible,
so that to high accuracy the electric field within and around the
trail should be electrostatic, i.e., curl-free (see
Sect.~\ref{discussion} ).

In order to improve modeling of non-specular trails and enhance our
knowledge of plasma column diffusion and fields, we revisit this
problem.  As in the majority of the previous studies, we restrict
ourselves to plasma columns that remain homogeneous along their
length and assume constant electron and ion temperatures.
Homogeneity along the trail implies that diffusion occurs only in
the plane perpendicular to the trail, though electron fluxes have
all three vector components. We further assume that the trail axis
is directed at a sufficient angle to the geomagnetic field (in this
paper, we restrict the analytical treatment to the purely orthogonal
case). The paper differs from Jones \cite{JonesDiffusion1991} in
that it includes the important effect of the background ionospheric
plasma. Finally, we consider only the case when there is no external
electric fields or strong neutral winds which often exist in real
E-region ionosphere, especially at high latitudes and in the
equatorial region.  We will describe these important effects in
future papers. In this paper, we present results of both numerical
simulations and analytical theory. The main result of this work is a
quantitative description of meteor trail evolution and the
polarization electric field associated with its ambipolar diffusion.
Results of our analytical model and simulations agree remarkably
well. We expect that applying our theoretical model to radar echoes
from trails should help researchers obtain useful information about
meteoroids and the surrounding atmosphere.

The paper is organized as follows. In Section \ref{formulation}, we
discuss qualitatively the ambipolar diffusion of meteor trails in
the E-region ionosphere, introduce the concept of the residual
potential and discuss some restrictions. In Section \ref{diffusion
equations}, we present the equations for meteor fields and diffusion
based on a simple two-fluid model model of a highly collisional
isothermal plasma and formulate proper boundary conditions. In
Section \ref{numerical analysis} we discuss results of our 2D
finite-element simulations, which give us useful insights into
finding the proper approach for the analytical treatment of the
problem. In Section \ref{analytical treatment}, we present our
analytical theory for the specific case of mutually orthogonal
meteor trail axis and the geomagnetic field. In Section \ref{Summary
of analytical results}, we summarize our analytical results. In
Section \ref{comparison of theory}, we compare our theory with
simulations quantitatively. In Section \ref{discussion}, we discuss
some caveats and implications of our theory. In Section
\ref{summary}, we give a summary of the paper. Appendices
\ref{self-similar solution} to \ref{effective integral} give
mathematical details of our analytical theory.

\section{Qualitative description of meteor trail diffusion\label{formulation}}

In this section, we discuss qualitative aspects of the meteor trail
diffusion which will allow us to formulate the problem and make
appropriate simplifications.

\subsection{Plasma physics conditions in E-region ionosphere
\label{E-region conditions}}

The majority of specular and non-specular radar echoes are observed in the
lower E region at altitudes 90--120~km, where the neutral atmosphere is many
orders of magnitudes denser than the ionospheric plasma. While the plasma left
behind a fast meteoroid in the form of a diffusing trail can be several orders
of magnitude denser than the background ionospheric plasma, it is usually much
less dense than the neutral atmosphere. The neutral atmosphere remains
essentially undisturbed by the plasma trail formation.

All characteristic spatial scales of the diffusion state are much larger than
the Debye length. The typical diffusion time scale is much longer that the
electron plasma period. Due to these, the quasi-neutrality holds to good
accuracy, $N_{e}\approx N_{i}\equiv N$, where $N_{e,i}$ are the electron and
ion densities, respectively.

In the lower E region, the charged particles collide predominantly with neutral
particles rather than between themselves. The following inequalities hold:
$\Omega_{i}\ll\nu_{in}$ and $\nu_{en}\ll\Omega_{e}$, where
$\Omega_{e,i}=eB_{0}/m_{e,i}$ are the electron and ion gyrofrequencies;
$\nu_{en}$ and $\nu_{in}$ are the electron-neutral and ion-neutral collision
frequencies respectively; $e$ is the elementary charge, $m_{e,i}$ are the
electron and ion masses, respectively, and $B_{0}=|\vec{B}_{0}|$ where
$\vec{B}_{0}$ is the geomagnetic flux density. The above inequalities express
the fact that electrons are strongly magnetized, while ions are demagnetized
due to frequent collisions with neutrals. In the equatorial region,
$B_{0}\simeq0.25\times10^{-4}$T, while at mid- and high latitudes
$B_{0}\simeq0.5\times10^{-4}$T. In the E-region ionosphere, O$_{2}^{{\ }+}$ and
NO$^{+}$ ions dominate: $m_{i}\simeq30\,m_{p}$, where $m_{p}$ is the proton
mass, so that $m_i/m_e\simeq 5.5\times 10^4$. Throughout the E region,
$\nu_{en}\simeq10\,\nu_{in}$.

Under conditions $\Omega_{i}\ll\nu_{in}$ and
$\nu_{en}\ll\Omega_{e}$, the electron and ion diffusion responses to
the external electric field differ significantly. The response of
unmagnetized ions is nearly isotropic, while the responses of
strongly magnetized electrons differ dramatically in the directions
parallel and perpendicular to the magnetic field. The ion isotropic
mobility $g_{i}$ (defined via $\vec{V}_{i}=g_{i}\vec{F}/e$, where
$\vec{F}$ is the external force) is given by $g_{i}\approx
e/(m_{i}\nu_{in})$. The corresponding electron mobilities represent
tensor components determined via similarly defined relations between
the electron velocity $\vec{V}_{e}$ and the external force
$\vec{F}$. In the diffusion approximation, the electron velocity
$\vec{V}_{e}$ is determined by the inertialess fluid momentum
equation (in the neutral atmosphere frame of reference),
\begin{equation}
0=-e\left(  \vec{E}+\vec{V}_{e}\times\vec{B}_{0}\right)  -\frac{\vec\nabla P_{e}%
}{N}-m_{e}\nu_{en}\vec{V}_{e}, \label{force}%
\end{equation}
where $\vec{E}$ is the electric field, $P_{e}\approx NT_{e}$ is the electron
pressure, and $T_{e}$ is the electron temperature (in energy units) assumed
constant. The mobility of magnetized electrons along the magnetic field is
$g_{e\parallel}\approx e/(m_{e}\nu_{en})$, while in the perpendicular
direction the electron mobilities are $g_{e\perp}^{\mathrm{Hall}}%
=e/(m_{e}\Omega_{e})=1/B$ (the Hall mobility), and $g_{e\perp}^{\mathrm{Ped}%
}\approx e\nu_{en}/(m_{e}\Omega_{e}^{2})$ (the Pedersen mobility).
While the parallel and Pedersen velocities, $\vec{V}_{e\parallel}$
and $\vec{V}_{e\perp }$, have the same directions as the
corresponding components of the electric field, the Hall velocity,
i.e., the $\vec{E}\times\vec{B}_{0}$ drift, is perpendicular to
$\vec{E}$.

\subsection{Ambipolar diffusion of plasma columns\label{ambipolar}}

At the earliest stage of trail plasma formation, kinetic processes
associated with ionization of ablated material dominate. After a
short time, however, the newly formed plasma cools down, typical
transport velocities become much smaller than the ion-acoustic
speed, and the trail spreads over distances at least several times
the characteristic mean free path \cite{Jones:Theory95}. This can be
considered as the beginning of the diffusion stage. In this paper,
we restrict our treatment to this stage.

The meteoroid velocity is usually much higher than typical diffusion
velocities. This means that the trail diffusion starts roughly simultaneously
over a sufficiently long distance along the trail. Considering a part of the
trail which is smaller than the typical variation scale of ionospheric and
neutral atmospheric parameters, we will neglect the spatial inhomogeneity and
assume approximate translational symmetry along the trail axis, making all
variations occur in the plane perpendicular to this axis.

\begin{figure}
\includegraphics[width=0.47\textwidth]{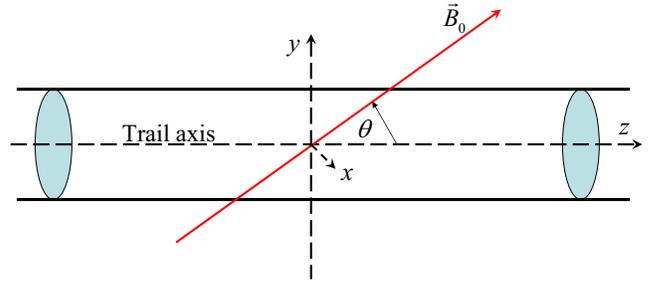}
\caption{\label{geometry}Geometry of the trail and magnetic field.}
\end{figure}

Assuming a non-zero angle between the trail axis and the magnetic
field, we will consider the diffusion of a 2D plasma trail with the
following geometry, see Fig.~\ref{geometry}. The trail density
depends on the $x$ and $y$ coordinates, while it is invariant along
the $z$ coordinate. The homogeneous magnetic field $\vec{B}_0$ lies
in the $y,z$ plane. In this geometry, the plasma density gradient
$\vec\nabla N$ and the polarization electric field $\vec{E}=-\
\vec\nabla\Phi$, where $\Phi$ is the electric potential, have only
$x$ and $y$ components, while the electron drift velocity may have
all three vector components.

In the special case when the trail axis is strictly parallel to
$\vec{B}_{0}$ \cite{OppEndtDyrGrl00:Electrodyn}, the ambipolar
diffusion is axially symmetric around the $z$-axis and its rate,
determined by the ratio between the ion and electron Pedersen
mobilities, roughly follows the lowest mobility. The collision
frequencies of both electrons and ions are proportional to the
neutral density which exponentially decreases with increasing
altitude. The ratio of the electron Pedersen mobility to the
isotropic ion mobility is given by the parameter
\begin{equation}
\psi\equiv\frac{g_{e\perp}^{\mathrm{Ped}}}{g_{i}}=\frac{\nu_{en}\nu_{in}
}{\Omega_{e}\Omega_{i}}.%
\label{psi}
\end{equation}
At higher altitudes (usually above 97\ km at the equatorial region
and above 94 km at high latitudes) where the parameter $\psi$ is
less than unity, the lowest is the electron Pedersen mobility, so
that the trail diffusion is determined by the slightly increased
electron Pedersen diffusion rate. At lower altitudes where $\psi>1$,
the lowest is the ion mobility, so that the trail diffusion is
determined by the slightly increased ion diffusion rate.

In the general case, the ambipolar diffusion is more complicated
because all components of the polarization electric field are
determined by the same scalar potential, $\Phi$. This means that the
diffusion in one direction necessarily affects the diffusion in
other directions. Quasineutrality requires the divergences of the
electron and ion fluxes to be equal, while the fluxes themselves may
differ. Nevertheless, vector components of the electron and ion
fluxes in any direction usually remain comparable. For 2D trail
diffusion, if the angle between the magnetic field and the trail
axis is not too small (see the conditions in
Sect.~\ref{restrictions}), there is a significant component of the
electron pressure gradient along the magnetic field. Accordingly,
there should be a significant component of the polarization electric
field in this direction,
$E_{\parallel}=-\vec\nabla_{\parallel}\Phi$. The electron mobility
along the magnetic field $\vec{B}_0$ (the `parallel' mobility) is
high, while the electron mobilities in the directions perpendicular
to $\vec{B}_0$ are much lower. The electron parallel mobility is
much greater than that of ions, ${g_{e\parallel}}/{g_{i}}\approx
{m_{i}\nu_{in}}/{m_{e}\nu_{en}} \simeq5500$ (Sect.~\ref{E-region
conditions}). Because the average fluid velocity of the highly
mobile electrons should be comparable to that of the low-mobile
ions, according to Eq.~(\ref{force}), the parallel component of the
polarization electric field should nearly cancel the electron
pressure
gradient,%
\begin{equation}
e\vec\nabla_{\parallel}\Phi\approx\frac{\vec\nabla_{\parallel}P_{e}}{N}=T_{e}%
\vec\nabla_{\parallel}\ln n,\label{parallel}%
\end{equation}
where $n(x,y)=N/N_{0}$ and $N_{0}$ is the undisturbed ionospheric
background density assumed constant and uniform. For isothermal
electrons, the electric field and pressure can be combined into one
force $\left(  T_{e}+T_{i}\right) \vec\nabla\phi_{\mathrm{res}}$,
where $\phi_{\mathrm{res}}$ is a dimensionless `residual' electric
potential defined as
\begin{equation}
\phi_{\mathrm{res}}\equiv\frac{e\Phi-T_{e}\ln n}{T_{e}+T_{i}}.%
\label{phiadd}
\end{equation}
Because in our geometry the magnetic field direction has a finite
$y$ component, while the trail and fields are homogeneous along $z$,
Eqs.~(\ref{parallel}) and (\ref{phiadd}) show that the residual
potential within and around the trail is essentially independent of
the $y$ coordinate,
\begin{equation}
\phi_{\mathrm{res}}(x,y)\approx\phi_{\mathrm{res}}^{0}(x).%
\label{Phi0add}
\end{equation}
The concept of the residual potential $\phi_{\mathrm{res}}$ is crucial for our
numerical and analytical treatment. Furthermore, the macroscopic force that
drives plasma instabilities is the total force acting on electrons, which is
just proportional to the gradient of $\phi_{\mathrm{res}}$.

\subsection{Restrictions on orientation of trail with $\vec{B}$
\label{restrictions}}

The effective cancelation of the electron pressure along the
magnetic field described by Eq.~(\ref{parallel}) should only occur
if the angle between the meteor trail axis and the magnetic field
$\theta$ is high enough,
\begin{equation}
\theta\gg\Theta_{0}\equiv\left(  \frac{g_{i}}{g_{e\parallel}}\right)
^{1/2}=\left(  \frac{m_{e}\nu_{en}}{m_{i}\nu_{in}}\right)  ^{1/2}%
\simeq1.35\times10^{-2}, \label{>theta0}%
\end{equation}
where the critical angle in degrees is
$\Theta_{0}\simeq0.8^{\circ}$. This means that there is a
significant component of the electric field along $\vec{B}_0$ to
move electrons. The opposite limiting case of $\theta\ll \Theta_{0}$
corresponds to the diffusion at nearly perfect alignment between the
trail axis and magnetic field when no parallel electron transport
exists \cite{Dyrud:anomalous01}. However the latter case, as well as
the most difficult for analysis intermediate case,
$\theta\sim\Theta_{0}$, are degenerate cases because of rather small
value of the critical angle $\Theta_{0}$. It is evident that the
overwhelming majority of meteors penetrating the Earth's atmosphere
satisfy the condition given by Eq.~(\ref{>theta0}). We will also
require that the electron Pedersen velocity component along the $y$
remains much smaller than the corresponding parallel velocity
component. This yields the condition
\begin{subequations}
\label{restrict}
\begin{equation}
\theta\gg\left(  \frac{g_{e\perp}^{\mathrm{Ped}}}{g_{e\parallel}}\right)
^{1/2}=\frac{\nu_{en}}{\Omega_{e}}=\Theta_{0}\sqrt{\psi}, %
\label{>psi*theta0}
\end{equation}
where the altitude dependent parameter $\psi$ was defined in Eq.~(\ref{psi}).
For further simplicity, we will also require%
\begin{equation}
\theta\gg\frac{\Omega_{i}}{\nu_{in}}=\frac{\Theta_{0}}{\sqrt{\psi}}.%
\label{>theta0/psi}
\end{equation}
\end{subequations}
The parameter $\psi$ decreases exponentially with increasing
altitude, see, e.g., Fig.~2 in Ref.~\cite{DimOppen2004:ionthermal1}.
At the altitude range where electrons are magnetized while ions are
unmagnetized, roughly between 80 and 120 km, the parameter $\psi$
varies between $10^{-3}$ and $10^{4}$, so that restrictions given by
Eqs.~(\ref{>psi*theta0}) and (\ref{>theta0/psi}) combined are
stronger than (\ref{>theta0}). In our major analytical treatment and
simulations, we will discuss the particular case of
$\theta=90^\circ$, but in Appendix~\ref{self-similar solution}, to
find explicit expressions for the self-similar solution, we will
consider a more general case restricted only by
Eq.~(\ref{restrict}).

\section{Diffusion equations\label{diffusion equations}}

If there is no significant ionization and recombination during the
diffusion stage, then the line plasma trail density along the $z$
axis remains nearly constant. The diffusion stage is adequately
described by fluid equations which include two inertialess momentum
equations for electrons and ions and two continuity equations. The
quasineutrality condition makes the Poisson equation for the
electric potential unnecessary. This set of equations can be readily
reduced to two coupled nonlinear partial differential equations
(PDEs) for the common plasma density and the electric potential,
\begin{subequations}
\label{initeq}%
\begin{align}
\partial_{t}n+\vec\nabla\cdot\vec{\Gamma}_{i} &  =0,\label{conti}\\
\vec\nabla\cdot\vec{\Gamma}_{i}-\vec\nabla\cdot\vec{\Gamma}_{e} &  =0,\label{divgegi}%
\end{align}
where for unmagnetized ions and strongly magnetized electrons the diffusion
flux densities, $\vec{\Gamma}_{e,i}$, are given by
\end{subequations}
\begin{subequations}
\label{Gamma}%
\begin{align}
\vec{\Gamma}_{i} &  =-\ n\ \frac{\vec\nabla\left(  e\Phi+T_{i}\ln
n\right)
}{m_{i}\nu_{in}},\label{gi}\\
\vec{\Gamma}_{e\parallel} &  =n\ \frac{\vec\nabla_{\parallel}\left(
e\Phi
-T_{e}\ln n\right)  }{m_{e}\nu_{en}},\label{geparal}\\
\vec{\Gamma}_{e\perp} &  =n\!\left[
\frac{\nu_{en}\vec\nabla_{\perp}( e\Phi-T_{e}\ln n)
}{m_{e}\Omega_{e}^{2}}+\frac{\hat{{b}}\times
\vec\nabla_{\perp}(e\Phi-T_{e}\ln n)  }{m_{e}\Omega_{e}}\right]\!\!
.\label{geperp}%
\end{align}
\end{subequations}
Here the subscripts ${\small \parallel}$ and ${\small \perp}$
pertain to the directions parallel and perpendicular to the magnetic
field $\vec{B}_0$, respectively, and $\hat{{b}}$ is the unit vector
in the $\vec{B}_0$ direction. The first term in the right-hand side
(RHS) of Eq.~(\ref{geperp}) describes the electron Pedersen flux,
while the second term describes the electron Hall flux. Here we
assume the isothermal approximation which is justified by frequent
collisions with the huge thermal reservoir of neutral atmosphere (in
a more general adiabatic process, we would have additional factors
$\gamma_{e,i}$ in front of $T_{e.i}\ln n$.).

Passing from $\Phi$ to the residual potential, Eq.~(\ref{Phi0add}),
we rewrite Eqs.~(\ref{initeq}) in the $x,y$ coordinates as
\begin{subequations}
\label{eqss}%
\begin{equation}
\partial_{t}n-D\left[  \vec\nabla^{2}n+\vec\nabla\cdot\left(  n\vec\nabla\phi
_{\mathrm{res}}\right)  \right]  =0,%
\label{eq2}
\end{equation}
\begin{eqnarray}
&\left(  1+\psi\right)  \partial_{x}\left(
n\partial_{x}\phi_{\mathrm{res}}\right)
+Q\partial_{y}\left(n\partial_{y}\phi_{\mathrm{res}}\right)\nonumber\\
&+\mu\left(
\partial_{x}\phi_{\mathrm{res}}\ \partial_{y}n-\partial
_{x}n\ \partial_{y}\phi_{\mathrm{res}}\right)  +\vec\nabla^{2}n=0,%
\label{eq1}
\end{eqnarray}
\end{subequations}
where we introduced the ambipolar diffusion coefficient,
\begin{equation}
D\equiv\frac{T_{e}+T_{i}}{m_{i}\nu_{in}}=\frac{\left( T_{e}+T_{i}\right)
\Theta _{0}}{eB_0\sqrt{\psi }}, \label{D}%
\end{equation}
a large dimensionless parameter
\begin{equation}
Q\equiv\frac{\sin^{2}\theta}{\Theta_{0}^{2}}=
\left(\frac{m_i\nu_{in}}{m_e\nu_{en}}\right)\sin^{2}\theta\gg
1,\psi,\frac{1}{\psi},
\label{Q}%
\end{equation}
and a dimensionless Hall parameter
\begin{equation}
\mu=\frac{\nu_{in}\cos\theta}{\Omega_{i}}=\frac{\sqrt{\psi}}{\Theta_{0}}
\cos\theta.%
\label{mu}
\end{equation}
Here we used the definitions of $\psi$, Eq.~(\ref{psi}), and
$\Theta_{0}$ in radians, Eq.~(\ref{>theta0}). We also used
Eq.~(\ref{>psi*theta0}) and neglected small additions to the large
parameter $Q$, which are associated with the ion and electron
Pedersen mobilities along $y$.

For the analytical treatment, it is convenient to rewrite
Eq.~(\ref{eqss}) in self-similar variables,
\begin{equation}
\xi=\frac{x}{\left(  Dt\right)  ^{1/2}},\qquad\zeta=\frac{y}{\left( Dt\right)
^{1/2}}. \label{selfvariab}
\end{equation}
As a result, we arrive at the following equations for
$n(\xi,\zeta,t)$ and $\phi_{\mathrm{res}}(\xi,\zeta,t)$:
\begin{subequations}
\label{Equ}%
\begin{align}
&  t\partial_{t}n-\frac{\xi\partial_{\xi}n+\zeta\partial_{\zeta}n}{2}%
-\vec\nabla^{2}n-\vec\nabla\cdot\left(
n\vec\nabla\phi_{\mathrm{res}}\right)
=0,\label{Equ1}\\
&  \left(  1+\psi\right)  \partial_{\xi}\left(  n\partial_{\xi}\phi
_{\mathrm{res}}\right)  +Q\partial_{\zeta}\left(  n\partial_{\zeta}%
\phi_{\mathrm{res}}\right) \nonumber\\
&  +\mu\left(  \partial_{\xi}\phi_{\mathrm{res}}\ \partial_{\zeta}%
n-\partial_{\xi}\phi_{\mathrm{res}}\ \partial_{\zeta}n\right)
+\vec\nabla^{2}n=0,%
\label{Equ2}
\end{align}
where the new $\vec\nabla$ operator pertains to the variables
defined by Eq.~(\ref{selfvariab}),
$\vec\nabla=(\partial_\xi,\,\partial_\xi)$. Our solution of meteor
trail diffusion applies this set of coupled PDEs for $\phi
_{\mathrm{res}}$ and $n$. Given homogeneous background plasma,
neutral atmosphere, and magnetic field, we assume the following
asymptotic boundary conditions,
\end{subequations}
\begin{equation}
n\rightarrow1\qquad\text{and}\qquad\phi_{\mathrm{res}}\rightarrow
0\qquad\text{as}\qquad x,y\rightarrow\pm\ \infty. \label{boucond}%
\end{equation}
We also note that our problem has the following symmetry,
\begin{subequations}
\label{symmetry}%
\begin{align}
n\left(  x,y\right)   &  =n\left(  -x,-y\right)  ,\label{Nsymm}\\
\phi_{\mathrm{res}}\left(  x,y\right)   &  =\phi_{\mathrm{res}}\left(
-x,-y\right)  . \label{Phisymm}
\end{align}
\end{subequations}
The following section describes a numerical solution of these equations and the
next section describes their analytic solution.

\section{Simulations of trail diffusion and fields\label{numerical analysis}}

In this section, we discuss results of our numerical solution of
Eq.~(\ref{eqss}) using a finite-element PDE solver FlexPDE
\footnote{URL: \ www.pdesolutions.com}. The challenge of these
simulations is to simultaneously resolve both the relatively small
scale of the trail density variations and the large scale of the
residual potential variations parallel to the magnetic field (along
$y$). This requires the box size along $x$ to be at least several
times the effective trail size in that direction, $\sigma_x$, while
the box size along $y$ should be several times $\Theta_0^{-1}$,
i.e., more than two orders of magnitude, larger than the box size
along $x$. FlexPDE uses an adaptive finite elements mesh in regions
with high gradients to resolve the fields and densities with high
precision, while uses a coarse mesh in regions where gradients
remain small.

We tested the effects of a finite simulation box on the solution by
varying its size. We also varied boundary conditions, setting either
the density disturbances, $\Delta n(t)\equiv n(t)-1$, and
$\phi_{\mathrm{res}}$, or the corresponding flux densities,
Eq.~(\ref{Gamma}), to zero on the boundaries. These tests
demonstrated that, for sufficiently large box sizes, the solution in
the inner region remained essentially unaffected by the choice of
boundary conditions.

\begin{figure*}
\includegraphics[width=0.47\textwidth]{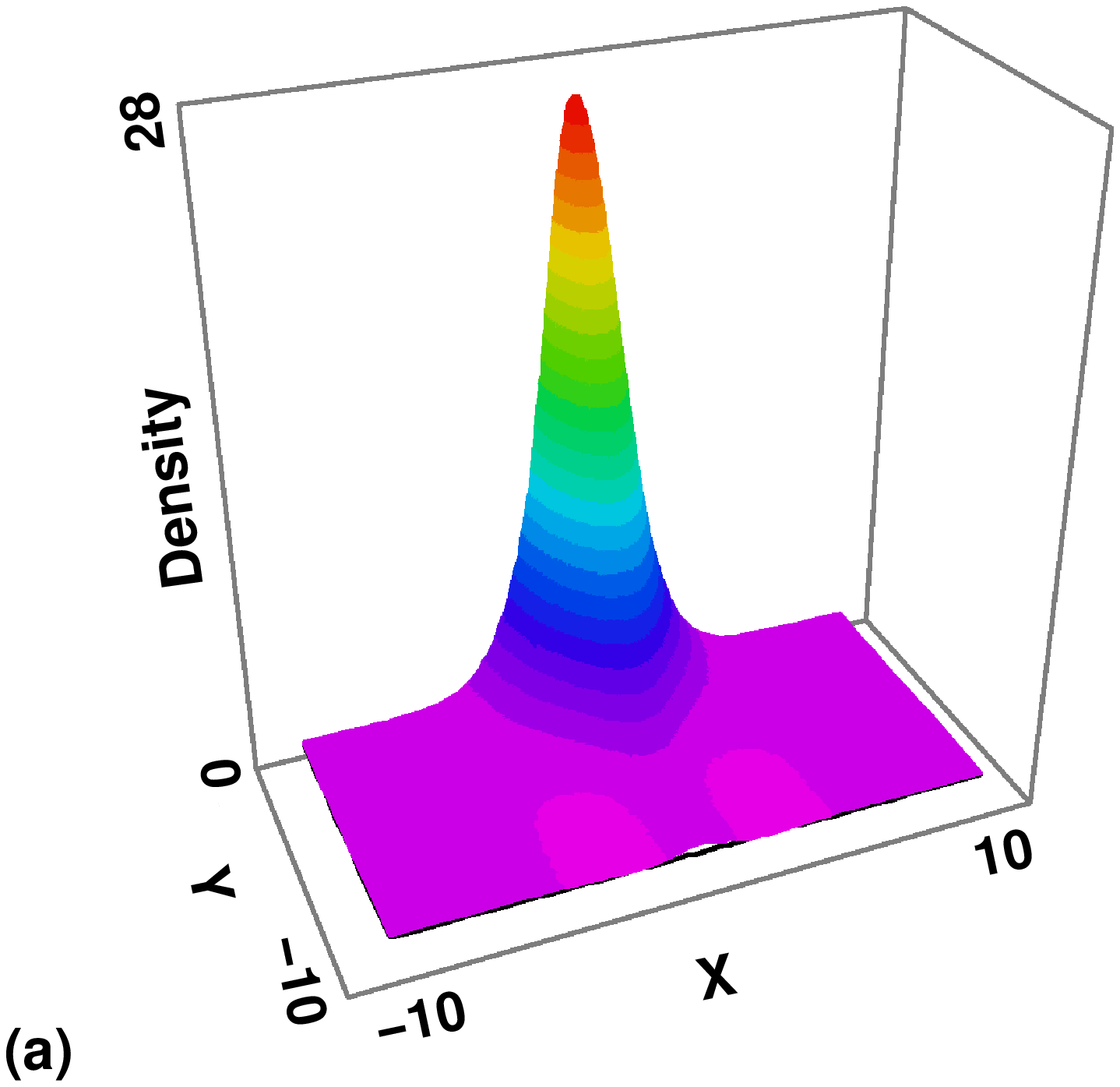} \hfill
\includegraphics[width=0.47\textwidth]{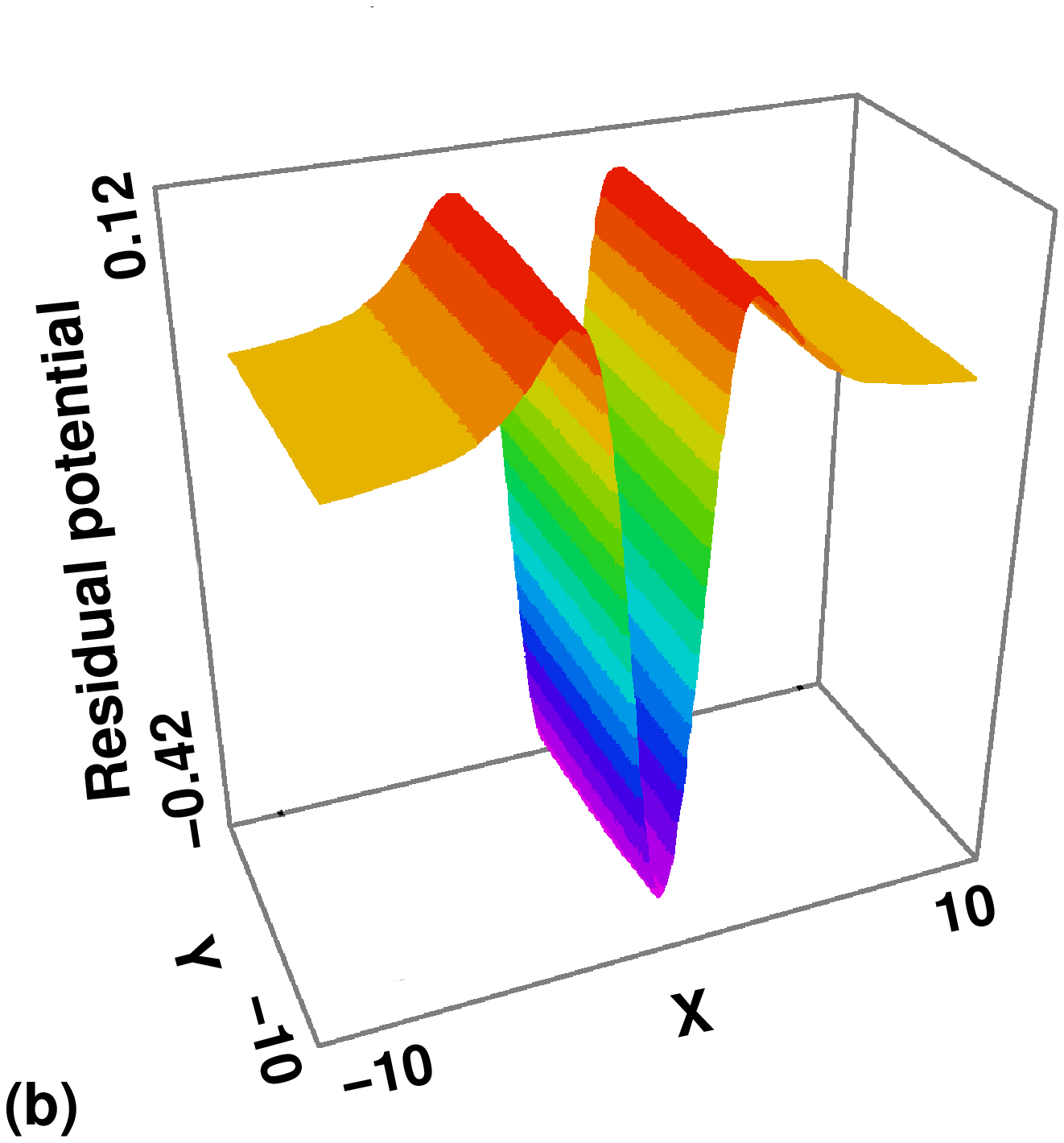}
{\caption{\label{Fig2} Typical structure of plasma density (a) and
residual potential (b). Here $\psi=0.2$, $\Delta n_0=100$, and
$t=2.5$ (the background density corresponds to $n=1$ and the initial
time for the self-similar solution $t_0=1$).}}
\end{figure*}

\begin{figure*}
\includegraphics[width=0.47\textwidth]{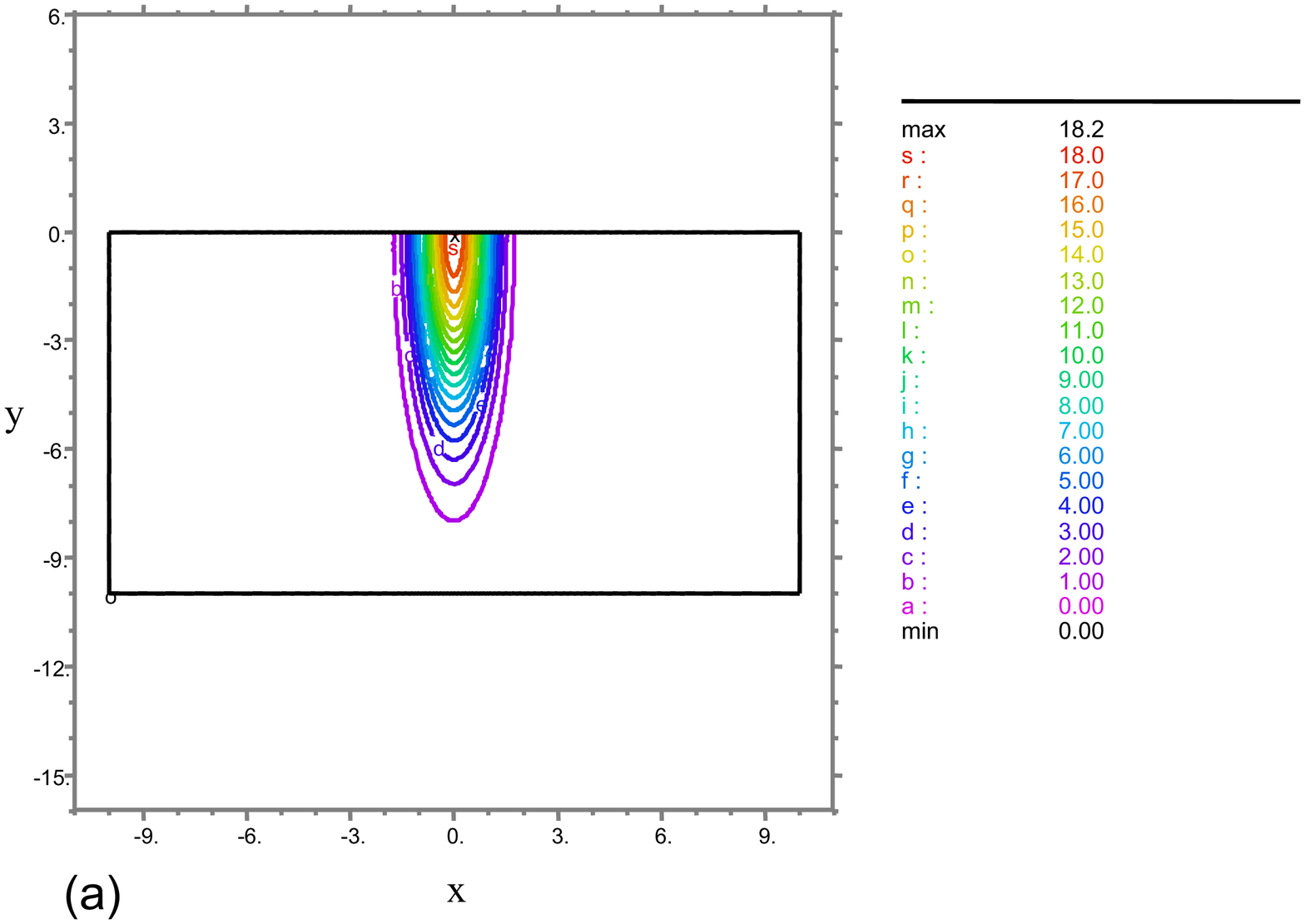} \hfill
\includegraphics[width=0.47\textwidth]{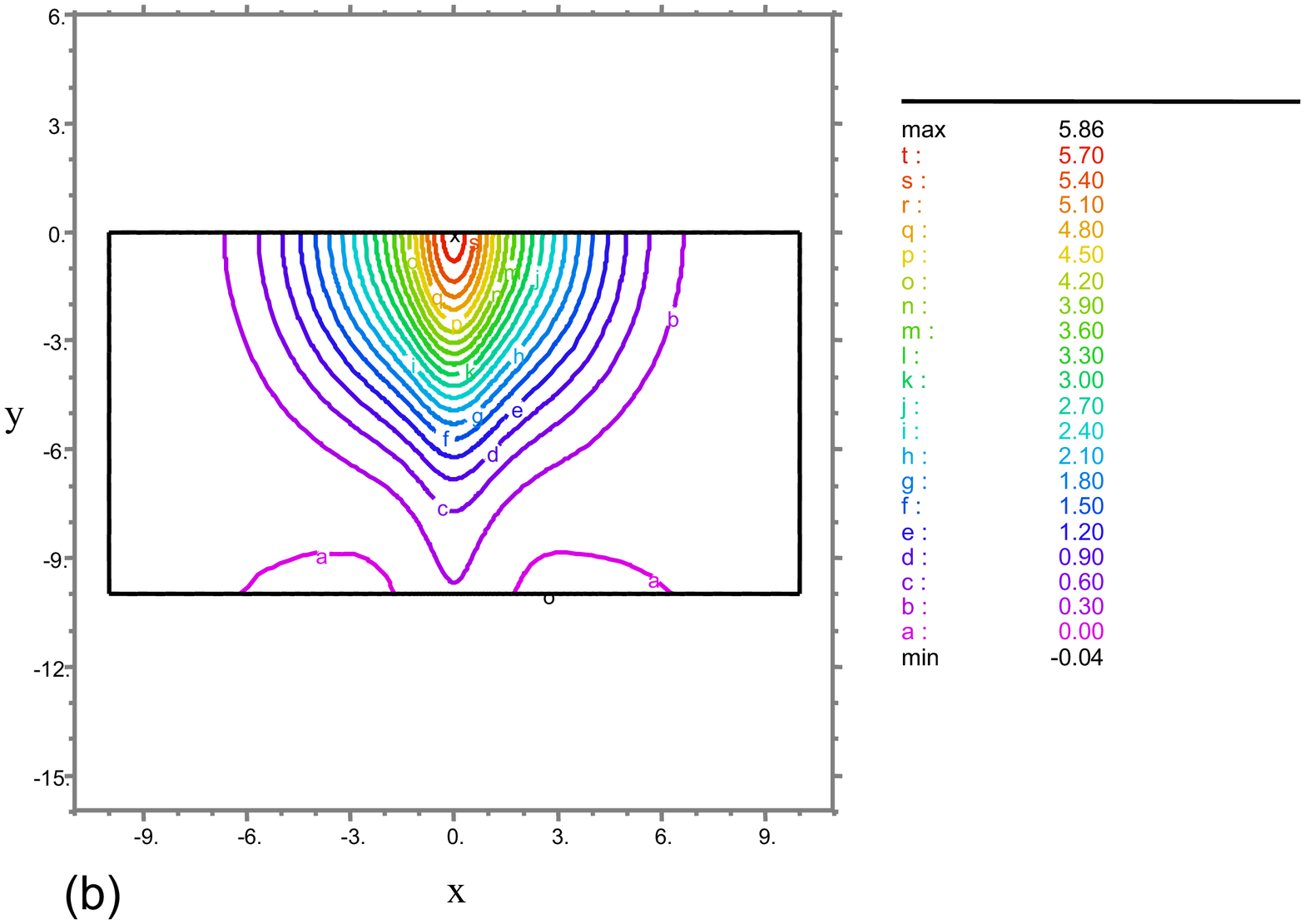}
{\caption{\label{Contours} Contours of trail density in the
$x,y$-plane for $\psi=0.05$, $\Delta n_0=100$, $t=5.5$: (a):
self-similar solution, (b): simulation.}}
\end{figure*}
\begin{figure}
\includegraphics[width=0.47\textwidth]{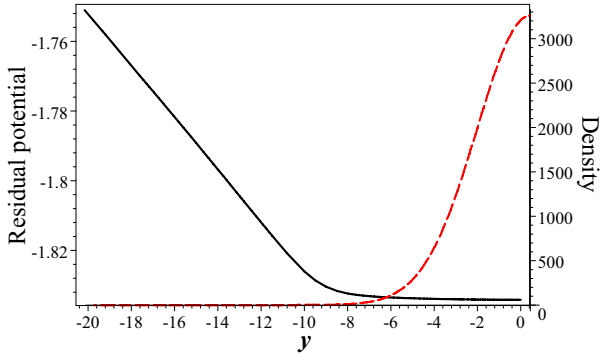}
\caption{\label{PotenDens}Typical variation  parallel to the
magnetic field (along $y$) at $x=0$ of residual potential (solid
curve, left scale) and trail density (dashed curve, right scale)
[for $\psi=0.2$, $\Delta n_0=10^4$, $t=3$]. In the near-trail
region, the density has exponentially strong variation along $y$,
while the residual potential varies only within several percent of
its maximum absolute value. At some point beyond the trail (here
about $y=-10$), the potential acquires a noticeable $y$-derivative
(electric field parallel to the magnetic field $\vec{B}_0$). This
derivative is always much less than the typical $x$-derivative
(electric field perpendicular to $\vec{B}_0$).}
\end{figure}

\begin{figure}
\includegraphics[width=0.47\textwidth]{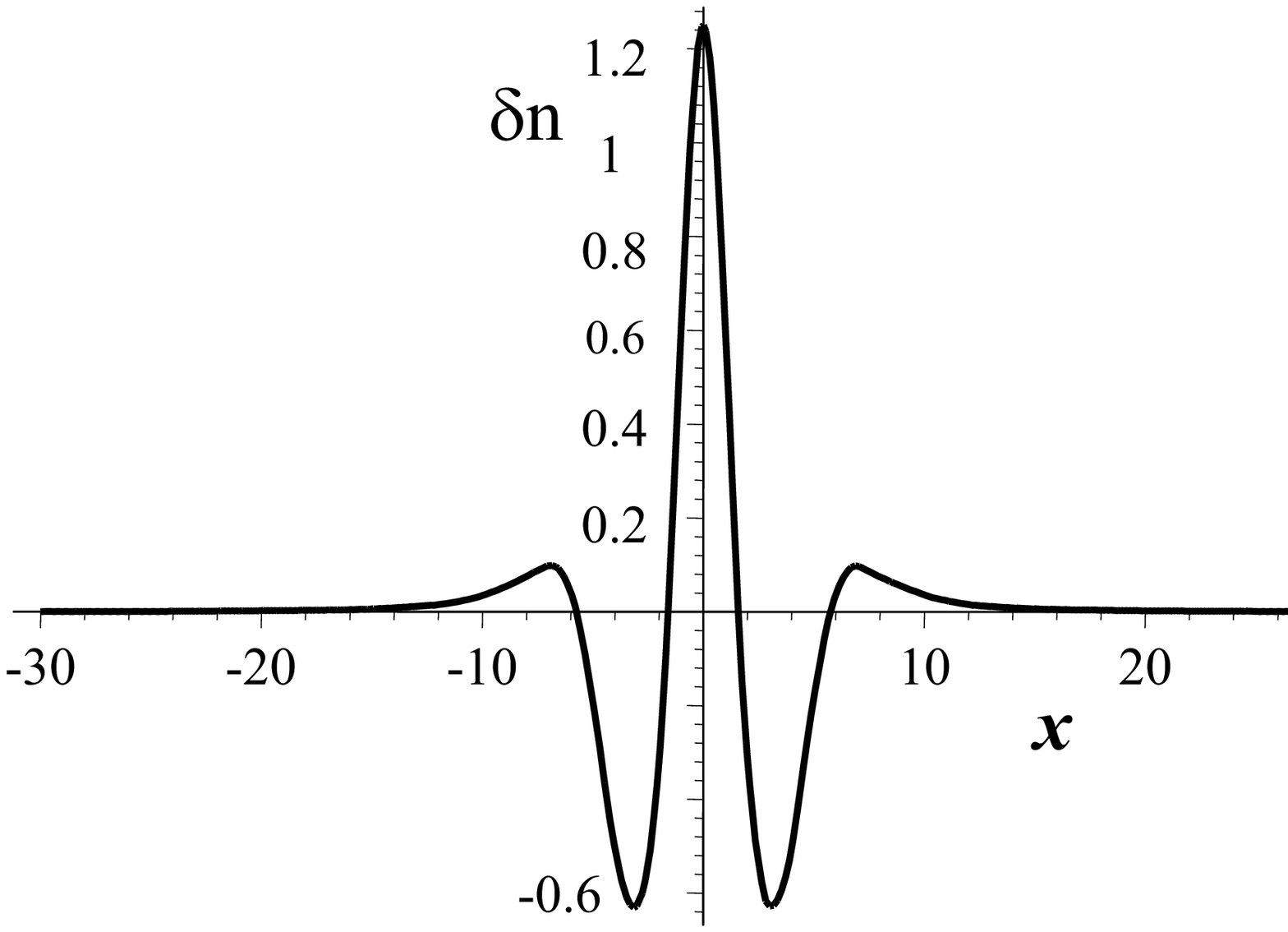}
\caption{\label{BackgrDensity}Disturbances of background density
beyond the trail perpendicular to the magnetic field (along $x$)
[for $\psi=0.2$, $\Delta n_0=10^4$, $t=3.5$ at the coordinate
$|y|=20$ located well beyond the trail].}
\end{figure}

In numerical simulations, as well as in the analytical theory
(Sect.~\ref{analytical treatment}), we have explored the strictly
perpendicular case of $\theta=90^\circ$ ($\mu=0$). In this case, we
solved Eq.~(\ref{eqss}) with $Q=5500$ corresponding to
$m_i/m_e=5.5\times 10^5$ and $\nu_{en}/\nu_{in}=10$, see
Sect.~\ref{E-region conditions}. As the initial condition at time
$t=t_{0}$, we chose a narrow and dense column of plasma, described
by the self-similar solution (SSS). We used normalized units where
the diffusion coefficient $D=1$ and the initial time for the SSS
solution $t_0=1$, so that the initial spatial distribution of the
total normalized plasma density $n=N/N_0$ was given by
\begin{equation}
n(x,y,1)= 1+\Delta n_{0}\exp\left[-\,\frac{(1+\psi)x^2}{4\psi}-
\frac{y^2}{4}\right]. %
\label{initSSS}
\end{equation}%
The characteristic sizes of the initial Gaussian density
distribution of the trail in the $x$ and $y$ directions,
$\sigma_{x0}=[2\psi/(1+\psi)]^{1/2}$ and $\sigma_{y0}= \sqrt 2$, are
nearly equal for $\psi\gg 1$ (lower altitudes), but differ
significantly for $\psi\ll 1$ (higher altitudes). To check the
effect of the initial conditions, we tried different initial
Gaussian density distributions corresponding to the same trail line
density. We have found that, after the time needed for the trail to
diffuse over a distance several times the original size of the
trail, the solution becomes virtually the same. Hence, it is only
weakly sensitive to the actual initial peak cross-section.

Equation~(\ref{eq1}) involves no time derivatives, so that formally
$\phi_{\mathrm{res}}$ needs no initial condition. However, the
FlexPDE application requires setting initial conditions for all
variables. In our simulations, we usually set up
$\phi_{\mathrm{res}}(t_0)=0$. We tested that after a rather short
time, the solver automatically sets up a time-dependent spatial
distribution of $\phi_{\mathrm{res}}$ which proves to be independent
of the initial condition for $\phi_{\mathrm{res}}$.

Bearing in mind the symmetry along the $y$ direction (i.e., parallel
to $\vec{B}_0$), we simulated a half of the entire space ($y\leq 0$)
with boundary conditions at $y=0$ given by zero derivatives of both
$n$ and $\phi_{\mathrm{res}}$. The box size, $|y|_{\max}=1000$, was
always at least two orders of magnitude larger than $\sigma_y(t)$,
but we needed to keep it that long because of the long-extended
residual potential.

In order to model different stages of trail diffusion we performed
several overlapping runs by varying the initial peak densities from
$\Delta n_0=10^4$ to $\Delta n_0=10$. We usually finished each run
at $t=40$, so that the box size $x_{\max}=30$ remained at least
several times $\sigma_{x}(t)$. Because of higher anisotropy and
stronger gradients, the cases of small $\psi$ are much more time and
resource consuming than those of $\psi\gtrsim 1$. On a 3GHz Pentium
4 Windows-based PC, our runs lasted from several hours (for
$\psi=10$) to more than a week (for $\psi=0.05$). Unfortunately,
FlexPDE intrinsic restrictions have not allowed us to simulate
$\psi<0.05$.

To monitor the accuracy of simulations, we used general relations
for the residual potential outlined in
Appendix~\ref{integrodifferential equations} and the exact property
of density disturbance integrals (DDIs) which is analytically
derived in Appendix~\ref{DDI}.

Figure \ref{Fig2} shows typical plasma density and residual
potential plotted over the small fraction of the simulation box
nearest the trail a short time after the simulation begins.
Figure~\ref{Contours} compares an example of the trail density
contours in the $x,y$-plane corresponding to the SSS (a) and to the
full simulation (b) after the trail has diffused to several of its
initial radius. Our simulations have revealed the following major
features:

\begin{enumerate}
\item%
\label{item1}%
Initially, the plasma density distribution within the trail closely
follows the anisotropic (for $\psi\lesssim 1$) SSS. At a later
stage, as the peak density falls, the trail remains nearly Gaussian
in each direction but becomes more isotropic than does the SSS, see
Fig.~\ref{Contours}, and it diffuses faster, with the isotropic
ambipolar diffusion rate given described by Eq.~(\ref{D}). The
transition from anisotropic to nearly isotropic diffusion usually
takes place while the peak density remains much larger than the
background density. Section~\ref{analytical treatment} discusses our
analytical model for this point.

\item%
\label{item3}%
The residual electric potential spreads along $y$ well beyond the
trail, see Fig.~\ref{PotenDens}, with spatial gradients along $y$
much less than those along $x$. Along $x$ it has a non-monotonic
symmetric structure with a deep minimum at the trail center and two
symmetric maxima around it, as shown in Fig.~\ref{Fig2}(b). At the
later stage, when the trail diffusion becomes nearly isotropic (see
Feature \ref{item1}), the residual electric field becomes much
smaller than that for the SSS.

\item%
\label{item4}%
Beyond the trail, where the exponentially small trail density is
much less than the background density, the residual electric field
extending along $\vec{B}_0$ may substantially disturb the density as
shown in Fig.~\ref{BackgrDensity}. The field evacuates plasma from
the region that maps along $\vec{B}_0$ the potential minimum back to
those about the potential maxima. As a result, the perturbation
density evolves in antiphase to the potential distribution along
with a maximum at $x=0$ and two symmetric minima roughly at $x$
where the residual potential has the two maxima, as visible on the
edge of Fig.~\ref{Fig2}(a). In our simulations, the maximum central
density compression more than doubled the undisturbed background
density, while the depletions reached about 80\% of that. These
strong disturbances were reached near the trail when the peak trail
density was at least several orders of magnitude larger than the
background density, see Fig.~\ref{BackgrDensity}. In all other cases
or locations, relative density disturbances were small. We estimate
them in Sect.~\ref{density disturbances}.
\end{enumerate}

We will discuss these features in Sects.~\ref{analytical treatment}
and \ref{comparison of theory}, when describing our analytical
theory and comparing it with simulations. Here we only give brief
explanations to some features.

\paragraph*{Feature \ref{item1}:}
At the later stage of trail diffusion, when the initially dense
trail density becomes much less dense (but remains much denser than
the background plasma), its gradual isotropization and acceleration
of diffusion to the ambipolar isotropic rate are due to the sharp
decrease with time in the residual electric field (Feature
\ref{item3}). In this case, the total force acting on electrons
becomes so small that diffusion is mainly determined by the total
pressure gradient acting on ions and, hence, is virtually unaffected
by the external magnetic field. The significant reduction of the
residual electric field is due to the response of the background
plasma which is not included in the SSS. To provide quasineutrality,
a less dense trail starts attracting more charged particles from the
background, thus reducing the need for strong ambipolar electric
field.

\paragraph*{Feature \ref{item3}:}
The extension of the residual potential along $y$, i.e., parallel to
the external magnetic field $\vec{B}_0$, is due to the high electron
mobility along $\vec{B}_0$. The deep potential minimum in the trail
center is formed by anisotropic ambipolar trail diffusion. These two
factors are included in the SSS and are independent of the existence
of background plasma. However, the two maxima seen in the residual
potential distribution along $x$, see Fig.~\ref{Fig2}(b),  are due
to background plasma. The electron and ion fluxes originating in the
trail extend into the background ionosphere with sharply anisotropic
and quite different patterns, while the divergences of the two
fluxes remain balanced. Beyond the trail, these patterns have a
quadrupole-like structures providing current closure. To drive the
return currents, the background ionosphere develops potential
gradients which oppose those within the trail, i.e., those
responsible for the trail diffusion. This gives rise to the two
symmetric potential bumps around the deep potential minimum and
draws background plasma into the trail edges.

The simulations allow us to determine diffusion rates and isotropy
for a range of altitude-dependent $\psi$. This will be presented in
conjunction with our analytical theory in Sect.~\ref{comparison of
theory}.

\section{Analytical theory \label{analytical treatment}}

Here we discuss our analytical theory for the `perpendicular' case
of $\theta=90^\circ$ ($\mu=0$), when a double mirror symmetry,
\begin{subequations}
\label{symmka}
\begin{align}
n\left(  \xi,\zeta\right)   &  =n\left(  \pm\xi,\pm\zeta\right),%
\label{n+-}\\
\phi_{\mathrm{res}}\left(  \xi,\zeta\right)   &  =\phi_{\mathrm{res}}\left(
\pm\xi,\pm\zeta\right),%
\label{phi+-}
\end{align}
\end{subequations}
takes place. The Hall velocity of electrons is directed strictly
along the trail axis, so that it does not affect the 2D diffusion of
the plasma trail. In the general case of $\theta\neq 90^\circ$, the
Hall mobility breaks the double mirror symmetry but keeps a
rotational (by $180^{\circ}$) symmetry around the trail axis,
Eq.~(\ref{symmetry}). For sufficiently large angles $\theta$,
distortions caused by the Hall mobility are relatively small (see
Appendix \ref{self-similar solution}), so that the double mirror
symmetry roughly takes place. However, the Hall mobility affects the
aspect ratio of typical scales along each axis. We reserve the more
general case for future work.

The trail diffusion and evolution of the ambipolar fields are
described by two coupled nonlinear PDEs for the plasma density and
the residual potential, see Eqs.~(\ref{eqss}) or (\ref{Equ}). Our
theory provides an approximate analytical solution of both
interrelated problems. The key point of the theory has been the
inclusion of the background plasma into consideration. In so doing,
even for $\mu=0$, one cannot find the exact analytical solution of
the coupled equations. However, based on the existence of the large
parameter $Q=\Theta_{0}^{-2}\simeq 5500$, Eq.~(\ref{Q}), and on the
insight from our simulations, see Sect.~\ref{numerical analysis}, we
have developed an approximate approach based on the perturbation
technique.

We have used the fact that the initially narrow and dense trail
keeps nearly Gaussian density distribution during the trail lifetime
even if the diffusion rate is changing with time, see
Sect.~\ref{numerical analysis}. Given the characteristic scale of
the Gaussian distribution, we have reduced Eq.~(\ref{Equ2}) to a
simpler equation for the residual potential. This one-dimensional
(1D) linear integrodifferential equation is the governing equation
of our theory. It has two equivalent forms that depend upon only one
dimensionless parameter $\rho$. This master parameter is
proportional to the the square of the characteristic Gaussian scale
along $x$, $\rho\propto\sigma_x^2(t)$, see Eq.~(\ref{rhosig}) below,
and, hence, monotonically grows with time. Approximate solutions of
the governing equation in different domains of $\rho(t)$ have
allowed us to obtain the spatial distribution of the residual
electric potential in the entire 2D space.

The trail diffusion is mainly determined by the parabolic behavior
of the potential along $x$ within the central region and is
practically insensitive to the potential behavior outside this
region. This has allowed us to find from Eq.~(\ref{Equ1}) the
explicit time dependence of $\rho(t)$ and thus close the solution.
As a result, we have obtained approximate analytic expressions
describing the trail diffusion and the evolution of the residual
electric fields. These expressions reasonably well agree with the
numerical results and can be used for quantitative predictions.

Our analytical theory consists of several major steps: (1) obtaining
explicit expressions for the self-similar solution; (2) solving an
equation for the electric potential in the `far zone' where the
density disturbances are relatively small; (3) obtaining the
governing equation for the residual potential in the `near zone'
depending on the master parameter $\rho$; (4) solution of this
equation in different ranges of $\rho$; (5) finding expressions for
$\rho(t)$ in the initial value problem, which gives the approximate
solution of the trail diffusion problem and closes the solution for
the residual potential; (6) estimating density disturbances of the
background plasma beyond the trail and obtaining a correction factor
for $\rho\gg 1$ at a later stage of trail diffusion. We have
implemented steps (1) to (3) in Appendices \ref{self-similar
solution} to \ref{near-zone potential}, while in
Sect.~\ref{near-zone outline} we will only explain the basic ideas
and discuss the resultant governing equation. We have implemented
steps (4) to (6) in Sect.~\ref{solution of equations} to
\ref{density disturbances}.

\subsection{Governing equation for near-zone residual potential%
\label{near-zone outline}}

Here we briefly describe the governing equation intended for the
solution of the residual potential problem. Given the trail density
spatial distribution, this approximate equation is derived from
Eq.~(\ref{Equ2}). The derivation requires separation of the entire
$\xi,\zeta$ plane into two overlapping regions: the near zone,
\begin{equation}
|\zeta| \equiv \frac{|y|}{(Dt)^{1/2}} \ll Q^{1/2},%
\label{nearzone}
\end{equation}
and the far zone,
\begin{equation}
|\zeta| \gg 1.%
\label{farzone}
\end{equation}
We emphasize that the terms `near' and `far' pertain to the
coordinate $\zeta$ only, so that the near zone includes the entire
$\xi$-axis (the $x$-axis).  The residual potential in the near zone
is $\xi$ dependent but remains nearly constant in the $\zeta$
direction, i.e., along $\vec{B}_0$, see Eq.~(\ref{Phi0add}), making
its calculation a 1D problem.

In the far zone, which includes the background ionosphere only, the
residual potential has a slow $\zeta$ dependence, but we may roughly
neglect there density disturbances (see Appendix \ref{far zone} and
Sect.~\ref{density disturbances}). The fact that divergences of the
electron and ion fluxes are nearly equal, as required by
quasi-neutrality, results in the Laplace equation in terms of some
renormalized coordinates. In those renormalized coordinates, the
entire near zone reduces to a thin cut which provides boundary
conditions for the Laplace equation in each of the two half-spaces,
$\zeta>0$ or $\zeta<0$. Because of the symmetry, it is sufficient to
consider only one of these half-spaces ($\zeta>0$). Given the
residual potential in the near zone $\phi_{\mathrm{res}}^{0}$, a
straightforward solution of the Laplace equation in integral form
yields the spatial distribution of the residual potential in the
entire 2D space, as described by Eq.~(\ref{solu}).

Using the Gaussian approximation of the trail plasma density,
Eq.~(\ref{Gauss}),
\begin{equation}
n_{\mathrm{Trail}}=\frac{\Delta
n_{0}t_{0}}{t}\sqrt{\frac{q_{0}}{q(t)}}\,\exp \left[-\
\frac{1}{4}\left( \frac{\xi ^{2}}{q(t)}+\zeta ^{2}\right) \right] ,%
\label{Gaussian}
\end{equation}
and integrating it over the near zone, we obtain the sought-for
boundary condition for the Laplace equation. This integration
involves $\phi_{\mathrm{res}}^{0}$ (with still unknown $\xi$
dependence) multiplied by the corresponding trail density integral
over the coordinate $\zeta$. Associating the upper half-space with a
complex plane in terms of renormalized coordinates and using
analytical properties of the potential in the far zone (see Appendix
\ref{far zone}), in Appendix \ref{near-zone potential} we obtain for
the the near-zone residual potential an integrodifferential equation
in two different, but equivalent, forms:
\begin{subequations}
\label{integrodiff}
\begin{equation}
\frac{2\rho}{\pi^{3/2}}\ e^{\frac{\eta^{2}}{4}}\mathcal{P}\!\!\int_{-\infty
}^{\infty}\frac{\varphi(\tau)  }{\tau-\eta}\,d\tau+\partial_{\eta
}\varphi(\eta)  =\eta,%
\label{Form2}
\end{equation}
\begin{equation}
\rho\varphi(\eta)  -\frac{1}{2\sqrt{\pi}}\
\mathcal{P}\!\!\int_{-\infty}^{\infty}\frac{\partial_{\tau}\varphi(\tau)}
{\tau-\eta }\,e^{-\tau^{2}/4}d\tau=S(\eta).%
\label{Form1}
\end{equation}
\end{subequations}
Here the renormalized near-zone
residual potential $\varphi$ and coordinate $\eta$ are defined by
\begin{subequations}
\label{Phieta}
\begin{equation}
\varphi(\eta)  =2(1+\psi)  \phi_{\mathrm{res}} ^{0}(\xi),%
\label{Phi}
\end{equation}
\begin{equation}
\eta
=\frac{\xi}{q^{1/2}}=\frac{x}{(qDt)^{1/2}}=x\sqrt{\frac{\gamma}{\rho{D}}}.
\label{eta}
\end{equation}
\end{subequations}
The coupled dimensionless parameters $q(t)$ and $\rho(t)$ are
proportional to the square of the Gaussian peak dispersion with
respect to $x$ coordinate, $\sigma_x^2$, Eq.~(\ref{sigmax}),
\begin{equation}
\rho(t)=\gamma q(t)t= \frac{\gamma \sigma_x^2}{2D} = \frac{\pi
\sigma_x^2N_0}
{\sqrt{1+\psi}\,\Theta_0 N_{\mathrm{lin}}}\,,%
\label{rho}
\end{equation}
where
\begin{equation}
\gamma  =\frac{1}{2\Theta _{0}\Delta n_{0}t_{0}\sqrt{\psi
}}=\frac{2\pi N_{0}\left(
T_{e}+T_{i}\right) }{\sqrt{\psi \left( 1+\psi \right) }\,N_{\mathrm{lin}}^{z}eB_0}%
\label{gamma}
\end{equation}
and $N_{\mathrm{lin}}$ is the line plasma density of the trail along
its axis $z$.

\begin{figure}
\includegraphics[width=0.47\textwidth]{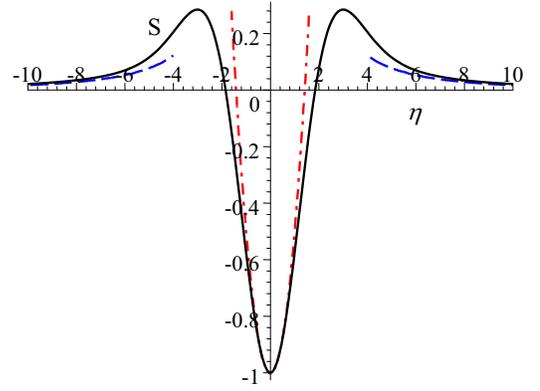}
\caption{\label{Fig.2(S)}Function $S(\eta)=\eta
e^{-\,\eta^{2}/4}\int_{0}^{\eta/2}e^{\tau^{2}} d\tau-1$,
Eq.~(\ref{S}) (solid curve), along with its parabolic approximation,
$-1+\eta^2/2,$ Eq.~(\ref{series}) (dash-dotted curve), and
large-$\eta$ asymptotics, $2/\eta^2$, Eq.~(\ref{asympt}) (dashed
curve).}
\end{figure}

The important function $S(\eta)=\eta
e^{-\,\eta^{2}/4}\int_{0}^{\eta/2}e^{\tau^{2}}d\tau-1$ in the RHS of
Eq.~(\ref{Form1}), see also Eq.~(\ref{S}), together with the
highest-order power-series approximations of $S(\eta)$,
Eqs.~(\ref{series}), (\ref{asympt}), is shown in
Fig.~\ref{Fig.2(S)}. In Sect.~\ref{solution of equations}, we will
see that this function is a good qualitative, and in some cases even
quantitative, representation of the residual potential in the near
zone.

Equivalent Eqs.~(\ref{integrodiff}a,b) are linear
integrodifferential equations in renormalized variables defined by
Eq.~(\ref{Phieta}). In these variables, these equations depend upon
only one dimensionless parameter $\rho$. We will use one of two
equivalent Eqs.~(\ref{integrodiff}a,b) depending upon the
convenience for the specific analysis. In Section
\ref{integrodifferential equations}, we discuss the general
analytical properties of the solution, which can be used for
monitoring the solution. In particular, under assumed
approximations, the integral of $\phi_{\mathrm{res}}^{0}(\xi)$ over
$\xi$ should be equal to zero. This explains mathematically the
non-monotonic, two-bump structure of the residual potential in the
near zone observed in our numerical calculations (see
Sect.~\ref{numerical analysis}, Feature 2).

The physical sense of the outlined mathematical procedure is that
different patterns of the electron and ion fluxes in the far zone
create a feedback to the near zone. The non-monotonic structure of
the residual potential (in each of the two half-spaces, $x>0$ and
$x<0$), can be explained as follows. The total force acting on
negatively charged electrons is proportional to
$e\vec\nabla\phi_{\mathrm{res}}^{0}$. Within the major trail, this
force via electron Pedersen diffusion pushes electrons outwards.
This requires the residual potential to have a minimum in the trail
center. Well beyond the trail, different electron and ion flux
patterns form return fluxes in the background ionosphere. On
approaching the trail, these fluxes are directed inwards which
requires oppositely directed gradients of the residual potential.
This gives rise to the two bumps of the residual potential.

\subsection{Solution of governing equation\label{solution of equations}}

Having solved equations for the residual potential in the near zone
and using formulas of Appendix \ref{far zone}, we can find the
ambipolar electric field and estimate the plasma density
disturbances both in the trail and in the background ionosphere.
Equivalent Eqs.~(\ref{integrodiff}a,b) have singular Cauchy kernels.
While general methods for solving singular integral equations exist
\cite{EstradaKanwal2000,Muskhelishvili}, we are unaware of such
methods for Cauchy-type integrodifferential equations. Even the
existence and uniqueness of the solutions of such equations is not
\textit{a priori} known and should be studied individually for each
specific equation. Nevertheless, for each value of $\rho$,
Eq.~(\ref{integrodiff}) has the unique solution. We will not dwell
here on the proof, but note that the uniqueness of the solution is
provided by the positive sign of $\rho$ [for negative $\rho$,
Eq.~(\ref{integrodiff}) would have infinite number of solutions].

While the equation for $\varphi(\eta)$ has the unique solution, it
is not possible to find it analytically in the general case. In this
section, we discuss approximate solutions of Eq.~(\ref{integrodiff})
for various ranges of $\rho$. We will proceed from the simplest case
to the more complicated ones.

\subsubsection{Late stage diffusion, $\rho\gg1$\label{late stage diffusion}}

The simplest limiting case of $\rho\gg1$ corresponds to a
sufficiently late stage of the meteor trail diffusion, when the peak
trail density exceeds the background not too significantly,
$(N_{\max}-N_0)/N_0\ll \psi^{1/2}/[2(1+\psi)\Theta_0]$
($\psi^{1/2}/\Theta_0 = \nu_{in}/\Omega_i\gg 1$). In this limiting
case, one can solve Eq.~(\ref{integrodiff}b) using a perturbation
technique. For $\rho\gg1$, the integral term in Eq.~(\ref{Form1}) is
small compared to the first term in the RHS. To the zeroth order, we
can neglect the integral term so that the zero-order solution is
given by $\varphi(\eta)\approx\varphi^{(0)}(\eta)$,
\begin{equation}
\varphi^{(0)}(\eta)=\frac{S(\eta)}{\rho}.%
\label{weak}
\end{equation}
In the central region, $|\eta|\ll 1$, the renormalized amplitude of
the electric field, $\partial_\eta\varphi\approx \eta^2/(2\rho)$,
proves to be much less than that for the self-similar solution,
$\partial_\eta\varphi^{SS}=\eta^2/2$.

The physical reason for much smaller residual electric field is as
follows. When electrons leave a sufficiently dense trail, mainly
along the magnetic field, slowly diffusing ions create a significant
ambipolar electric field to retard electrons. If the trail is not
dense then there are enough background electrons to substitute for
those leaving the trail, so that no strong residual electric field
is needed. The integral term in Eq.~(\ref{Form1}), corresponding to
the term $\partial_\eta\varphi$ in Eq.~(\ref{Form2}), describes the
electric field associated with the trail electrons, while the
remaining terms in the left-hand sides (LHSs) include the effect of
background electrons. In the case of $\rho\gg1$, the latter terms
dominate.

\begin{figure}
\includegraphics[width=0.47\textwidth]{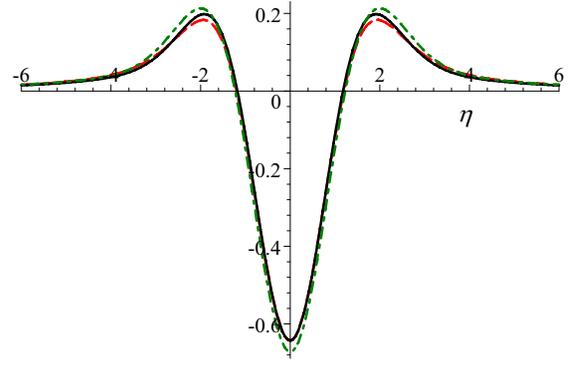}
\caption{\label{comparison1}First-order corrections in $\varphi^{(1)}(\eta)$,
Eqs.~(\ref{weak1}) and (\ref{pero}), multiplied by $-\rho^2$. Solid curve:
$J(\eta)$, Eq.~(\ref{Jexact}); dashed curve: $\tilde{J}=lS(p\eta)$,
Eq.~(\ref{Jprime}); dot-dashed curve: the second term (multiplied by $-\rho^2$)
in the RHS of Eq.~(\ref{pero}) for $\rho\gg 1$.}
\end{figure}

To verify that the neglected integral term is really small, we need to obtain
the next-order approximation. To this end, we substitute the zeroth-order
expression $\varphi^{(0)}(\eta)$ to the integral term in Eq.~(\ref{Form1}).
Using Eqs.~(\ref{J})--(\ref{Jprime}), we obtain the first-order approximation,
\begin{equation}
\varphi^{(1)}(\eta)=\frac{S(\eta)}{\rho} - \frac{J(\eta)}{\rho^2}
\approx\frac{S(\eta)}
{\rho}-\frac{lS(p\eta)}{\rho^2}, %
\label{weak1}
\end{equation}
where $J(\eta)$ is given by Eq.~(\ref{Jexact}) and $l\approx 0.643$,
$p\approx 1.546$, Eq.~(\ref{lp}). Figure \ref{comparison1} shows the
function $J(\eta)$ and its approximate counterpart
$\tilde{J}(\eta)=lS(p\eta)$. The two functions practically perfectly
agree everywhere except the two symmetric maxima. While the
approximate expression in the RHS of Eq.~(\ref{weak1}) is less
accurate, it is much simpler for analysis. Equation (\ref{weak1})
clearly shows that since the second term is small compared to the
first one the perturbation approach employed here is consistent.

As seen from Eq.~(\ref{weak1}), for $\rho\gg 1$ the residual
potential is approximately a linear combination of two $S$-functions
with different arguments. The function $S(\eta)$ has major
qualitative properties of the residual potential shown in
Fig.~\ref{Fig2}(b). One can also check that this function
automatically satisfies the general analytical properties of the
solution given by Eqs.~(\ref{int=0}), (\ref{hypot}), and
(\ref{constr}).

\subsubsection{Intermediate case, $\rho\sim1$\label{intermediate case}}

The perturbation technique of the previous subsection fails for
$\rho\sim 1$, and especially for $\rho\ll 1$. In these cases, we
have found no rigorous analytical solution. However, we have
developed an approximate approach for solving the
integrodifferential equations using analytical fitting with
iterations. This approach is similar to the numerical shooting
method. Assuming a reasonable initial approximation for a zero-order
iteration $\varphi(\eta)\simeq\varphi^{(0)}(\eta)$ with adjustable
parameters, we may put this function into the integral term of one
of the integrodifferential equations and find the next-order
iteration. Comparison between the two iterations at the critical
central region allows us to adjust the parameters in order to find
the best fit.

The two forms of the integrodifferential equation for the near-zone
potential, Eqs.~(\ref{integrodiff}a,b), are equivalent in rigorous
mathematical sense, but this is not so if we seek the
trial-and-error approximate solutions. For the intermediate case of
$\rho\sim 1$, it is more convenient to use Eq.~(\ref{Form1}). This
allows us to disregard the correct asymptotic behavior,
Eq.~(\ref{hypot}), of the initial trial function [this behavior is
required in Eq.~(\ref{Form2}) to eliminate the diverging effect of
the exponentially growing factor in front of the integral].

We start by choosing a simple trial function, $\varphi^{(0)}(\eta)$,
which satisfies the condition (\ref{int=0}) but not necessarily
(\ref{hypot}). Putting it into the integrodifferential term in
Eq.~(\ref{Form1}), we calculate the first-order iteration,
\begin{equation}
\varphi^{(1)}(\eta)=\frac{1}{\rho}\left(  \frac{1}{2\sqrt{\pi}}\ \mathcal{P}%
\!\!\int_{-\infty}^{\infty}\frac{\partial_{y}\varphi^{(0)}(y)}{y-\eta}%
\,e^{-y^{2}/4}dy+S(\eta)\right).%
\label{Phi1}
\end{equation}
It can be readily shown that $\varphi^{(1)}(\eta)$ will
automatically satisfy Eqs.~(\ref{int=0}) and (\ref{hypot}). In
principle, the same procedure could be continued further to
calculate next-order approximations $\varphi^{(2)}(\eta)$,
$\varphi^{(3)}(\eta)$, but because of increasing complexity the
analytical calculation of the integrals is hardly possible.

To make calculations simpler, we choose as the initial function
$\varphi ^{(0)}(\eta)$ the even $\eta$-derivative of the odd
function $-A\eta\exp(-\lambda\eta^{2})$,
\begin{equation}
\varphi^{(0)}(\eta)=A\left(  2\lambda\eta^{2}-1\right)  \exp\left(  -\lambda
\eta^{2}\right).%
\label{maychoose}
\end{equation}
Because the generating function $-A\eta\exp\left(
-\lambda\eta^{2}\right)$ tends to zero as $\eta\rightarrow\infty$,
the trial function with $\phi(\eta)=\varphi^{(1)}(\eta)$
automatically satisfies Eq.~(\ref{int=0}). Besides, in the central
part it qualitatively looks like the expected residual potential
around the major trail.

Now we find $\varphi^{(1)}(\eta)$, Eq.~(\ref{Phi1}). Direct calculation yields
\begin{eqnarray}
\varphi^{(1)}(\eta) &=&\frac{S(\eta)}{\rho}
-\frac{2\lambda A}{\rho\sqrt{1+4\lambda}%
}\nonumber\\
&\times&\!\!\left[  \frac{4\lambda}{1+4\lambda}+\left(
3-2\lambda\eta^{2}\right)
S(\eta\sqrt{1+4\lambda})\right]\!.%
\label{pero}
\end{eqnarray}
The major residual potential is located in the central region, where
it is nearly parabolic. It is natural then to fit the parabolic
behavior of the two functions, $\varphi^{(0)}(0)$ and
$\varphi^{(1)}$, at small $|\eta|$. For $\rho>0.1$, this will
provide a good fit in the entire central region and, due to
Eq.~(\ref{int=0}), a reasonable fit in the nearby region as well.
Fitting the two parabolas,
\begin{equation}
\varphi^{(0)}(0)=\varphi^{(1)}\left(  0\right)  ,\qquad\partial_{\eta\eta}^{2}
\varphi^{(0)}(0)=\partial_{\eta\eta}^{2}\varphi^{(1)}\left(  0\right),%
\label{fittingCa}
\end{equation}
we obtain two equations for the two unknown parameters, $\lambda$
and $A$. Using the Taylor expansion for the function $S(\eta)$ at
small $\eta$, Eq.~(\ref{series}), and similar for
$\varphi^{(0)}\left( \eta\right) $, Eq.~(\ref{maychoose}), we obtain
\begin{subequations}
\label{rhoA}%
\begin{align}
\rho & =\frac{8\lambda^{2}\left(  5+16\lambda\right)  }{\left(  1-6\lambda
\right)  \left(  1+4\lambda\right)  ^{3/2}},\label{rhorho}\\
A  & =\frac{\left(  1-6\lambda\right)  \left(  1+4\lambda\right)  ^{3/2}%
}{2\lambda\left(  3+10\lambda+16\lambda^{2}\right)  }.\label{CCC}%
\end{align}%
\end{subequations}
These expressions give an implicit dependence of the two fitting
parameters, $A$ and $\lambda$, on $\rho$. The parameter $\lambda$
varies in the range from $0$ ($\rho\rightarrow0$,
$A\rightarrow\infty$) to $1/6$ ($\rho\rightarrow\infty$,
$A\rightarrow0$). We cannot explicitly express parameters $A$ and
$\lambda$ in terms of $\rho$. However, we can determine the
asymptotic behavior of $A(\rho)$ and $\lambda(\rho)$ at large and
small values of $\rho$ and construct interpolating formulas that
would provide a reasonable approximation to the exact solution. We
choose the following interpolation formulas,
\begin{subequations}
\label{lamA}%
\begin{align}
\lambda & \approx\lambda_{\mathrm{int}}(\rho)=\left(  \frac{\rho+c\rho^{2}%
}{40+b\rho+d\rho^{2}}\right)  ^{1/2},\label{intLambda}\\
A  & \approx A_{\mathrm{int}}(\rho)=\left(  \frac{10+N\rho}{9\rho+G\rho
^{2}+N\rho^{3}}\right)  ^{1/2},\label{intC}%
\end{align}
\end{subequations}
where $b\approx 122.7$, $c\approx 1.52$, $d\approx 54.74$, $N\approx
14.67$, and $G\approx 29.7$. These approximations deviate from the
exact functions $\lambda(\rho)$ and $A(\rho)$ only several percent
at worst (when $\rho$ is small).

For large $\rho$, the first-order solution $\varphi^{(1)}(\eta)$
matches the corresponding solution obtained in the previous
subsection. Indeed, the major terms in Eqs.~(\ref{weak1}) and
(\ref{pero}) are the same, $S(\eta)/\rho$. Furthermore, for $\rho\gg
1$ we have $\lambda\approx 1/6$ and $A\approx 1/\rho$. In this case,
the smaller second term in the RHS of Eq.~(\ref{pero}) approximately
agrees with the second term in Eq.~(\ref{weak1}) with the largest
mismatch near the two maxima of $\varphi^{(1)}(\eta)$,
$|\eta|\approx 2$. Figure \ref{comparison1} shows that the relative
mismatch there is about 10\% for exact $\varphi^{(1)}(\eta)$ in
Eq.~(\ref{weak1}) and is roughly twice as large for the approximate
expression [in terms of $S(p\eta)$]. In all other locations, the
agreement is much better. Furthermore, because this is the mismatch
between the two minor terms, the relative mismatch between the two
full solutions $\varphi^{(1)}(\eta)$, Eqs.~(\ref{weak1}) and
(\ref{pero}), is $\rho$ times smaller.

\begin{figure*}
\includegraphics[width=0.47\textwidth]{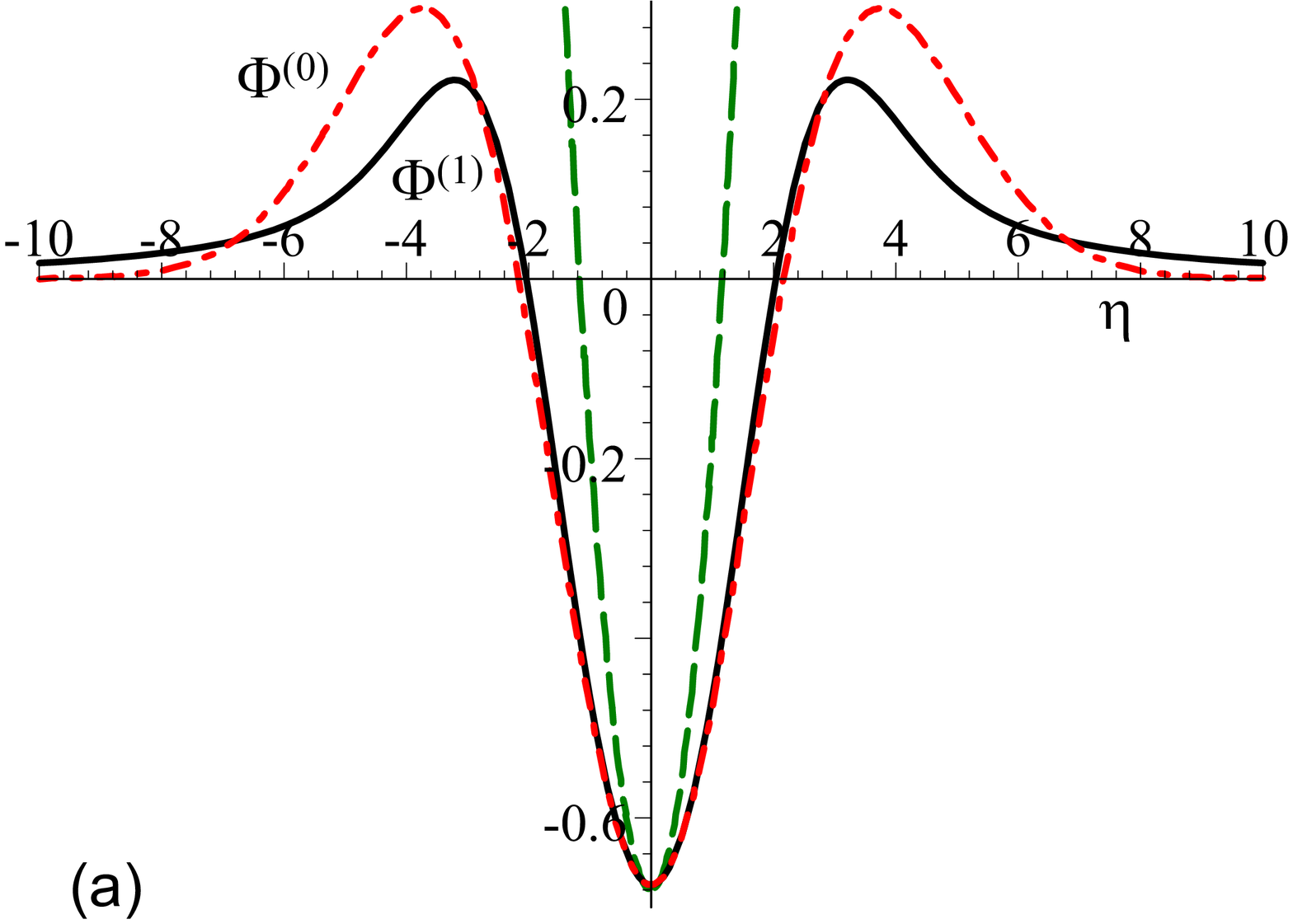} \hfill
\includegraphics[width=0.47\textwidth]{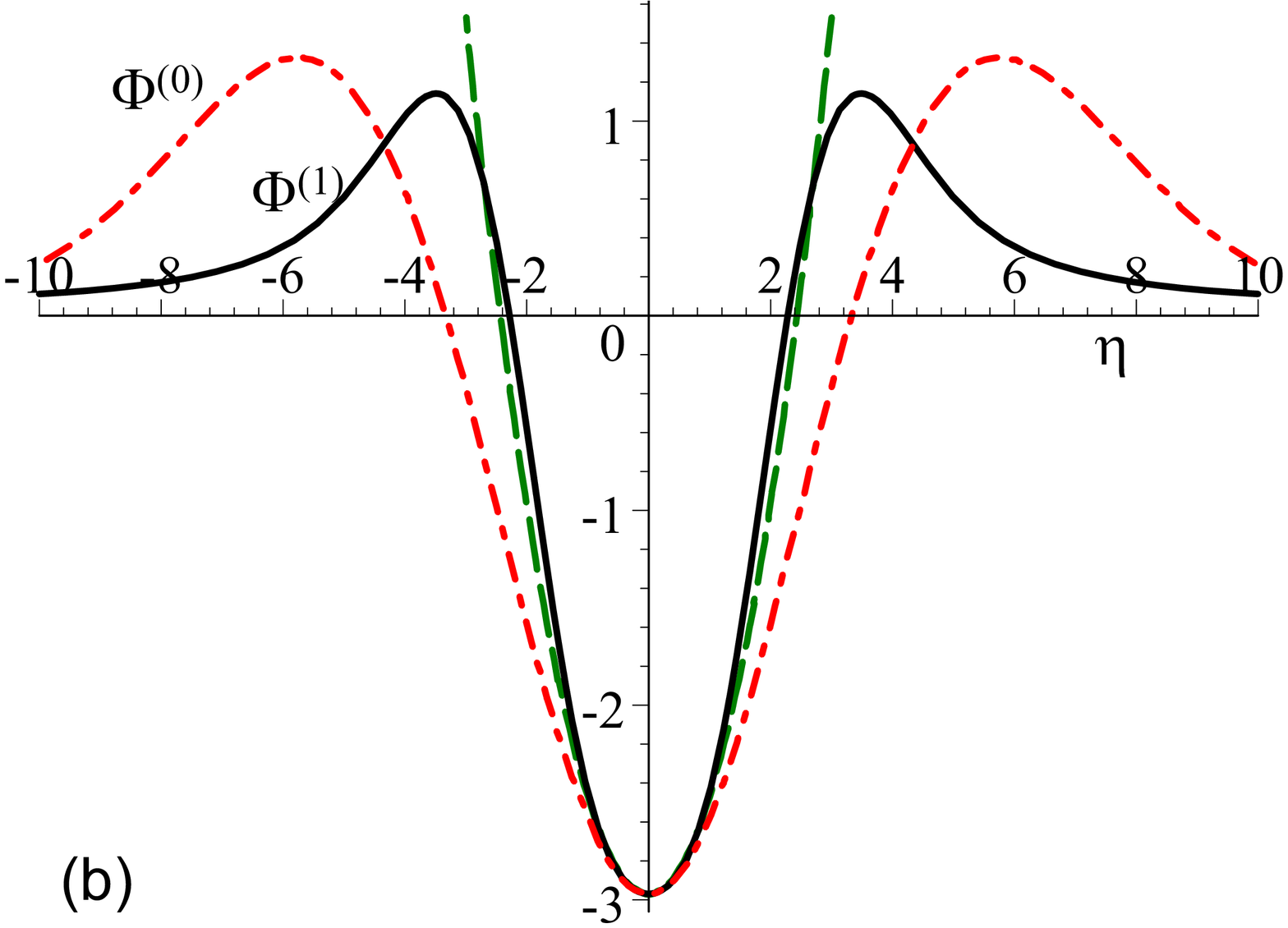}
\caption{\label{approxRho=1}Zero-order, $\varphi^{(0)}(\eta)$, and
first-order, $\varphi^{(1)}(\eta)$, approximations: (a) $\rho=1$
($\lambda=0.107$, $A=0.675$); (b) $\rho=0.1$
($\lambda=4.52\times10^{-2}$, $A=2.97$). The dashed parabolic curves
in the middle show the self-similar solution,
$\varphi^{\mathrm{ss}}(\eta) = \varphi(0)+\eta^2/2$.}
\end{figure*}

Figure~\ref{approxRho=1} shows the fitting solution for $\rho=1$ and
$\rho=0.1$ and the parabolic SSS for the residual potential,
$\varphi^{\mathrm{ss}}(\eta) = \varphi(0)+\eta^2/2$. From
Fig.~\ref{approxRho=1}(a) we see that for $\rho=1$ the two
approximations, $\varphi^{(0)}(\eta)$ and $\varphi^{(1)}(\eta)$, are
reasonably close to each other in the entire central region, but the
self-similar solution deviates significantly from both.
Figure~\ref{approxRho=1}(b) shows that for small $\rho$ the two
approximations start deviating from each other even in the central
region. The reason for this is that the lowest parabolic term
$\propto\eta^2$ in the Taylor expansion for $\varphi^{(1)}(\eta)$
becomes so small that the higher-order terms $\propto\eta^{4}$ start
playing the major role, even at sufficiently small $\eta$. At the
same time, the first-order approximation $\varphi^{(1)}(\eta)$
closely approaches the self-similar solution in the central region,
as is should for small $\rho$ (see the following Section).

\subsubsection{Early stage diffusion, $\rho\ll1$\label{early stage diffusion}}

The limiting case of $\rho\ll1$ deserves special attention because
it usually applies when the meteor trail is first detected. This
case corresponds to the early diffusion stage of a sufficiently
dense plasma trail, when the trail density evolution follows the
SSS, so that $q\approx q_0=\psi/(1+\psi)$, see Eq.~(\ref{q0}) below.
According to Eqs.~(\ref{rho}) and (\ref{gamma}), if $\rho\ll 1$ then
\[
n_{\max}\equiv\frac{N_{\max}}{N_{0}}\gg\frac{\psi^{1/2}}{2\left(
1+\psi\right)  \Theta_{0}}=\frac{\nu_{in}}{2\left(  1+\psi\right)
\Omega_{i} }.
\]
This case is the most difficult case for the analysis. Being unable
to obtain a rigorous solution in the entire range of $\varphi\left(
\eta\right) $, we can construct a reasonable approximation.
Qualitatively, the solution has the same basic form as for
$\rho\gtrsim 1$, viz., a deep potential minimum surrounded by two
symmetric maxima. Smallness of $\rho$ suggests neglecting the
integral term in the RHS of Eq.~(\ref{Form2}), so that we obtain the
approximate expression
\begin{equation}
\varphi\left(  \eta\right)  \approx\varphi\left(  0\right)  +\frac{\eta^{2}}{2}
\label{strong}%
\end{equation}
corresponding to the parabolic SSS with the unknown $\varphi(0)$.
However, this is only valid within a restricted range of $\eta$,
where the exponentially growing factor in front of the integral in
the LHS of Eq.~(\ref{Form2}) is of no importance. As $\rho
\exp(\eta^{2}/4)$ becomes of order unity, i.e., the value of
$|\eta|$ approaches a critical value
$\eta_{\mathrm{0}}\simeq2\sqrt{\ln\left( 1/\rho\right)  }$, the fast
growing exponential factor starts overpowering the small parameter
$\rho$, so that for sufficiently large $\left\vert \eta\right\vert
$, the integral term in the LHS of Eq.~(\ref{Form2}) becomes more
important. The critical coordinates $|\eta|=\eta_{\mathrm{0}}$ are
located slightly beyond the two maxima of the residual potential,
where the parabolically growing potential of the major trail starts
transforming to a decreasing potential beyond the trail.

The parameter $\eta_{\mathrm{0}}$ is of major interest for us
because it determines the rough boundary for the parabolic potential
and hence for the maximum residual electric field responsible for
the electron drift. Because the above heuristic estimate for
$\eta_{\mathrm{0}}$ was based on a simple order-of-magnitude
comparison, we need a more accurate and consistent estimate. We will
construct a piece-wise approximation for $\varphi(\eta)$, which
would be close to the self-similar approximation given by
Eq.~(\ref{strong}) at $|\eta|\lesssim\eta_{\mathrm{0}}$ and
proportional to $\eta^{-2}$, Eq.~(\ref{asympt}), at
$|\eta|>\eta_{\mathrm{0}}$. We will do this by iterations. As a
zero-order approximation, $\varphi^{(0)}(\eta)$, we will set
\begin{equation}
\varphi^{(0)}(\eta)=\left\{
\begin{array}
[c]{ccc}%
\varphi_{0}+\frac{\eta^{2}}{2} & \text{if} & |\eta|<\eta_{0},\\
\left(  \varphi_{0}+\frac{\eta_{0}^{2}}{2}\right) \frac{\eta_{0}^{2}}{\eta^{2}}
& \text{if} & |\eta|>\eta_{0},
\end{array}
\right.
\label{simple}
\end{equation}
with an unknown constant $\varphi_{0}$. The function $\varphi^{(0)}(\eta)$ is a
continuous function, while its derivative,
$\partial_{\eta}\varphi^{(0)}(\eta)$, is discontinuous at $|\eta|=\eta_{0}$. As
a next-order approximation, we will
construct a smooth function $\varphi^{(1)}(\eta)$,%
\begin{equation}
\varphi^{(1)}(\eta)=\left\{
\begin{array}
[c]{ccc}%
\varphi_{0}+\frac{\eta^{2}}{2}+\delta\varphi^{(1)}\left(  \eta\right)   &
\text{if}
& |\eta|<\eta_{0},\\
\left(  \varphi_{0}+\frac{\eta_{0}^{2}}{2}+\delta\varphi^{(1)}\left(  \eta
_{0}\right)  \right)  \frac{\eta_{0}^{2}}{\eta^{2}} & \text{if} & |\eta
|>\eta_{0},
\end{array}
\right.  \label{nextiter}%
\end{equation}
where the deviation from the self-similar solution,
$\delta\varphi^{(1)}(\eta)$, can be found from Eq.~(\ref{F2}) equivalent to
Eq.~(\ref{Form2}). Substituting
$\varphi^{(0)}(\tau)$ for $\varphi(\tau)$ in the integral term, we obtain%
\begin{equation}
\frac{d\delta\varphi^{(1)}}{d\eta}\approx-\ \frac{2\rho e^{\frac{\eta^{2}}{4}}%
}{\pi^{3/2}}\ \mathcal{P}\!\!\int_{-\infty}^{\infty}\frac{\varphi^{(0)}(\tau
)}{\tau-\eta}\,d\tau=-\ \frac{\rho e^{\frac{\eta^{2}}{4}}I(\eta)}{\pi^{3/2}%
},\label{deriva}%
\end{equation}
where%
\begin{eqnarray}
I(\eta) &=&\left(  \eta_{0}^{2}-\eta^{2}\right)
\frac{\eta_{0}^{2}+2\varphi _{0}+\eta^{2}}{\eta^{2}}\ln\left\vert
\frac{\eta_{0}+\eta}{\eta_{0}-\eta }\right\vert\nonumber\\
&-&\frac{2\eta_{0}\left(  \eta_{0}^{2}-\eta^{2}+2\varphi_{0}\right)
}{\eta}.\label{I}%
\end{eqnarray}
The function $\delta\varphi^{(1)}(\eta)$ is negative with the roughly
exponentially growing absolute value. At $|\eta|=\eta_{0}$, we have%
\begin{equation}
\left.  \frac{d\delta\varphi^{(1)}}{d\eta}\right\vert _{\eta=\eta_{0}}%
=\frac{4\rho e^{\frac{\eta_{0}^{2}}{4}}\varphi_{0}}{\pi^{3/2}},\label{ddelF1}%
\end{equation}
which allows us to match the derivative of $\varphi^{(1)}$ at both sides of
$|\eta|=\eta_{0}$,
\begin{equation}
\eta_{0}+\frac{4\rho e^{\frac{\eta_{0}^{2}}{4}}\varphi_{0}}{\pi^{3/2}}%
=-\ \frac{2}{\eta_{0}}\left(  \varphi_{0}+\frac{\eta_{0}^{2}}{2}+\delta\varphi
^{(1)}\left(  \eta_{0}\right)  \right)  .\label{derivconti}%
\end{equation}
Integrating Eq.~(\ref{deriva}), we obtain%
\begin{equation}
\delta\varphi^{(1)}(\eta)=-\ \frac{\rho}{\pi^{3/2}}\int_{0}^{\eta}e^{\frac
{\tau^{2}}{4}}I(\tau)d\tau,\label{ddeltaPhi}%
\end{equation}
where, notwithstanding the singular denominators in $I(\eta)$, the
integrand at $\tau\rightarrow0$ behaves regularly,
$I(\tau)\propto\tau$. Assuming sufficiently large $\eta_{0}^{2}$, to
the first-order accuracy in the $\eta_{0}^{-2}$ expansion, we obtain
\begin{eqnarray}
\delta\varphi^{(1)}(\eta_{0})  &\approx&\frac{8\rho e^{\frac{\eta
_{0}^{2}}{4}}}{\pi^{3/2}\eta_{0}}\left[\left(
1+\frac{6-2\tilde\gamma-4\ln \eta_{0}}{\eta_{0}^{2}}\right)
\varphi_{0}\right.\nonumber\\
&+&\left. 4-2\tilde\gamma-4\ln\eta_{0}\right],%
\label{deltaF}
\end{eqnarray}
where $\tilde\gamma\approx0.5772$ is the Euler constant.
Substituting this to Eq.~(\ref{derivconti}), we obtain the relation
between $\rho$, $\eta_{0}$, and $\varphi_{0}$. To exclude the
unknown constant $\varphi_{0}$ and determine $\rho$ as a function of
$\eta_{0}$, we need one more equation. We will use the general
relation $\int_{0}^{\infty}\varphi\left(  \eta_{0}\right)  d\eta=0$,
Eq.~(\ref{int=0}), which is necessary for consistency of the Laplace
equation in the far zone, see Appendix \ref{far zone}. Integrating
the major parabolic part of the potential at $|\eta|<\eta_{0}$ in
Eq.~(\ref{nextiter}) is trivial, but integrating
$\delta\varphi^{(1)}(\eta)$, Eq.~(\ref{ddeltaPhi}), is complicated.
Note, however, that integrating the major part gives rise to an
additional large factor $\sim\eta_{0}$ and every next integration of
$\delta \varphi^{(1)}(\eta)$\ gives rise to an additional reduction
factor $\sim\eta _{0}^{-2}$. While the derivative of
$\delta\varphi^{(1)}$ at $|\eta|=\eta_{0}$, Eq.~(\ref{deriva}), is
comparable to $\eta_{0}$, the relative contributions of
$\delta\varphi^{(1)}(\eta_{0})$, and especially of its integral, are
small compared to those from the major part. bearing this in mind,
we obtain to the first-order accuracy with respect to
$\eta_{0}^{-2}$,
\begin{subequations}
\label{at_eta0}%
\begin{align}
\rho &  \approx\frac{\pi^{3/2}e^{-\ \frac{\eta_{0}^{2}}{4}}}{\eta_{0}}\left(
1+\frac{2}{\eta_{0}^{2}}\right)  ,\label{rho_eta0}\\
\varphi_{0} &  \approx-\ \frac{\eta_{0}^{2}}{3}+\frac{4}{3},\label{Phi_eta0}%
\end{align}
\end{subequations}
Application of other general relations, Eq.~(\ref{constr}), instead
of Eq.~(\ref{int=0}), yields somewhat different values of the
first-order corrections. In Eq.~(\ref{rho_eta0}), we will disregard
these corrections and obtain the solution of $\eta_{0}(\rho)$ in
terms of the Lambert W-function, $W(x)$ \footnote{E.~W. Weisstein,
"Lambert W-Function," from MathWorld-A Wolfram Web Resource,
http://mathworld.wolfram.com/LambertW-Function.html}:
\begin{equation}
\eta_{0}=\left[  2W\left(  \frac{\pi^{3}}{2\rho^{2}}\right) \right]
^{1/2}\label{Lambert}%
\end{equation}
To logarithmic accuracy, we obtain
\begin{equation}
\eta_{0}\simeq2\left\{  \ln\left[  \frac{\pi^{3/2}}{2\rho\sqrt{\ln\left(
\pi^{3/2}/\rho\right)  }}\right]  \right\}  ^{1/2}\simeq2\left(  \ln\frac
{1}{\rho}\right)  ^{1/2},%
\label{eta0(rho)}
\end{equation}
in good agreement with the above heuristic estimate.

\subsection{Trail diffusion\label{trail diffusion}}

Given the approximate expressions for the residual potential, we
will solve now the problem of trail diffusion described by
Eqs.~(\ref{eq2}) or (\ref{Equ1}). While this treatment will require
a number of approximations, the comparison with simulations
described in Sect.~\ref{comparison of theory} shows that the
analytical theory developed below is valid to a good accuracy.

For the strictly perpendicular case, $\theta=90^{\circ}$, the
self-similar solution (SSS) obtained in Appendix \ref{self-similar
solution} is given by
\begin{eqnarray}
n^{\mathrm{ss}}(\xi,\zeta,t)&\equiv&\frac{N^{\mathrm{ss}}}{N_0}
\approx\frac{n_{\mathrm{lin}}}{4\pi Dt}
\left(\frac{1+\psi}{\psi}\right)^{1/2}\nonumber\\
&\times&\exp\left[ -\ \frac{1}{4}\left(
\frac{1+\psi }{\psi}\ \xi ^{2}+\zeta ^{2}\right) \right],%
\label{SelfSo}
\end{eqnarray}
where $n_{\mathrm{lin}}\equiv N_{\mathrm{lin}}/N_0$ is the trail
line density along the axis $z$ normalized to the background plasma
density. Under assumption that the initial maximum density is well
above the background plasma density, $N_{\max}\gg N_0$, the
self-similar Gaussian profile can be used as a reasonable initial
condition for the density disturbance $\Delta n(t_0)=n(t_0)-1$, as
we did in our simulations,
\begin{eqnarray}
\Delta n(t_0)\approx n^{\mathrm{ss}}(t_0)&=&\Delta n_{0}\,\exp
\left[ -\ \frac{1}{4}\left( \frac{1+\psi }{\psi }\ \xi ^{2}+\zeta
^{2}\right)
\right], \nonumber\\
\Delta n_{0}&=&\frac{n_{\mathrm{lin}}}{4\pi
Dt_0}\left(\frac{1+\psi}{\psi}\right)^{1/2}. %
\label{nSS}
\end{eqnarray}%
Our numerical computations, starting from the initial condition $n_{\mathrm{%
Trail}}\left( t_{0}\right) =n^{\mathrm{ss}}(t_{0})$ with $n_{0}\gg
1$, show that, for some time, the trail density approximately
follows the SSS. As the peak density decreases with increasing time,
the solution starts to gradually deviate from the SSS. However,
within the  trail density peak, it keeps the nearly Gaussian form,
\begin{equation}
n_{\mathrm{Trail}}=\frac{\Delta
n_{0}t_{0}}{t}\sqrt{\frac{q_{0}}{q(t)}}\,\exp \left[-\
\frac{1}{4}\left( \frac{\xi ^{2}}{q(t)}+\zeta ^{2}\right) \right],%
\label{Gau}
\end{equation}
where the characteristic diffusion scale in the $x$ direction ($\xi
=x/\sqrt{Dt}$) is determined by a time-dependent coefficient $q(t)$
with the SSS initial condition,
\begin{equation}
q(t_{0})=q_{0}\equiv \frac{\psi }{1+\psi }.%
\label{q0}
\end{equation}
In the original coordinates $x$ and $y$, the nearly Gaussian peak can be
expressed in terms of the $x,y$-dispersions, $\sigma_{x,y}$,
\begin{subequations}
\label{sigmae}
\begin{eqnarray}
\sigma _{x}(t)&=&\left\langle x^{2}(t)\right\rangle\\%
\label{sigmax}%
&\equiv& \left. \int_{-\infty }^{\infty
}x^{2}n_{\mathrm{Trail}}(x,0,t)dx\right/ \int_{-\infty }^{\infty
}n_{\mathrm{Trail}}(x,0,t)dx,\nonumber%
\end{eqnarray}
\begin{eqnarray}
\sigma _{y}(t)&=&\left\langle y^{2}(t)\right\rangle\\%
\label{sigmay}%
&\equiv& \left. \int_{-\infty }^{\infty
}y^{2}n_{\mathrm{Trail}}(0,y,t)dy\right/ \int_{-\infty }^{\infty
}n_{\mathrm{Trail}}(0,y,t)dy,\nonumber%
\end{eqnarray}
\end{subequations}
as
\begin{equation}
n_{\mathrm{Trail}}= \frac{n_{\mathrm{lin}}}{2\pi\sigma_x\sigma_y}\,
\exp\left(-\,\frac{x^2}{2\sigma_x^2} - \frac{y^2}{2\sigma_y^2}
\right).
\label{insigmas}
\end{equation}
The small residual electric field in the $y$ direction (i.e., along
$\vec{B}_{0}$) practically does not affect the SSS,
$\sigma_y^2=2Dt$, because the parallel residual electric field is
small, so that the diffusion along $\vec{B}_{0}$ is determined by
the isotropic diffusion rate $D$, Eq.~(\ref{D}). At the same time,
the strong residual electric field along $x$ affects the diffusion
in this direction, especially at altitudes above 93-97 ($\psi
\lesssim 1$). The deviation of the characteristic trail size from
that determined by isotropic diffusion is described in
Eq.~(\ref{Gau}) by the coefficient
\begin{equation}
q(t)=\frac{\sigma _{x}^{2}(t)}{2Dt}. %
\label{qusig}
\end{equation}
The coefficient $q(t)$ slowly grows with time as the trail diffuses
faster than it does in the SSS. As the electric field falls well
below that predicted by the SSS (see Sec.~\ref{solution of
equations}), the diffusion becomes nearly isotropic, $q(t)\simeq 1$,
determined in the two directions roughly by the same isotropic
diffusion rate $D$.

The coefficient $q(t)$ is determined by the residual electric
potential in the near zone. In the central part of the near zone,
the potential along $\xi $, $\phi _{\mathrm{res}}^0(\xi)$, always
has a nearly parabolic behavior similar to the SSS,
Eq.~(\ref{phiSSanz}),
\begin{equation}
\phi _{\mathrm{res}}^0-\phi _{\mathrm{res}}^0(0)\approx \frac{B_{xx}\xi ^{2}}{4}.%
\label{paraB}
\end{equation}
Unlike Eq.~(\ref{SSsolution}), however, the coefficient $B_{xx}$ in
Eq.~(\ref{paraB}) is no longer a constant but changes with time. Our
simulations show that the central region described by the nearly
parabolic $\xi$-dependence is broad enough to include the entire
trail peak. It is the parabolic behavior of the potential that
imposes the Gaussian shape of the trail.

Substituting Eqs.~(\ref{Gau}) and (\ref{paraB}) into the continuity
equation (\ref{Equ1}), we obtain a differential equation for $q(t)$,
\begin{equation}
t\ \frac{dq}{dt}=-\left( 1+B_{xx}\right) q+1,%
\label{dPdT}
\end{equation}
which is not yet closed because the parabolic coefficient $B_{xx}$
is itself an unknown function of $q(t)$. To determine it, we will
use the solutions for the residual potential found in
Sect.~\ref{solution of equations}. These solutions are expressed in
terms of renormalized variables $\varphi $ and $\eta$ defined by
Eqs.~(\ref{Phieta}). In these variables, the parabolic behavior of
Eq.~(\ref{paraB}) is described by
\begin{equation}
\varphi(\eta) - \varphi(0)\approx \frac{F\eta^{2}}{2}. %
\label{viaF}
\end{equation}
Here the coefficient $F(\rho)$ is similar to $B_{xx}$. Unlike the
latter, it depends upon one parameter $\rho$ related to $q$ by
Eq.~(\ref{rho}), $\rho (t) =\gamma q(t)t$. Comparing
Eqs.~(\ref{paraB}) and (\ref{viaF}) using (\ref{Phieta}) yields the
relation $B_{xx}q=F(\rho )/(1+\psi)$, which reduces Eq.~(\ref{dPdT})
to a closed nonlinear differential equation for $\rho (t)$,
\begin{equation}
\frac{1}{\gamma }\frac{d\rho }{dt}=-\ \frac{F(\rho )}{1+\psi }+1.%
\label{drot}
\end{equation}
Integrating Eq.~(\ref{drot}) for $t(\rho)$ with the initial
condition at $t=t_0$ given by $\rho (t_{0})\approx\rho
_{0}\equiv\gamma q_{0}t_{0}$,
\begin{equation}
\rho _{0}= \frac{\gamma{\psi}t_0}{1+\psi}=\frac{\sqrt{\psi
}}{2(1+\psi)
\Theta_{0}n_{0}}\ll 1,%
\label{rho00}
\end{equation}
Eq.~(\ref{rho0}), we obtain for the general $F(\rho)$:
\begin{equation}
\gamma ( t-t_{0}) =\int_{\rho _{0}}^{\rho }\frac{d\varrho }{1-F(\varrho
)/(1+\psi )}.%
\label{solo1}
\end{equation}
During the initial time when $\rho (t)\ll 1$, both the trail density
and the parabolic approximation of the potential follow the SSS,
$B_{xx}\approx 1/\psi$, $q\approx q_0$, $F\approx 1$,
\begin{equation}
\qquad \rho (t)\approx \frac{\gamma \psi t}{1+\psi }.
\label{initial}
\end{equation}
Using this, in Eq.~(\ref{solo1}) we can set $F=1$ for $t\leq t_{0}$
and then replace $t_{0}$ and $\rho _{0}$ by zeroes. This corresponds
to diffusion of the initially infinitely dense and thin
($\delta$-function like) trail. The solution of $\rho (t)$ is
determined by reversing Eq.~(\ref{solo1}). The RHS of
Eq.~(\ref{solo1}) is always positive, so that this reversal is
unambiguous.

In the course of the trail diffusion, when the parameter $\rho$
becomes of order unity and greater, the monotonically decreasing
function $F(\rho )$ starts deviating from unity. To obtain the
explicit expression for $F(\rho)=(1/\eta)(d\varphi/d\eta)$, we will
use the solutions found in see Sect.~\ref{solution of equations}.
While we have no simple analytical expression for $F(\rho )$ for all
$\rho $, we can approximate it by interpolating between $F\approx 1$
for $\rho \ll 1$, Eq.~(\ref{strong}), and $F\approx 1/\rho$ at $\rho
\gg 1$, Eq.~(\ref{weak}). The simplest monotonic interpolation is
\begin{equation}
F(\varrho )\approx \frac{1}{1+\varrho }.  \label{interF}
\end{equation}%
Substituting this into Eq.~(\ref{solo1}) allows us to obtain the
solution for $t(\rho)$:
\begin{equation}
\gamma t= \rho +\frac{1}{1+\psi }\,\ln \left( 1+\frac{1+\psi
}{\psi }\,\rho \right).%
\label{tinterp}
\end{equation}%
Reversing this relation, we obtain an explicit expression for $\rho
(t)$ in terms of the Lambert $W$-function,
\begin{equation}
\rho (t)= \frac{W\left( \psi e^{\left( 1+\psi \right) \gamma t+\psi
}\right) -\psi }{1+\psi }.%
\label{rhotinterp}
\end{equation}%
This general expression describes the transition from the SSS for
$\gamma t\ll 1$, Eq.~(\ref{initial}), to
\begin{equation}
\rho (t) \approx \gamma t-\frac{1}{1+\psi }\ln \left( \frac{1+\psi
}{\psi }\, \gamma t\right)%
\label{appr}
\end{equation}
at large $\gamma t$. According to Eqs.~(\ref{rho}), (\ref{sigmax}),
and (\ref{qusig}),
\begin{equation}
\sigma _{x}^{2}(t)=\frac{2D\rho (t)}{\gamma }=\frac{ \sqrt{1+\psi
}\,\Theta
_{0}N_{\mathrm{lin}}}{\pi N_{0}}\,\rho (t),%
\label{rhosig}
\end{equation}
so that the first term on the RHSs of Eqs.~(\ref{appr}) corresponds
to diffusion over $x$ with the isotropic rate $D$, while the second
logarithmic term describes a small time delay. Since the trail
diffusion along $y$ has the same rate, $\sigma_y^2\approx 2Dt$, the
above expressions predict an evolution from anisotropic to isotropic
diffusion.

The diffusion starts changing its character from the self-similar,
sharply anisotropic diffusion to the nearly isotropic one roughly at
a time $t=t_{\mathrm{cr}}$ when $\rho (t_{\mathrm{cr}})=1$.
According to Eqs.~(\ref{gamma}) and (\ref{tinterp}), the
corresponding critical time $t_{\mathrm{cr}}$ is given by
\begin{eqnarray}
t_{\mathrm{cr}} & =&\frac{N_{\mathrm{lin}}eB_{0}\mathcal{K}(\psi
)}{2\pi \left( T_{e}+T_{i}\right) N_{0}}\approx9.235\times
10^{-2}\mathrm{s}\left( \frac{B_{0}}{5\times 10^{4}\mathrm{nT}}\right)   \notag \\
&\times& \left(
\frac{N_{\mathrm{lin}}}{10^{14}\mathrm{m}^{-1}}\right) \left(
\frac{10^{11}\mathrm{m}^{-3}}{N_{0}}\right) \left(
\frac{1000\mathrm{K }}{T_{e}+T_{i}}\right) \mathcal{K}(\psi ),%
\label{tcri}
\end{eqnarray}%
where
\begin{equation}
\mathcal{K}(\psi )=\sqrt{\frac{\psi }{1+\psi }}\left[ 1+\psi +\ln \left(
\frac{1+2\psi }{\psi }\right) \right]   \label{K(psi)}
\end{equation}%
The critical transition time from anisotropic to isotropic diffusion
is mostly relevant for high altitudes with $\psi\ll 1 $, where the
anisotropy of diffusion is clearly pronounced. In this case, the
function $\mathcal{K}(\psi )\approx \sqrt{\psi } [1+\ln (1/\psi )]$.
According to Eq.~(\ref{psi}), $\sqrt{\psi }\propto N_nB_{0}^{-1}$.
From Eq.~(\ref{tcri}) we see that $t_{\mathrm{cr}}$ is practically
independent of the magnetic field, making the high and low latitude
diffusion evolve similarly. The altitudinal dependence of
$t_{\mathrm{cr}}$, however, depends strongly on the background
plasma density $N_0$ and even more so on the neutral density $N_n$.

\subsection{Density disturbances beyond the trail\label{density disturbances}}

The residual electric potential originated within the trail and
extended along the magnetic density creates disturbances in the
background plasma, as we observed in our numerical calculations, see
Sect.~\ref{numerical analysis}, Feature 3. These density
disturbances occur due to the fact that the extended residual
potential attracts ions from the surrounding ionosphere into the
central part with the deep potential minimum around $x=0$. This
results in plasma compression in the central potential minimum and
depletion in the adjacent regions where the residual potential has
two symmetric maxima.

In this section, we will estimate density disturbances in the
background plasma beyond the trail. One motivation for this is that
when obtaining the equation for the residual potential in the near
zone, Eq.~(\ref{integrodiff}), we have completely neglected these
disturbances. This has allowed us to reduce the equation for the
residual potential in the far region to the Laplace equation in
renormalized coordinates, see Eq.~(\ref{Laplace}) in
Appendix~\ref{far zone}. To estimate the effect of density
disturbances, we will consider the strongest case of dense trail
described in Sect.~\ref{early stage diffusion}. Another motivation
is to find a proper adjustment for the parameter $\rho$, which is
needed for the residual potential at sufficiently large state of
trail diffusion.

\subsubsection{Strongest case, $\rho\ll 1$: justification of analytical
approach\label{justi}}

To justify the neglect of density disturbances beyond the trail, see
Appendix~\ref{far zone}, we will make a simple analytical estimate
of density disturbances beyond the trail. The largest disturbances
beyond the trail occur in the near zone in the early stage of
dense-trail diffusion, $\rho\ll 1$.

Given the spatial distribution of the residual potential
$\phi_{\mathrm{res}}$, the dynamics of density perturbations is
described by Eq.~(\ref{eq2}). If we assumed a stationary regime,
$\partial_{t}n=0$, then imposing boundary conditions at infinity
where $n\rightarrow1$ and $\phi_{\mathrm{res}}^0\rightarrow0$ we
would obtain the local Boltzmann distribution of plasma density,
$n=\exp(-\phi_{\mathrm{res}}^0)$. In the strongest case of
$\rho\ll1$, according to Eq.~(\ref{Phi_eta0}), the potential minimum
is $\varphi_0\simeq-\eta_0^2/3$, so that according to
Eq.~(\ref{Phi}) the corresponding potential
$\phi_{\mathrm{res}}\approx
\phi_{\mathrm{res}}^0=\varphi_0/2(1+\psi)\simeq-\eta_{0}^{2}/[6(1+\psi)]$.
For sufficiently small $\rho$, the parameter $\eta_{0}$ depends
logarithmically on $\rho$ and may reach several units,
Eq.~(\ref{eta0(rho)}). In this case, the local Boltzmann
distribution would yield disturbances of the background plasma near
the trail much greater than the background density itself, $n\sim
1/\rho\gg 1$, making the assumption underlying our theoretical
approach to be invalid. Fortunately, our numerical simulations, even
in the cases as strong as $n_{\max}=10^4$, have shown that the
background density disturbances are at worst of order unity in the
central near zone and are much less beyond it, as described in
Sect.~\ref{numerical analysis}. The physical reason for this is that
the trail diffusion is a non-stationary process with a diffusion
prehistory. Plasma density disturbances caused by the potential
expanding with time prove to be noticeably smaller than those caused
by the stationary potential.

For a simple estimate of plasma density disturbances outside the
trail in the non-stationary process, we linearize Eq.~(\ref{eq2})
for relatively small density disturbances, $\delta n=\delta
N/N_{0}\equiv n-1$. In accord with our theory and simulations, we
also assume that the major gradients beyond the trail are
perpendicular to the magnetic field, i.e., are directed mainly along
the $x$ coordinate. As a result, we obtain from Eq.~(\ref{eq2}) a
simpler equation,
\begin{equation}
\partial_{t}\delta n-D\partial_{xx}^{2}\delta n=D\partial_{xx}^{2}%
\phi_{\mathrm{res}}\left(  x,t\right)  .%
\label{urva}
\end{equation}
Solving this linear equation via the proper Green function and
performing integration by parts, we obtain
\begin{align}
\delta n(x,t) &  =\sqrt{\frac{1}{16\pi
D}}\int_{t_{0}}^{t}\frac{d\tilde{t}}{\left(
t-\tilde{t}\right)  ^{3/2}}\nonumber\\
&  \times\!\int_{-\infty}^{\infty}(\tilde{x}-x) \exp\left[ -\
\frac{(\tilde{x}-x)^{2}}{4D(t-\tilde{t}) }\right]
\partial_{\tilde{x}}\phi_{\mathrm{res}}(\tilde{x},\tilde{t})
d\tilde{x}\,.%
\label{SolGreen}
\end{align}
This expression shows that general density disturbances are
determined by the entire distribution of the residual electric field
over $\tilde x$ from all previous times $\tilde t$. It is
straightforward to check that for stationary $\phi _{\mathrm{res}}$
and $t\gg t_{0}$, Eq.~(\ref{SolGreen}) reduces to dependence $\delta
n(x,t)=-\phi_{\mathrm{res}}(x,t)$ corresponding to the local
Boltzmann distribution for $\phi_{\mathrm{res}}\ll 1$.

Before applying a specific model for
$\partial_{\tilde{x}}\phi_{\mathrm{res}}(\tilde{x},\tilde{t})$ we
note the following. When the local coordinate $\tilde{x}$ varies
from $0$ to $\infty$, the residual electric field
$-\partial_{\tilde{x}}\phi_{\mathrm{res}}(\tilde{x},\tilde{t}) $
changes its sign. In the central region it draws plasma to the
center, while outside the central region it repels it. At any
previous time $\tilde{t}<t$, the local density disturbances are
determined by integral contributions from the two competing regions
that generally do not balance each other. In the absence of the
exponential factor in the RHS of Eq.~(\ref{SolGreen}), according to
Eq.~(\ref{Exdx=0}), the two contributions would exactly compensate
each other. The exponential factor, however, breaks the balance. For
example, in the center, $x=0$, the attraction always dominates, so
that the total density disturbance there is positive (plasma
compression). In the adjacent regions, the situation is opposite, so
that the repulsion dominates there, resulting in plasma depletion.

The strongest density disturbances are within the near zone, where
the potential reaches its absolute minimum. For simple estimate of
the potential $\varphi$ in the early stage of trail diffusion,
$\rho\ll 1$, we choose its zero-order approximation,
$\varphi^{(0)}$, Eq.~(\ref{simple}), so that
$\phi_{\mathrm{res}}\approx
\phi_{\mathrm{res}}^0\approx\varphi^{(0)}/2(1+\psi)$. Recalling the
definition of the renormalized coordinate $\eta$, Eq.~(\ref{eta}),
and renormalizing the local time variable $\tilde{t}$, we obtain
\begin{equation}
\partial_{\tilde{x}}\phi_{\mathrm{res}}\left(  \tilde{x},\tilde{t}\right)
=\frac{1}{2\psi}\times\left\{
\begin{array}
[c]{ccc}%
\frac{\omega}{\sqrt{\tau}} & \text{if} & \left\vert \frac{\omega}%
{\sqrt{\tau}}\right\vert <\eta_{1}\\
-\ \frac{\eta_{1}^{4}}{3}\left(  \frac{\sqrt{\tau}}{\omega}\right)  ^{3} &
\text{if} & \left\vert \frac{\omega}{\sqrt{\tau} }\right\vert >\eta_{1}
\end{array}
\right.,%
\label{model}
\end{equation}
where%
\begin{equation}
\tau\equiv\frac{\tilde{t}}{t},\qquad\eta_{1}\equiv\sqrt{\frac{\psi}{1+\psi}}
\,\eta_{0}.%
\label{sigmu}
\end{equation}
We restrict our analytic estimate to the center, $x=0$, where we
expect the largest positive density disturbances, and assume $t\gg
t_{0}$. As a result, performing the integration over space, we
obtain from Eqs.~(\ref{SolGreen}) and (\ref{model})
\begin{equation}
\delta n(0,t)\approx\frac{1}{8\psi\sqrt{\pi}}\int_{0}^{1}\left(  I_{1}%
-I_{2}\right)  d\tau,\label{cherezII}%
\end{equation}
where
\begin{subequations}
\label{II}%
\begin{eqnarray}
I_{1} &  =&\frac{2\sqrt{\pi}}{\tau}\operatorname{erf}\left(
\frac{\eta _{1}\sqrt{\tau}}{2\sqrt{1-\tau}}\right)\nonumber\\
&-&\frac{2\eta_{1}}{\sqrt {\tau\left( 1-\tau\right)  }}\exp\left(
-\ \frac{\eta_{1}^{2}\tau
}{4\left(  1-\tau\right)  }\right) ,\label{I1}\\
I_{2} &  =&\frac{\eta_{1}^{4}\tau\sqrt{\pi}}{6\left(  1-\tau\right)
^{2} }\left[  \operatorname{erf}\left(
\frac{\eta_{1}\sqrt{\tau}}{2\sqrt {1-\tau}}\right)
-1\right]\nonumber\\
&+&\frac{\eta_{1}^{3}\tau^{1/2}}{3\left( 1-\tau\right)
^{3/2}}\exp\left(  -\ \frac{\eta_{1}^{2}\tau}{4\left(
1-\tau\right)  }\right)  \label{I2}%
\end{eqnarray}
\end{subequations}
are positive functions of $\tau$ and $I_{2}(\tau)<I_{1}(\tau)$.

For small $\eta_{1}$, which for $\eta_{0}\sim3$--$5$ is possible
only at sufficiently high altitudes where $\psi\ll\eta_{0}^{-2}$,
the two functions are mainly localized near $\tau=1$
($\tilde{t}\simeq t$). As a result, we obtain
$\int_{0}^{1}I_{1}d\tau\approx\eta_{1}^{2}\sqrt{\pi}/2$ and
$\int_{0}^{1} I_{2}d\tau\approx\eta_{1}^{2}\sqrt{\pi}/3$, so that in
this case $\delta n(0,t)\simeq\eta_{0}^{2}/50$. Under real physical
conditions, the value of $\eta_{0}$ can reach several units, so that
the maximum density disturbance of the background plasma in the near
zone is of order unity. The largest value of density disturbances
observed in our simulations was $\delta n(0,t)\simeq 2$, which is
larger than those following from the linearized Eq.~(\ref{urva}) but
is of the same order of magnitude.

At lower altitudes, $\psi\gtrsim1$, according to Eq.~(\ref{sigmu}),
the value of $\eta_{1}$ is not small but is always less than
$\eta_{0}$. As the value of $\eta _{1}>1$ grows, the maxima of
$I_{1,2}$ shift to smaller values of $\tau$. This means that the
density disturbances at a given time $t$ become more affected by
electric fields at past time, $\tilde{t}<t$. In addition, the ratio
$I_{2}/I_{1}$ becomes smaller, which means that the effect of the
oppositely directed electric field beyond the major region of the
potential distribution, $\omega/\sqrt{\tau}>\eta_{1}$,
Eq.~(\ref{model}), becomes less important. For $\eta_{1}\gg1$,
$\int_{0}^{1}I_{1}d\tau\simeq4\sqrt{\pi}\ln\eta_{1}$. The value of
$\eta_{1}<\eta_{0}$, however, can reach 3--4 as a maximum (at
sufficiently large $\psi$ corresponding to lower altitudes). In this
case, the
integral $\int_{0}^{1}\left(  I_{1}-I_{2}\right)  d\tau\lesssim\int_{0}%
^{1}I_{1}d\tau$ can reach several units at most, so that $\delta
n(0,t)$ should remain small.

These simple estimates confirm our numerical observations that even
in the strongest cases the relative plasma density disturbances in
the near zone are of order unity at worst, see
Fig.~\ref{BackgrDensity}. In all other situations or locations, such
as for $\rho\gtrsim 1$, $t\sim t_{0}$, etc., and especially in the
far zone where the major current closure takes place, the density
disturbances are much smaller. This shows that our basic theoretical
scheme which relies on the Laplace equation for the electric
potential in the far zone and neglects density disturbances is a
reasonable approximation.

\subsubsection{Weakest case, $\rho\gg 1$: adjustment of $\rho$ for residual
potential\label{weakest case}}

The simulations show that Eq.~(\ref{rhotinterp}) describes well the
evolution of the simulated trail density for all parameters $\rho$,
while the residual potential solution described in
Sect.~\ref{solution of equations} shows a discrepancy for the late
diffusion stage when $\rho$ is large. To fit the evolution of the
simulated potential, the expression for the master parameter $\rho$
given by Eq.~(\ref{rhotinterp}) needs an adjustment
$\rho(t)\rightarrow\tilde\rho(t)$ corresponding to an effective time
lag. We argue that small background density disturbances beyond the
trail are responsible for this discrepancy.

To estimate background plasma density disturbances in the later
stage of trail diffusion, we will use the zeroth-order approximation
for the residual potential given by Eq.~(\ref{maychoose}). In the
limit of $\rho\gg 1$, when $\rho(t)\simeq\gamma t$, we have in the
near zone
\begin{equation}
\phi_{\mathrm{res}}^0(x,t)\simeq\frac{I(x/\sqrt{Dt})}{2(1+\psi)\rho(t)},%
\label{fio}
\end{equation}
where
\[
I(z)=-\ e^{-\ \frac{z^{2}}{6}}\left(1-\frac{z^{2}}{3}\right) .
\]%
Substituting Eq.~(\ref{fio}) for $x=0$ into Eq.~(\ref{SolGreen}), we
obtain
\begin{equation}
\delta n(x,t)\simeq\frac{1}{2(1+\psi)\rho(t)}.%
\label{dnsmall}
\end{equation}
Thus for $\rho\gg 1$ the background density disturbances are small,
justifying the use of Eqs.~(\ref{urva}) and (\ref{SolGreen}).

We will use Eq.~(\ref{dnsmall}) to reconcile the discrepancy
described above. One of the key factors in the derivation of the
governing equation for the near-zone potential,
Eq.~(\ref{integrodiff}), has been the calculation of the density
disturbance integral over the effective $\zeta$-region in the near
zone, $\Delta N_{\zeta_1}$, as described in Appendix~\ref{near-zone
potential}. This effective region includes the trail with a nearly
constant $\varphi_{\mathrm{res}}$ and an adjacent region where a
noticeable $\zeta$-derivative of $\varphi_{\mathrm{res}}$ builds up,
as seen in Fig.~\ref{PotenDens} at $y\simeq -10$. This derivative
defines the fields and current closure structure in the far zone.
The effective integral $\Delta N_{\zeta_1}$ should include both the
trail density and the disturbances of the background plasma. The
calculation of Appendix~\ref{near-zone potential} took into account
only the former and completely neglected the latter. For a dense
trail with $\rho\ll 1$, this proves to be justified because the
relative contribution of the background plasma disturbances to the
integral proves to be small compared to the contribution of the
trail density. On the contrary, for a much less dense trail with
$\rho\gg 1$, the contribution of background density disturbances
becomes comparable to that from the trail density.

In Appendix~\ref{effective integral}, we have obtained the effective
integral $\Delta N_{\zeta_1}(\xi)$ and the relation between
$\tilde\rho$ and $\rho$ in terms of two coupled parameters of order
unity, $\beta(\rho)$ and $\tilde\beta(\rho)$, Eqs.~(\ref{B}) and
(\ref{tilderho}). Being unable to obtain these parameters
analytically, we have used our simulations for $\psi\geq 0.05$ to
obtain approximate Eq.~(\ref{betaemp}). For smaller $\psi$, we
proposed Eq.~(\ref{betaemp1}), which represents a conjecture and
needs a special study.

\section{Summary of analytical results\label{Summary of analytical results}}

In this Section, we summarize our major analytical results which can
be directly applied in comparisons with observations. To use the
theoretical results for practical applications, we need to pass from
the renormalized variables of Sect.~\ref{analytical treatment} to
the original ones. Where appropriate, we will use simplified
versions of analytic expressions.

The original residual potential $\phi_{\textrm{res}}(x,y,t)$ is
defined in Eq.~(\ref{phiadd}). According to Eq.~(\ref{Phieta}), the
original residual potential in the near zone,
$\phi_{\textrm{res}}^{0}(x,t)\equiv \phi_{\textrm{res}}(x,0,t)$, in
terms of the actual coordinate $x$ and time $t$, is given by
\begin{equation}
\phi_{\textrm{res}}^{0}(x,t) =
\frac{1}{2(1+\psi)}\,\varphi\left(\sqrt{\frac{\gamma}{D\rho(t)}}\,\,x,\,
\rho(t)\right),%
\label{poten}
\end{equation}
where $\gamma$ is defined by Eq.~(\ref{gamma}). Assuming the
E-region conditions, $\nu_{en}\simeq 10\nu_{in}$ and $m_i\simeq 30
m_p$, we write $\gamma$ in a form convenient for comparison with our
FlexPDE simulations, Sect.~\ref{numerical analysis} where
$n_0=N_{\mathrm{Trail}}(t_0)/N_0$ and $D=1$,
\begin{subequations}
\begin{equation}
\gamma =\frac{1}{2\Theta _{0}\Delta n_{0}t_{0}\sqrt{\psi }}\simeq
\frac{37.1}{\Delta n_0t_0\sqrt{\psi}}.
\end{equation}
In terms of the trail line density $N_{\mathrm{lin}}$, background
plasma density $N_0$, temperatures $T_{e,i}$, and geomagnetic field
$B_0$, we express $\gamma$ and the diffusion coefficient $D$ as
\begin{eqnarray}
\gamma &=& \frac{2\pi N_{0}\left( T_{e}+T_{i}\right) }{\sqrt{\psi
\left( 1+\psi \right) }\,N_{\mathrm{lin}}eB}\nonumber\\
&\simeq &\frac{1.08\times
10^{-2}\,}{\sqrt{\psi(1+\psi)}}\left(\frac{T_e+T_i}
{1000\mathrm{K}}\right)\left(\frac{0.5\times 10^{-4}\mathrm{T}}{B_0}\right)\nonumber\\
&\times& \!\!\left(\frac{N_0}{10^{10} \mathrm{m}^{-3}}\right)\!\!
\left(\frac{10^{14}
\mathrm{m}^{-1}}{N_{\mathrm{lin}}}\right)\mathrm{s}^{-1},
\end{eqnarray}
\begin{equation}
D=\frac{T_{e}+T_{i}}{m_{i}\nu_{in}}\simeq\frac{23.2}{\sqrt{\psi}}\left(
\frac{T_{e}+T_{i}}{1000\mathrm{K}}\right) \left(  \frac
{0.5\times10^{-4}\mathrm{T}}{B_{0}}\right)\mathrm{m}^{2}\mathrm{s}^{-1}.
\label{Dpract}
\end{equation}
\label{pract}
\end{subequations}
The only time dependence in $\phi_{\textrm{res}}^{0}$ is associated
with the dimensionless parameter $\rho(t)$. The function
$\varphi(\eta,\rho)$ has different approximate expressions depending
upon the range of $\rho$, as described in Sect.~\ref{solution of
equations}. This parameter monotonically varies from small values in
the early diffusion stage to large ones in the later diffusion
stage. According to Eqs.~(\ref{tinterp}) and (\ref{rhotinterp}), the
time dependence of $\rho$ can be approximately determined by
\begin{equation}
\gamma t = \rho +\frac{1}{1+\psi }\,\ln \left( 1+\frac{1+\psi }{\psi }\,
\rho \right),%
\label{tinterp1}
\end{equation}%
or, in the explicit form,
\begin{equation}
\rho (t)= \frac{W\left( \psi e^{\left( 1+\psi \right) \gamma t+\psi }\right)
-\psi }{1+\psi }\,,
\label{paramerho}
\end{equation}
where $W(x)$ is the Lambert W-function.

To calculate the residual potential for sufficiently large $\rho$,
the parameter $\rho$ needs an adjustment,
$\rho\rightarrow\tilde{\rho}< \rho$,
\begin{equation}
\tilde\rho(\rho)=\frac{\rho}{2}\left(1-\tilde\beta(\rho) +
\sqrt{1-2\tilde\beta(\rho)}\right),
\label{rhotilde}
\end{equation}
where
\begin{subequations}
\label{Bem}
\begin{equation}
\tilde\beta(\rho) \approx
\frac{0.92\rho}{(1+\psi)(2.2+\rho)}\,\sqrt{\frac{\gamma t}{\pi
\rho}},\qquad\mathrm{for}\ \psi\geq 0.05,%
\label{Bemp}
\end{equation}
\begin{equation}
\tilde\beta(\rho) \approx \frac{\rho}{2(2.2+\rho)}\,
\sqrt{\frac{\gamma t}{\rho}}\,, \qquad\mathrm{for}\ \psi< 0.05,
\label{Bemp1}
\end{equation}
\end{subequations}
and the relationship between $\gamma t$ and $\rho$ is given by
Eq.~(\ref{tinterp1}). Equation~(\ref{Bemp}) includes the
approximation for $\tilde\beta(\rho)$, Eq.~(\ref{betaemp}), obtained
by fitting FlexPDE runs for $\psi\geq 0.05$, while Eq.~(\ref{Bemp1})
is an extrapolation to smaller $\psi$.

In the earlier stage of dense-trail diffusion when $\rho(t)\approx
\psi\gamma t/(1+\psi)\ll 1$, a simple approximation for the
near-zone potential is given by the piece-wise function,
Eq.~(\ref{simple}),
\begin{equation}
\phi_{\mathrm{res}}^0(x,t)\simeq\frac{1}{2(1+\psi)}\times\left\{
\begin{array}
[c]{ccc}%
-\,\frac{\eta_0^{2}}{3}+\frac{\eta^{2}}{2} & \text{if} & |\eta(x,t)|<\eta_{0},\\
\frac{\eta_{0}^{4}}{6\eta^{2}} & \text{if} & |\eta(x,t)|>\eta_{0},
\end{array}
\right.
\label{piece-wise}
\end{equation}
where $\eta(x,t)=x[\gamma/D\rho(t)]^{1/2}$ and $\eta_0 \simeq
2\left[\ln(1/\rho(t))\right]^{1/2}$.

For simple estimates of the residual potential near the trail at a
later stage, $\rho(t)\sim\gamma t\gtrsim 1$, one can use the
zero-order Eq.~(\ref{maychoose}). To compare with simulations,
however, we should use the more complicated, but more accurate,
first-order Eq.~(\ref{pero}), and replace $\rho$ by $\tilde\rho$,
Eq.~(\ref{rhotilde}), and $\eta$ by
$\tilde\eta(x,t)=x[\gamma/D\tilde\rho(t)]^{1/2}$,
\begin{eqnarray}
\phi_{\mathrm{res}}^0(x,t)&=&\frac{1}{2(1+\psi)\tilde\rho(t)}\nonumber\\
&\times& \left\{S(\tilde\eta)-\frac{2\lambda(\tilde\rho)
A(\tilde\rho)}{\sqrt{1+4\lambda(\tilde\rho)} }\left[
\frac{4\lambda(\tilde\rho)}{1+4\lambda(\tilde\rho)}\right.\right.\nonumber\\
&+&\!\!\left.\left.\left(3-2\lambda(\tilde\rho)\tilde\eta^{2}\right)
S(\tilde\eta\sqrt{1+4\lambda(\tilde\rho)})\right]\right\}\!,
\label{perone}
\end{eqnarray}
where the functions $S$, $\lambda$, and $A$ are defined by
Eqs.~(\ref{S}) and (\ref{lamA}). Given
$\phi_{\textrm{res}}^{0}(x,t)$, the residual potential in all
locations is given by Eq.~(\ref{solu}). At large distances from the
trail, $x^2+\alpha^2y^2\gg Dt$, according to Eqs.~(\ref{quadr}) and
(\ref{rho}), the residual potential has a 2D quadrupole structure,
\begin{equation}
\phi_{\mathrm{res}}(x,y,t)\approx \frac{C_1(\rho)\,\rho D(x^2 -
\alpha^2y^2)}{2(1+\psi)\gamma(x^2+\alpha^2y^2)^{2}}\,,
\label{fardi}
\end{equation}
where $C_1(\rho)=\lim_{|\eta|\rightarrow\infty}[\eta^2\varphi(\eta,\rho)]$,
Eq.~(\ref{asympto}).

The expression for the trail density is much simpler than those for
the potential. According to Eq.~(\ref{Gau}) with $q=\rho/\gamma t$,
Eq.~(\ref{rho}), for all values of $\rho$ it has a nearly Gaussian
form,
\begin{equation}
\Delta n_{\mathrm{Trail}}(x,y,t) \approx \Delta n_{0}
\sqrt{{\frac{t_{0}\rho_{0}}{t\rho(t)}}}\exp \left[-\
\frac{1}{4D}\left(\frac{\gamma
x^{2}}{\rho(t)}+\frac{y^{2}}{t}\right) \right]\!,
\label{GauGau}
\end{equation}
where $\rho_0\equiv\rho(t_0)\approx \gamma\psi t_0/(1+\psi)$,
Eq.~(\ref{rho00}), and no adjustment for all $\rho$.

Eliminating $t$ from Eqs.~(\ref{tinterp1}) and (\ref{GauGau}), we
express the peak Gaussian density, $\Delta
n_{\mathrm{Peak}}\equiv\Delta n_{\mathrm{Trail}}(0,0,t)$, in terms
of $\rho(t)$,
\begin{equation}
\Delta n_{\mathrm{Peak}}=\left\{2\Theta_{0}\sqrt{\rho\left[
\left(1+\psi\right) \rho+\ln\left(1+\frac{1+\psi}{\psi}\,\rho\right)
\right] }\right\} ^{-1}\!\!,
\label{ntra}
\end{equation}
while the same quantity in the self-similar solution is given by
\begin{equation}
\Delta
n_{\mathrm{Peak}}^{\mathrm{SS}}=\left\{2\Theta_{0}\sqrt{\psi}\left[
\rho+\frac{1}{1+\psi }\ln\left(1+\frac{1+\psi}{\psi}\,\rho\right)
\right] \right\} ^{-1}\!\!.
\label{nSSS}
\end{equation}
These expressions have a universal form independent of the initial
conditions, provided $t> 2t_0$ and $\Delta n(t_0)\gg 1$.

\section{Comparison of theory and simulations\label{comparison of theory}}

\begin{figure*}
\includegraphics[width=0.47\textwidth]{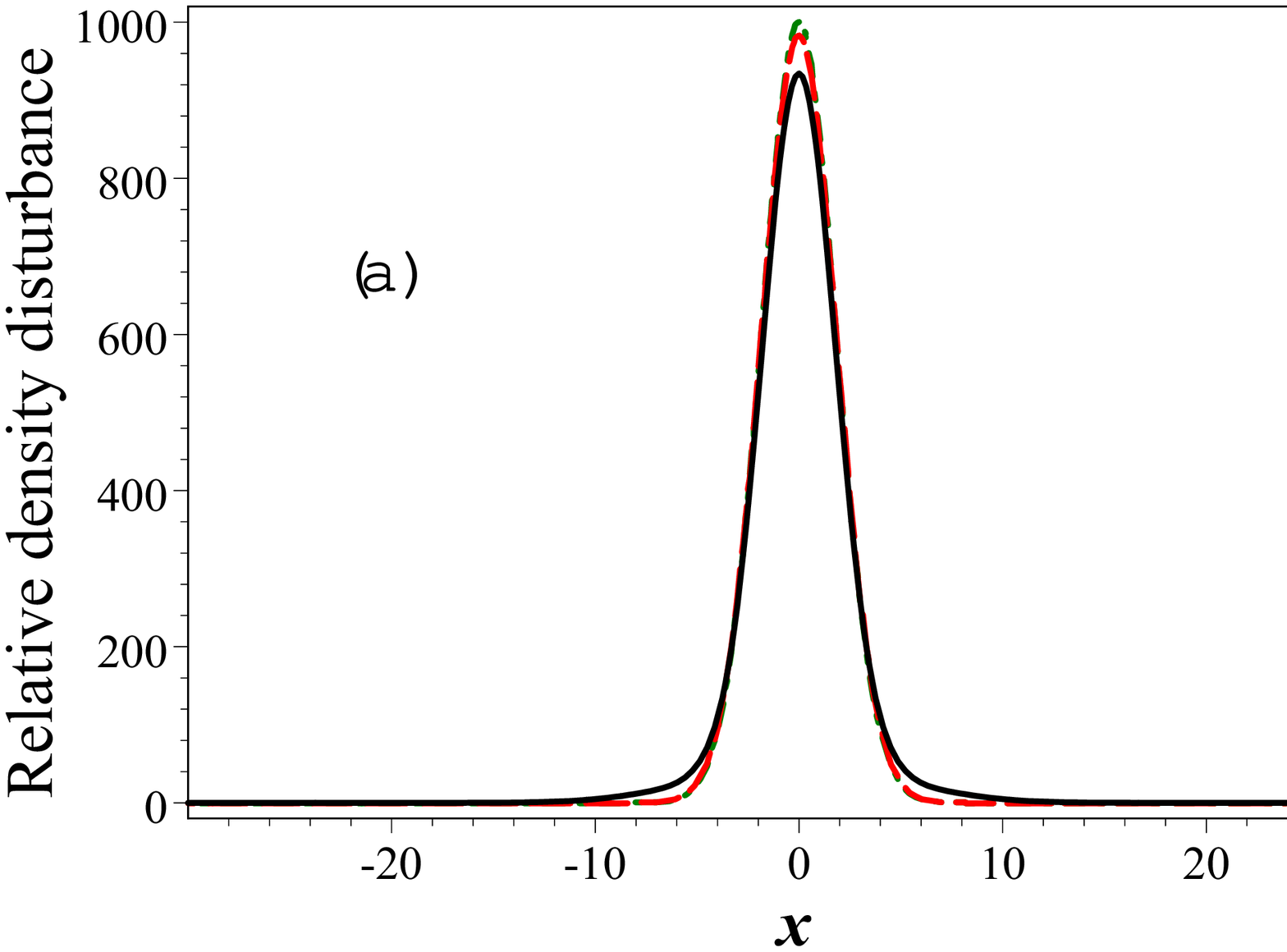}\hfill
\includegraphics[width=0.47\textwidth]{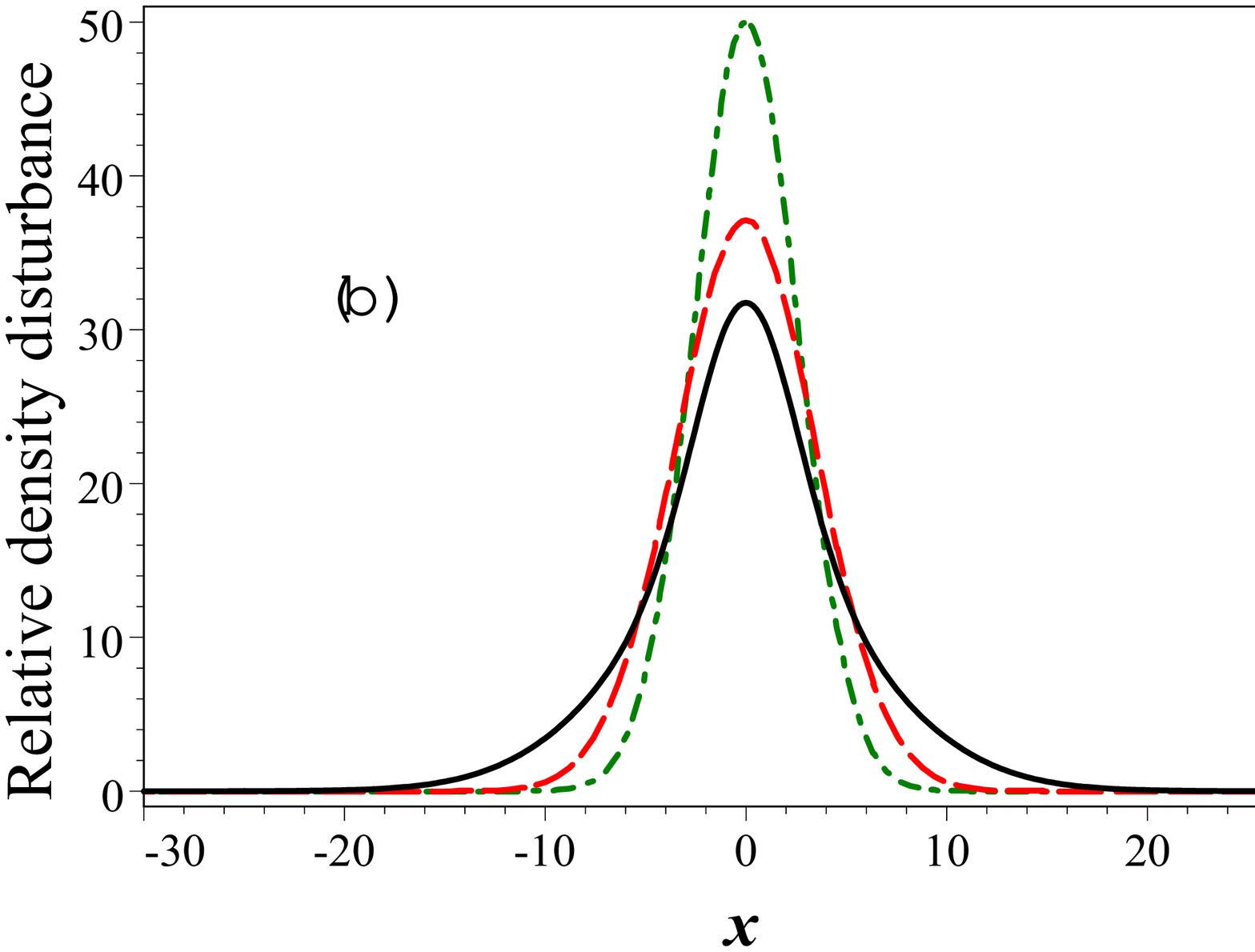}\vspace{18mm}
\includegraphics[width=0.47\textwidth]{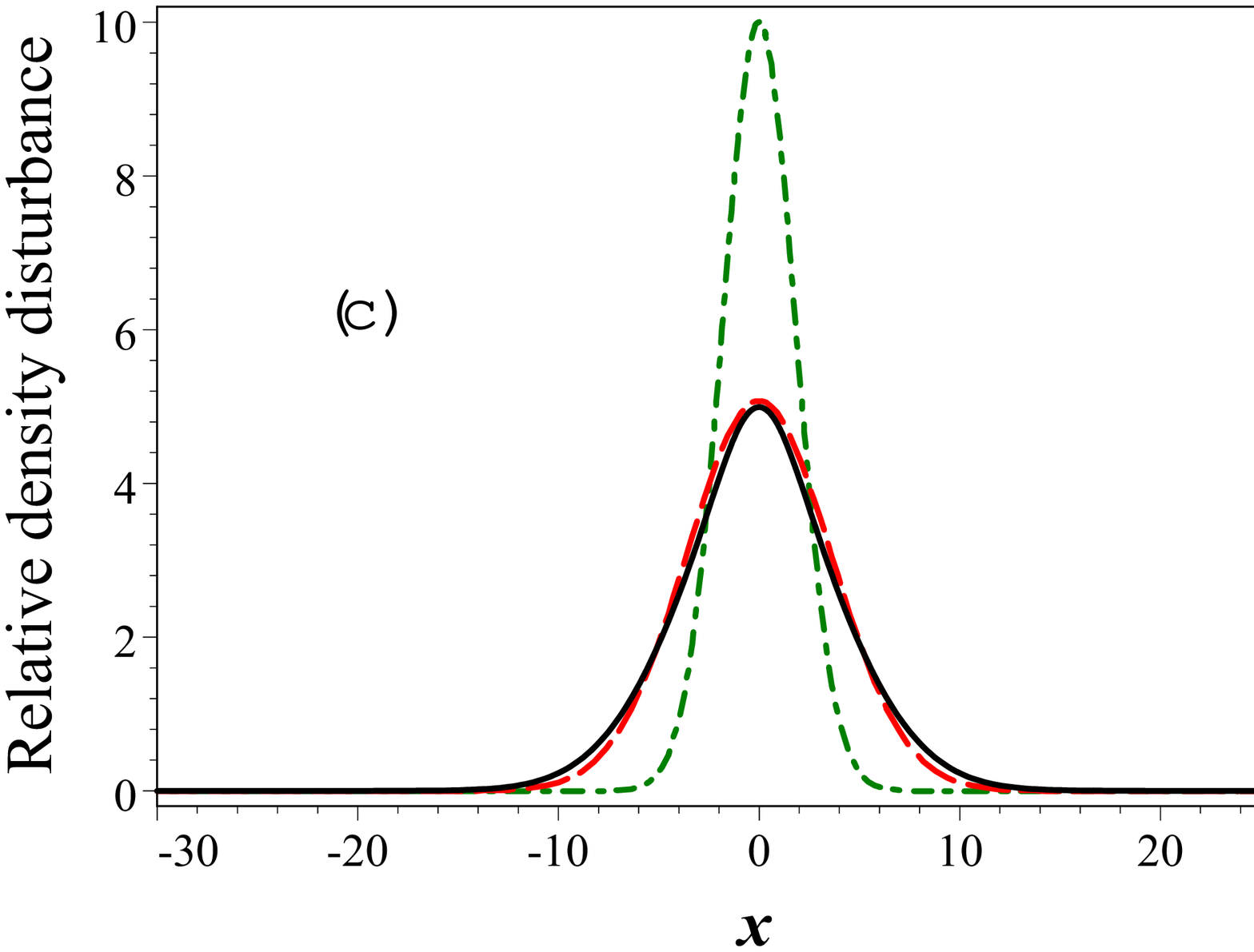}\hfill
\includegraphics[width=0.47\textwidth]{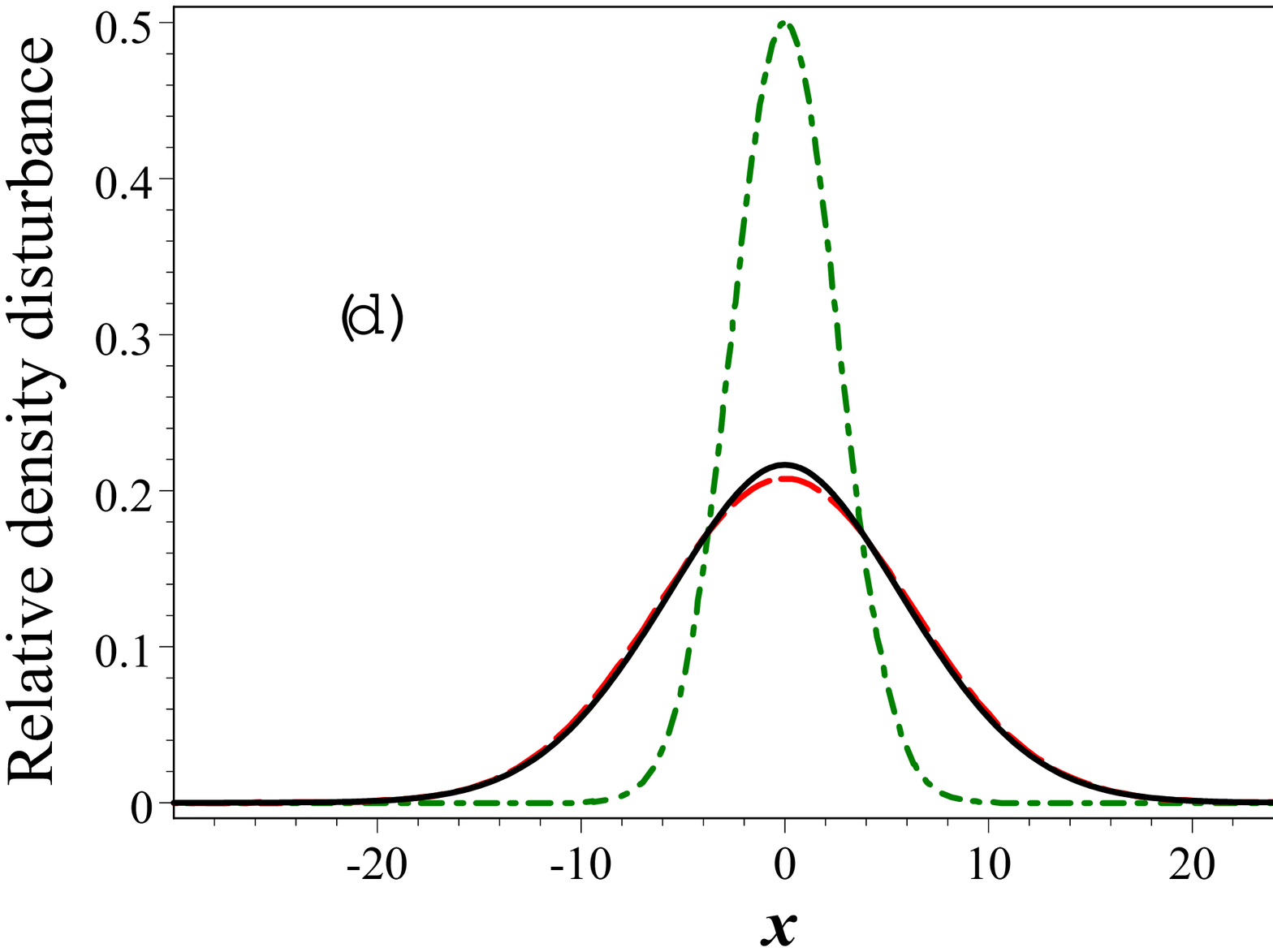}
\caption{\label{densities} Trail density distribution along $x$ for
$\psi=0.2$. (a): $\Delta n_0 =10^4$, $t=10$, $\rho\approx 1.4\times
10^{-3}$. (b): $\Delta n_0 =1000$, $t=20$, $\rho\approx 0.5$. (c):
$\Delta n_0 =100$, $t=10$, $\rho\approx 5.37$. (d): $\Delta n_0
=10$, $t=20$, $\rho\approx 160$. Solid curves: the numerical
solution; dot-dashed curves: the self-similar solution,
Eq.~(\ref{SelfSo}); dashed curves: the analytical solution,
Eq.~(\ref{GauGau}).}
\end{figure*}

\begin{figure}
\includegraphics[width=0.47\textwidth]{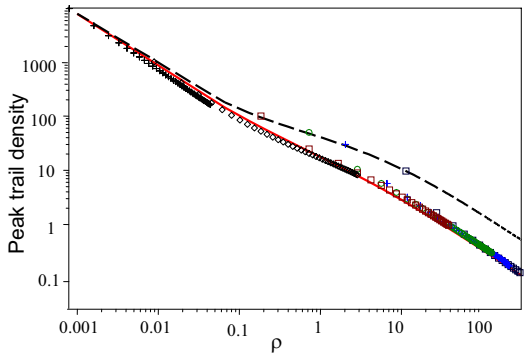}
\caption{\label{densityvsrho} Peak trail density vs. parameter
$\rho$ for $\psi=0.05$. Points shown as alternating crosses,
diamonds, boxes, and circles represent the numerical solution from
several runs: $\Delta n_0 =10^4, 10^3, 100, 50, 30, 10$. For each
run, the consecutive points (from left to right) correspond to
$t/t_0=1, 2, 3\ldots$. Solid curve shows the analytical solution
given by Eq.~(\ref{ntra}). Dashed curve shows the self-similar
solution given by Eq.~(\ref{nSSS}). }
\end{figure}

\begin{figure*}
\includegraphics[width=0.45\textwidth]{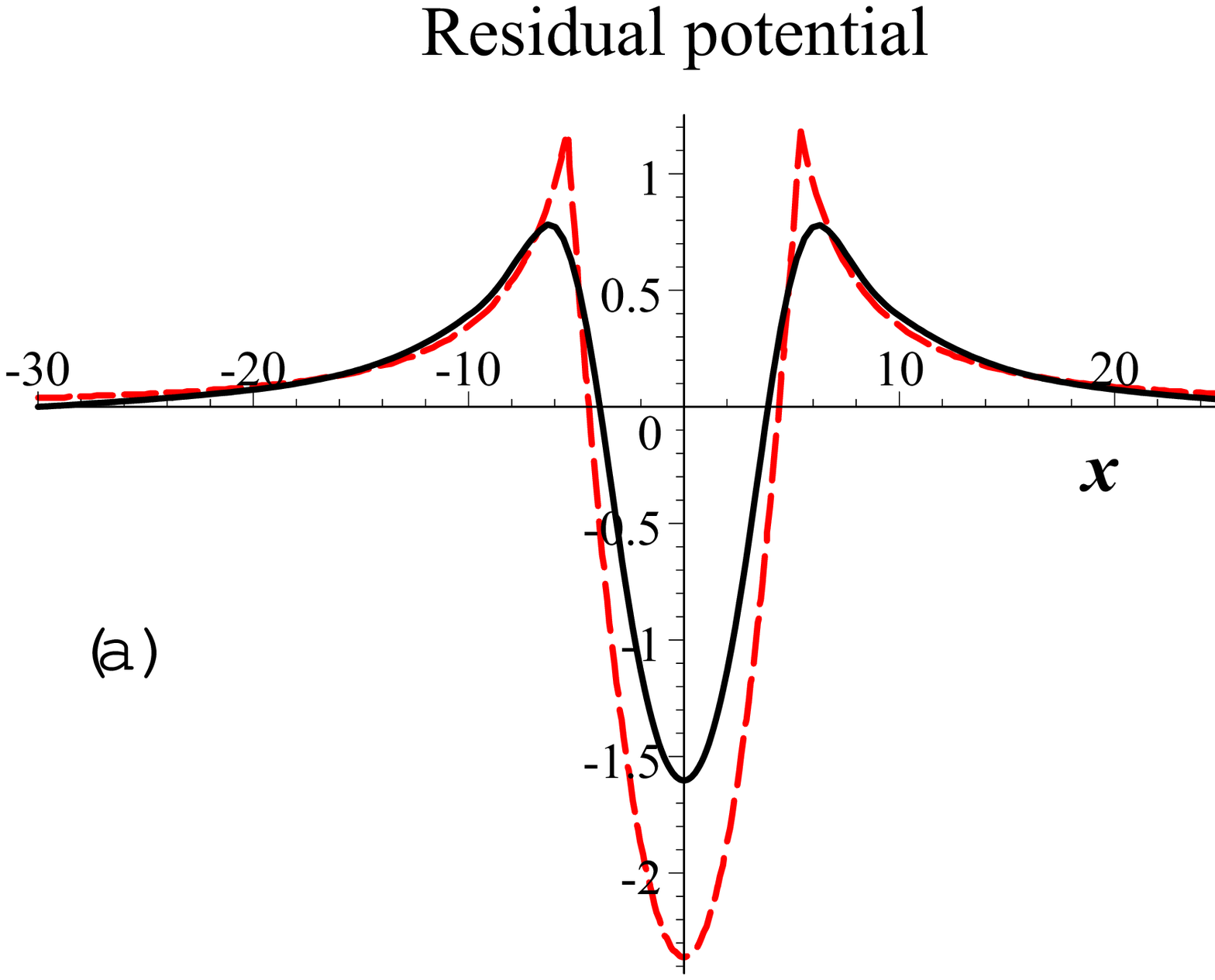}\hspace{10mm}
\includegraphics[width=0.45\textwidth]{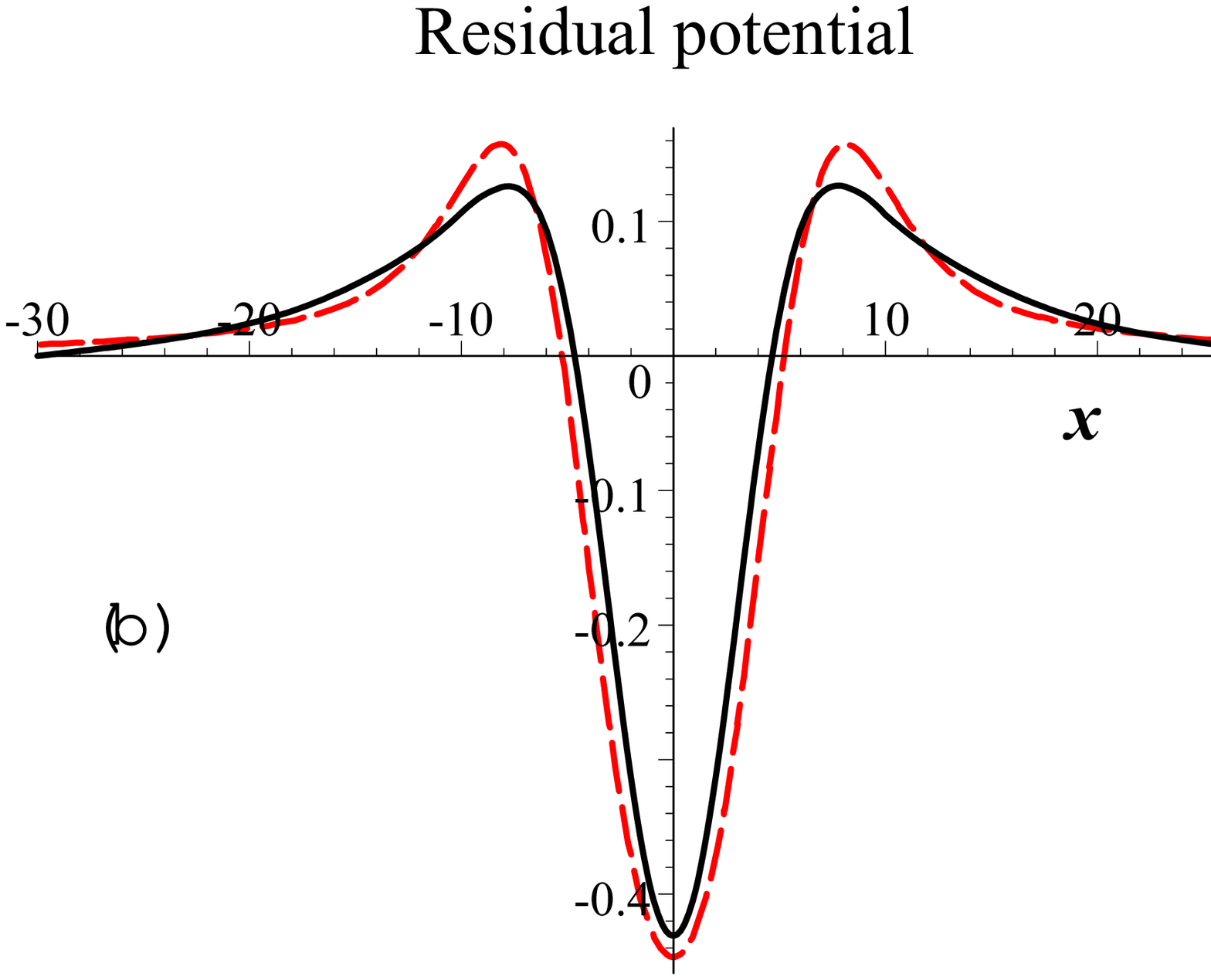}\vspace{18mm}
\includegraphics[width=0.45\textwidth]{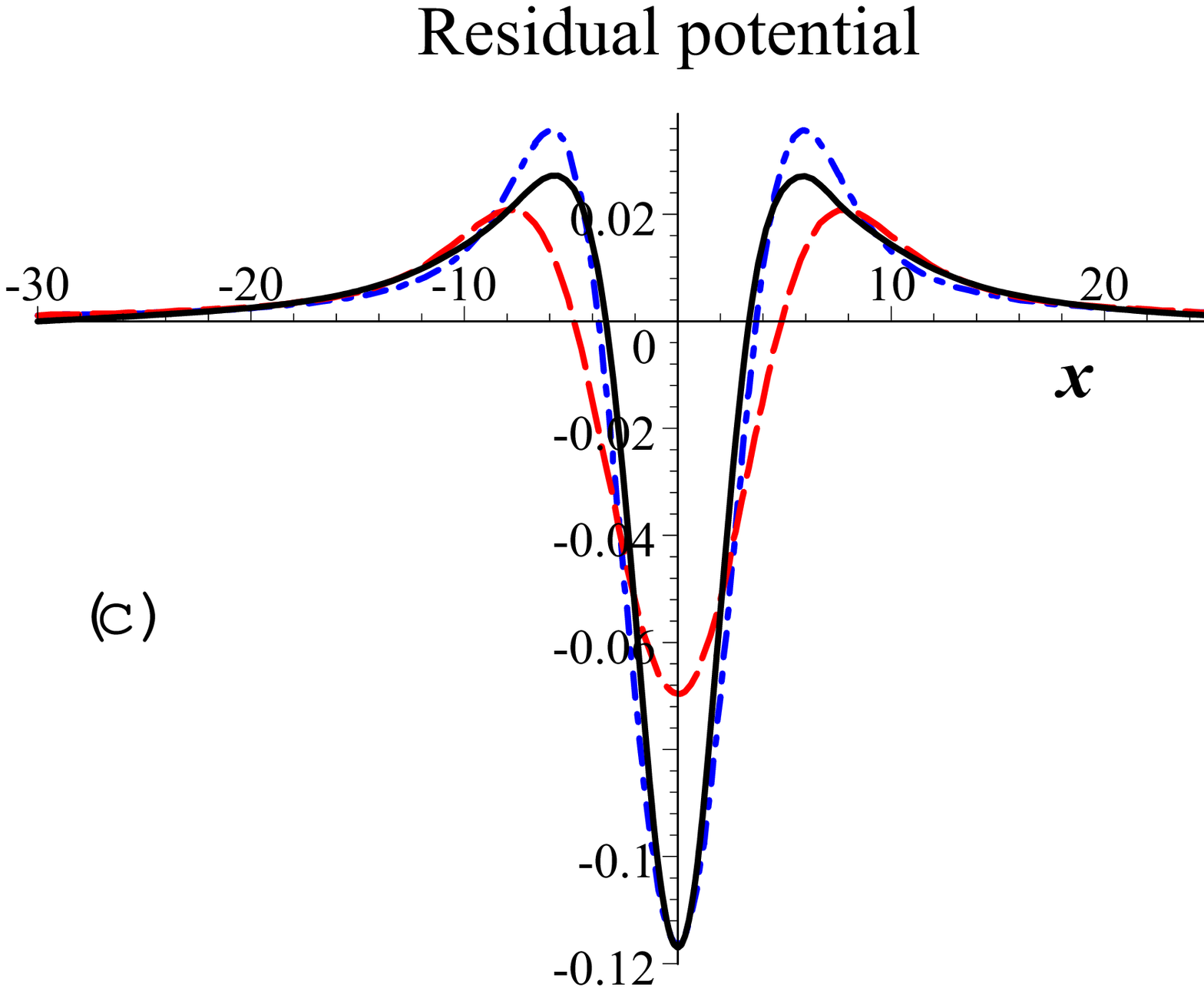}\hspace{10mm}
\includegraphics[width=0.45\textwidth]{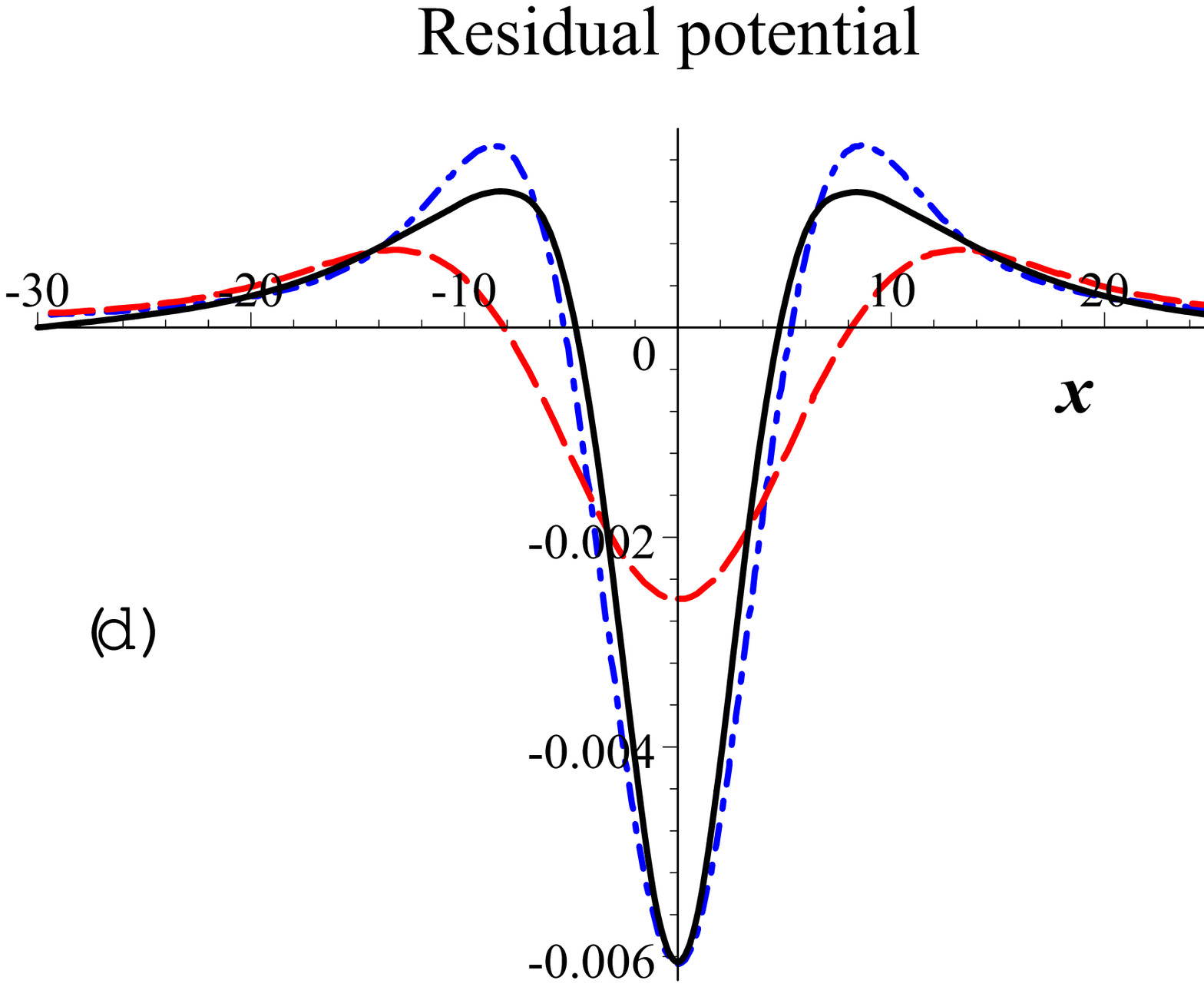}
\caption{\label{potentials} Residual potentials at $y=0$,
$\phi_{\mathrm{res}}^0(x,t)$, corresponding to density distributions
in Fig.~\ref{densities}. Solid curves show the numerical solutions,
while dashed and dot-dashed curves show the analytical solutions
according to different equations. (a): $\Delta n_0 =10^4$, $t=10$,
$\rho\approx 1.4\times 10^{-3}$, dashed curve corresponds to
Eq.~(\ref{piece-wise}). (b): $\Delta n_0 =1000$, $t=20$,
$\rho\approx 0.5$, dashed curve corresponds to Eq.~(\ref{perone})
with $\tilde\rho\rightarrow\rho$. (c): $\Delta n_0 =100$, $t=10$,
$\rho\approx 5.37$, dashed curve corresponds to Eq.~(\ref{perone})
with $\tilde\rho\rightarrow\rho$, dot-dashed curve corresponds to
Eq.~(\ref{perone}) with $\tilde\rho\approx 3$. (d): $\Delta n_0
=10$, $t=20$, $\rho\approx 160$, dashed curve corresponds to
Eq.~(\ref{perone}) with $\tilde\rho\rightarrow\rho$, dot-dashed
curve corresponds to Eq.~(\ref{perone}) with $\tilde\rho\approx
68$.}
\end{figure*}

In this Section, we compare our analytical theory with simulations
outlined in Sect.~\ref{numerical analysis}. We start with comparison
of our simulations with the theoretical expression for the trail
density, Eq.~(\ref{GauGau}). In the early stage of dense-trail
diffusion, $\rho\ll 1$, the nearly Gaussian peak in simulations
closely follows the self-similar solution (SSS), Eq.~(\ref{SelfSo}),
though it shows a slightly faster diffusion, as seen in
Fig.~\ref{densities}(a). The density predicted by Eq.~(\ref{GauGau})
is closer to the SSS curve than to the numerical one because the
interpolation expression for $\rho(t)$ given by
Eq.~(\ref{rhotinterp}) is less accurate for $\rho\ll 1 $ than it is
for $\rho\gtrsim 1$.

When $\rho\sim 1$, the analytic solution for the density peak given
by Eq.~(\ref{GauGau}) starts deviating from the SSS and becomes
closer to the numerical solution, as illustrated in
Fig.~\ref{densities}(b). When $\rho$ becomes larger, $\rho\gg 1$,
the analytical theory shows an excellent agreement with simulations,
while the SSS predicts a noticeably slower diffusion, as shown in
Figs.~\ref{densities}(c) and (d).

Figure~\ref{densityvsrho} shows the peak trail densities vs.~$\rho$
taken from the simulations, analytical theory, and self-similar
solution. The numerical solution is shown by separate groups of
points taken from several different runs (each group has its own
point shape). Each run started at $t_0=1$ with different values of
the initial peak trail density. In each group, consecutive points
correspond to equidistant moments of time: $t=1, 2, 3,\dots$. The
first point of the group always lies on the dashed curve
corresponding to Eq.~(\ref{nSSS}) because the SSS was the initial
condition for each run. However, starting from $t=2$, the numerical
points closely approach the theoretical curve given by
Eq.~(\ref{ntra}), while the SSS solution given by Eq.~(\ref{nSSS})
remains noticeably offset. The theoretical curve overlays the
numerical points for $\rho\gtrsim 1$, while for $\rho\ll 1$ it shows
a slight deviation from the numerical points (see the beginning of
this Section). The transition from an anisotropic diffusion to a
more isotropic one occurs near the inflection point about $\rho= 1$.
Notice that this takes place when the trail peak density remains
well above the background plasma density, $\Delta n_{\max}\sim
(\Theta_0\sqrt\psi)^{-1}\simeq 80$ for $\psi=0.05$.

Now we compare with simulations the theoretical expressions for the
residual potential in the near zone. Figure \ref{potentials} shows
the residual potential along $x$ for the same conditions as in
Fig.~\ref{densities}. Because the parameter $\rho$ spans a broad
range of values from small to large ones we will apply either
Eq.~(\ref{piece-wise}) or Eq.~(\ref{perone}).
Figure~\ref{potentials}(a) shows that for $\rho\ll1$ the simple
piece-wise approximation agrees reasonably with simulations in all
areas not too close to the two positive bumps of the potential.
Indeed, while there is a significant difference between the values
of the potential minimum at $x=0$, the potential derivative (the
residual electric field) is the same in the inner region
characterized by the parabolic dependence and occupied by the trail,
see Fig.~\ref{densities}(a). On the other hand, a good agreement
also exists well beyond the trail, where the residual potential
decreases with increasing $|x|$ and the corresponding electric field
changes its sign. The zero-order piece-wise approximation is rough
in the transitional zone near the two potential maxima, where it has
a discontinuity in the electric field. The maximum electric fields
in the piece-wise formula are reached near the discontinuity points,
approaching them from inside. The maximum electric fields in
simulations are reached at some locations in the inner region closer
to the center and hence have smaller values. Thus the simple
analytical formula yields nearly correct electric fields everywhere
except the transitional zones between the inner and outer regions,
where it overestimates the electric field magnitude. We have
attempted modeling the transitional electric field with higher-order
interpolations to provide a smooth transition, but this
underestimated the field. The error for the maximum electric field,
however, remains within the range of tens percent for all our
simulations.

For $\rho\sim 1$, the theoretical expression given by
Eq.~(\ref{perone}) with unadjusted $\rho$
($\tilde\rho\rightarrow\rho$), agrees well with simulations
practically in all locations, as seen in Fig.~\ref{potentials}(b).
As $\rho$ becomes large, Eq.~(\ref{perone}) with unadjusted $\rho$
shows a significant discrepancy, as seen in Fig.~\ref{potentials}(c)
and (d). However, if we adjust the parameter $\rho$,
$\rho\rightarrow\tilde\rho$ based on matching of the potential
minima for the two solutions and apply Eq.~(\ref{perone}) then the
discrepancy practically disappears. It is important that the
analytical solution with only one adjusted parameter
$\rho\rightarrow\tilde\rho$ causes the theoretical
$\phi_{\mathrm{res}}^0$ to match the numerical solution well not
only near the potential minimum but everywhere. The relation between
$\rho$ and $\tilde\rho$ based on our simulations results in the
empirical $\rho$-dependent adjustment coefficient $\tilde\beta$,
Eqs.~(\ref{rhotilde}) and (\ref{Bem}). We emphasize that the
adjusted parameter $\tilde\rho$ is only needed for the residual
potential and not for the trail density, as described in
Sect.~\ref{density disturbances}.

In this Section, we showed the comparison of theory and simulations
mainly for small $\psi$. For large values of $\psi$, the trail
diffusion in all stages of trail diffusion was nearly isotropic with
small values of the residual electric field, in full accord with the
analytical theory.

\section{Discussion\label{discussion}}

In this section, we will start by discussing some caveats, then we
will estimate the induction electric field in the plasma trail, and
finally we will dwell on plasma instabilities.

\subsection{Caveats}

In our theory, we have made a number of assumptions which are not
perfectly valid. Among those were assumptions about constant
electron and ion temperatures and about one sort of ions. However,
the initially hot temperatures of the newly produced meteor trail
plasma need some time for cooling. Also,     this plasma includes
material different from the ambient atmosphere, so that there may be
at least two kinds of ions with different masses. We will include
these factors in future work.

We also note that our full analytical theory has been developed only
for the particular case of a mutually orthogonal meteor trail axis
and magnetic field, $\theta=90^{\circ}$. A more general situation
occurs when the angles between the trail axis and $\vec{B}_0$,
$\theta$, are arbitrary but satisfy restrictions given by
Eq.~(\ref{restrict}). It is only discussed for the self-similar
solution described in Appendix~\ref{self-similar solution}. If we
ignore the effect of electron Hall currents into the process of the
meteor trail diffusion then we can apply all our results where
according to Eq.~(\ref{Q}) the quantity $\alpha^{-1}$,
Eq.~(\ref{alpha}), and the numerator of $\rho$, Eq.~(\ref{rho}),
acquire an additional factor $\sin\theta$. Our preliminary numerical
computations show that the electron Hall current at
$\theta\neq90^{\circ}$ affects the meteor trail diffusion in such a
way that the trail density contours become more isotropic and
rotated at a small angle in the $x,y$ plane, as the SSS does (see
Appendix \ref{self-similar solution}).

\subsection{Induction electric field}

Now we estimate the induction electric field associated with the
meteor plasma trail currents and show that, except for
extraordinarily high-density meteors, this field is negligible. This
is of importance for the electrostatic field approximation employed
in our approach. The induction electric field may only occur due to
disturbances of the magnetic field, $\delta\vec{B}$, which in turn
are due to the electric current, $\vec{j}$, caused by the plasma
trail diffusion. Note that the magnetic field disturbances, even in
spite of their extreme weakness, can be observable using a
sophisticated measurement technique for diagnostics purposes.
Furthermore, the initial formation of the trail current may have
caused ELF/VLF signals observed on the ground and correlated with
the meteor showers \cite{PriceBlum2000,TrautnerKoshchny2002}.

According to the Maxwell equations, we have
\begin{subequations}
\label{MaxwellEq}%
\begin{align}
\vec\nabla\times\vec{E} &  =-\ \partial_{t}\delta\vec
{B},\label{Faraday}\\
\vec\nabla\times\delta\vec{B} &  =\mu_{0}\vec{j},\label{Ampere}%
\end{align}
\end{subequations}
where $\mu_{0}$ is the permeability of free space and we neglected
the displacement current (all characteristic speeds in the trail
diffusion are much less than the speed of light, etc.). The net
electric current is due to the fact that electrons and ions have
different responses to the external force. Being proportional to the
plasma density, the net electric current is mainly concentrated
within the trail. It is predominantly formed by the unbalanced
$\vec{E}\times\vec{B}$ drift of electrons. In the strongest case,
the trail density behaves roughly in accord with the self-similar
solution, so that for a simple estimate we can use the corresponding
expressions for the current given by Eq.~(\ref{currentdens}). This
current with the density $j_{z}=(e\Omega_{e}x\sin\theta/2\nu_{en}t)
N^{\mathrm{ss}}$, $N^{\mathrm{ss}}=\left(  C/t\right)  \exp\left\{
-\ \left[  \left( 1+\psi\right)  x^{2}/\psi+y^{2}\right]
/4Dt\right\}  $, Eqs.~(\ref{nSSanz}) and (\ref{SSsolution}), is
directed along the trail axis and flows in opposite directions in
the two halves of the trail, $x>0$ and $x<0$ (in the actual 3D,
spatially inhomogeneous, and restricted trail, the current forms a
closed loop). Setting as above $\theta=90^{\circ}$ and taking a
typical scale along $x$ as $\Delta x=\left[4\psi Dt/\left(
1+\psi\right)  \right]  ^{1/2}$ and $N^{\mathrm{ss}}\sim N_{\max}$,
according to Ampere's law, Eq.~(\ref{Ampere}), we estimate the
typical magnetic field disturbance
as
\begin{equation}
\delta B\sim\frac{2e\mu_{0}\Omega_{e}D}{\nu_{en}}\left(  \frac{\psi}{1+\psi
}\right)  N_{\max}=\frac{2\mu_{0}(T_{e}+T_{i})}{\left(  1+\psi\right)  B_{0}%
}\ N_{\max},\label{deltaB}%
\end{equation}
where in the last equality we used the definitions of $D$, Eq.~(\ref{D}), and
$\psi$, Eq.~(\ref{psi}). The relative magnetic field disturbance is%
\begin{equation}
\frac{\delta B}{B_{0}}\sim\frac{\beta_{0}N_{\max}}{\left(  1+\psi\right)
N_{0}},\label{delBrel}%
\end{equation}
where $\beta_{0}$ is the ratio of the undisturbed total plasma pressure,
$N_{0}(T_{e}+T_{i})$, to the magnetic pressure, $B_{0}^{2}/2\mu_{0}$,%
\[
\beta_{0}\approx1.39\times10^{-6}\!\left(\frac{N_{0}}{10^{11}\mathrm{m}^{-3}%
}\right)\!\! \left(\frac{T_{e}+T_{i}}{1000\mathrm{K}}\right)\!\!
\left( \frac{0.5\times10^{-4}\mathrm{T}}{B_{0}}\right)^{2}\!\!.
\]
In the E-region ionosphere, $N_{0}\simeq\left(  10^{9}\text{--}10^{11}\right)
\mathrm{m}^{-3}$, $T_{e}\sim T_{i}\simeq300\mathrm{K}$, $B_{0}\simeq\left(
0.25\text{--}0.6\right)  \times10^{-4}\mathrm{T}$, so that the relative
disturbance of the magnetic field is small, unless the plasma trail is
extremely dense, $N_{\max}/N_{0}\gtrsim10^{6}$--$10^{8}$.

To estimate the contribution of the induction component into the
total electric field, we can estimate the ratio of $\
|\vec\nabla\times\vec{E}|$ to
$|\vec\nabla\cdot\vec{E}|\gtrsim(T_{e}+T_{i})/e(\Delta x)^{2}$.
According to
Eqs.~(\ref{Faraday}) and (\ref{deltaB}), after simple algebra we obtain%
\begin{equation}
\left\vert \frac{\vec\nabla\times\vec{E}}{\vec\nabla\cdot\vec{E}%
}\right\vert \lesssim\frac{4\Theta_{0}\sqrt{\psi}}{\left(  1+\psi\right)
^{2}}\left(  \frac{\delta B}{B_{0}}\right)  .\label{ratio}%
\end{equation}
This ratio has an a additional factor in front of $\delta B/B_{0}$,
which is always small since according to Eq.~(\ref{>theta0}),
$4\Theta_{0}\sqrt{\psi}/(1+\psi)^{2}\leq
0.325\Theta_{0}\simeq0.44\times10^{-2}$. Because $\delta B/B_{0}$ in
the regular meteor trail is small, the induction electric field
proves to be even smaller. This justifies our initial assumption
that the total electric field is nearly curl-free, $\vec
{E}=-\vec\nabla\varphi$.

\subsection{Dynamics of electric field and plasma instabilities}

The ambipolar electric field associated with trail diffusion may
drive plasma instabilities responsible for observable non-specular
radar echoes. In this Section, we will make simple estimates of the
driving field and instability threshold.

The driving field is determined by the total external force acting
on electrons, $\vec {E}_{\mathrm{res}}=-(T_{e}+T_{i})
\vec\nabla\phi_{\mathrm{res}}$, where $\phi_{\mathrm{res}}$ is the
residual potential defined by Eq.~(\ref{phiadd}). Equations
(\ref{piece-wise}) to (\ref{fardi}) give approximate analytical
expressions for the residual potential if the magnetic field and the
meteor trail axis are mutually orthogonal. These expressions depend
upon the dimensionless parameter $\rho(t)$ given by
Eq.~(\ref{paramerho}) in terms of the effective rate $\gamma$,
Eq.~(\ref{gamma}). The latter parameter is proportional to an
effective trail-background interaction cross-section
$\sigma_{\mathrm{eff}}\equiv N_{\mathrm{lin}}/N_0$. After the
critical time given by Eq.~(\ref{tcri}), the diffusion process
becomes more isotropic and the residual potential decreases
drastically. The critical time $t_{{\mathrm{cr}}}$ varies depending
upon the ionospheric conditions and meteor parameters. Nighttime
conditions with low $N_0$ and given $N_{\mathrm{lin}}$ are
equivalent to daytime conditions with much larger $N_0$ and the
proportionally increased column line density $N_{\mathrm{lin}}$. For
example, assuming equatorial day-time ionosphere,
$N_{0}\sim10^{11}\mathrm{m}^{-3}$, and a typical linear trail
density, $N_{\mathrm{lin}}\sim10^{14}\mathrm{m}^{-1}$
\cite{Bronshten:Physics83,Ceplecha:Meteor98}, we obtain a critical
time of tens of milliseconds. For mid- or high latitudes, night-time
conditions, $N_{0}\sim10^{9}\mathrm{m}^{-3}$, and
$N_{\mathrm{lin}}\gtrsim10^{15}\mathrm{m}^{-1}$ (or faster and/or
bigger meteoroids), we obtain that the critical time may reach tens
of seconds.

The strongest electric field occurs in the early diffusion stage,
$\rho\ll1$, and at the edge of the nearly parabolic region of the
potential, Eq.~(\ref{strong}). Using Eqs.~(\ref{simple}),
(\ref{eta0(rho)}), and Eq.~(\ref{Phieta}), we obtain the maximum
value of the residual electric field,
$|E_{x}^{\max}|\simeq[D\gamma\ln(
1/\rho)/\rho]^{1/2}m_{i}\nu_{in}/e(1+\psi)$, where the diffusion
coefficient $D$ is defined by Eq.~(\ref{D}). In the later stage of
trail diffusion when $\rho\gg1$, Eq.~(\ref{weak}) gives
$[\partial_{\eta}\varphi (\eta)]^{\max}\simeq0.7/\rho$, ignoring the
adjustment of $\rho$ described in Sect.~\ref{density disturbances}.
In the original variables, we obtain $|E_{x}^{\max}|\simeq
0.35(D\gamma/\rho)^{1/2}m_{i}\nu_{in}/ e(1+\psi)\rho$. Interpolating
between these limiting expressions, we can write a simple formula,
\begin{equation}
|E_{x}^{\max}|\simeq\frac{m_{i}\nu_{in}}{e(1+\psi)}\left[
\frac{D\gamma}{2\rho}\ \ln\left( 1+\frac{0.3}{\rho^{2}}\right)
\right]
^{1/2},%
\label{interpol}
\end{equation}
roughly valid in the entire domain of $\rho$.

\begin{figure*}
\includegraphics[width=0.47\textwidth]{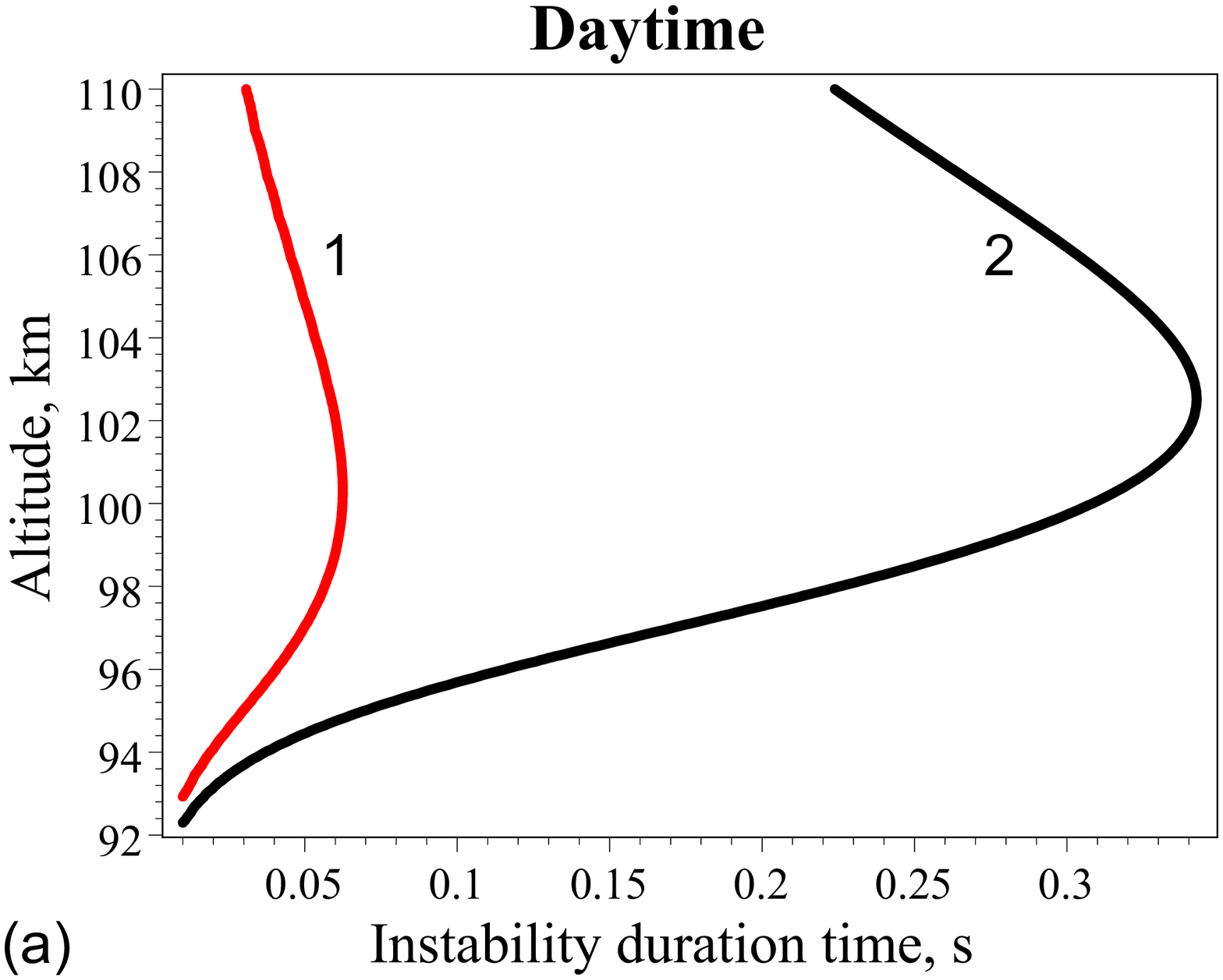} \hfill
\includegraphics[width=0.47\textwidth]{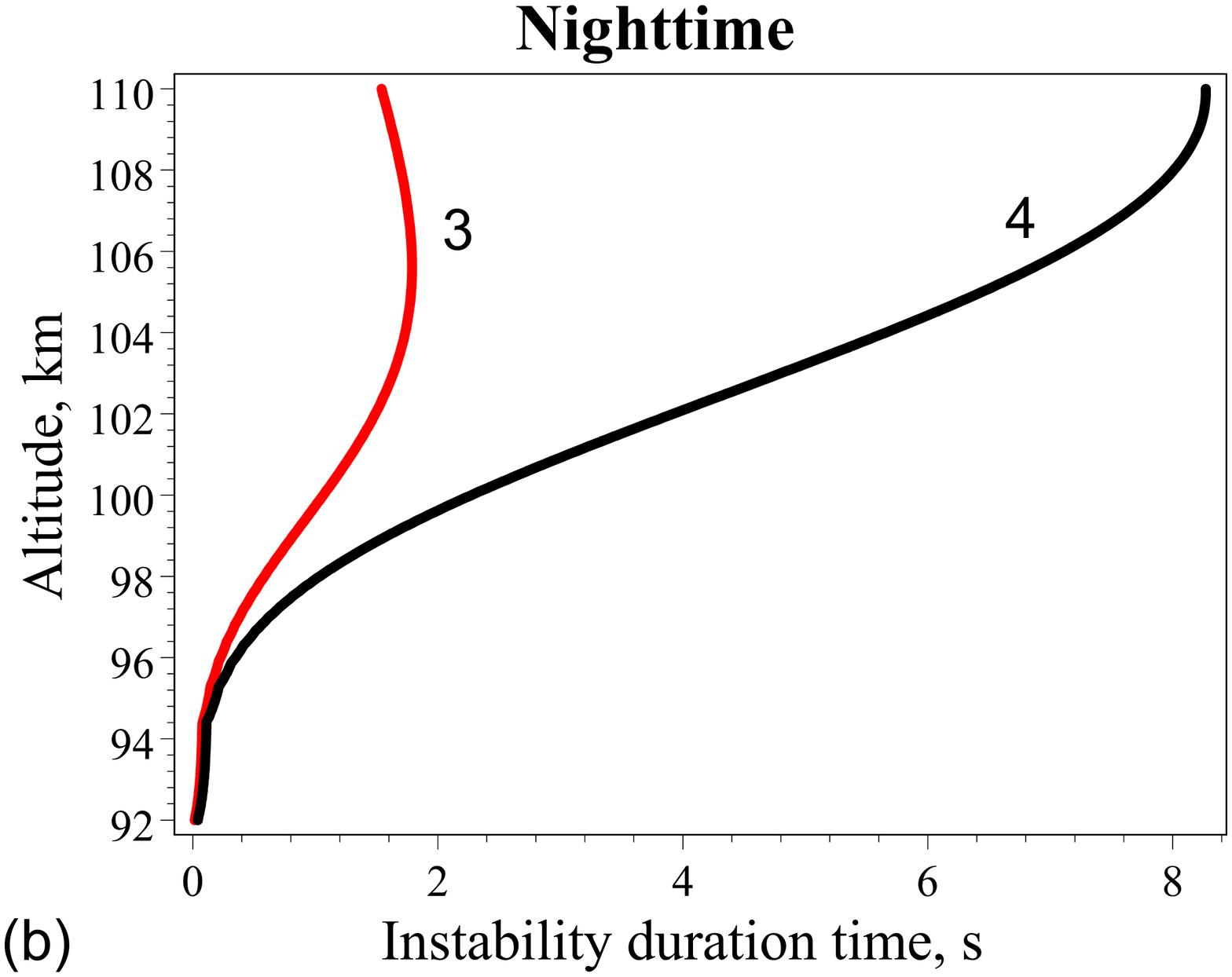}
{\caption{\label{dimantFig13} Altitudinal dependence of FB
instability duration $t_{\mathrm{FB}}$ for the equatorial E region
during (a) daytime ($N_0=10^{11}\mathrm{m^{-3}}$), where curve 1 is
for $N_{\mathrm{lin}}= 10^{14}\mathrm{m}^{-1}$ and curve 2 is for
$N_{\mathrm{lin}}= 10^{15}\mathrm{m}^{-1}$, and (b) nighttime
($N_0=10^{9}\mathrm{m^{-3}}$), where curve 3 is for
$N_{\mathrm{lin}}= 10^{14}\mathrm{m}^{-1}$ and curve 4 is for
$N_{\mathrm{lin}}= 10^{15}\mathrm{m}^{-1}$. Curves 1, 2, 3, and 4
correspond to $\sigma_{\mathrm{eff}}=10^3,\,10^4,10^5$, and
$10^6\mathrm{m}^{2}$, respectively.}}
\end{figure*}

Now we estimate the Farley-Buneman (FB) instability criterion
\cite{Farley:Ejet63,Buneman:Ejet63}. For a homogeneous plasma, the
simplest FB instability criterion, obtained using the two-fluid
plasma model for sufficiently long-wavelength waves, is given by
$|\vec{V}_{0}|>(1+\psi)C_{s}$ \cite{Farley:Theory85}, where
$\vec{V}_{0}=\vec{E}_{\mathrm{res}}\times\vec{B}_{0}/B_{0}^{2}$ is
the $\vec{E}_{\mathrm{res}}\times\vec{B}_{0}$ drift velocity and
$C_{s}\equiv[(T_e + T_i)/m_i]^{1/2} = (D\nu_{in})^{1/2}$ is the
ion-acoustic speed. Applying this criterion to the maximum field
given by Eq.~(\ref{interpol}), expressing the corresponding drift
speed as $|\vec{V}_{0}|=e|E_{x}^{\max }|/m_{i}\Omega_{i}$, and using
Eqs.~(\ref{psi}) and (\ref{>theta0}), we write the FB instability
criterion as $|E_{x}^{\max }|
> (1+\psi)(D\nu_{in})^{1/2}m_i\Omega_i/e$. Expressing $E_{x}^{\max }$ in
$N_{\mathrm{lin}}$, we reduce the FB instability criterion to
\begin{equation}
\frac{1}{\rho}\ln\left(1+\frac{0.3}{\rho^{2}}\right)  >P, %
\label{FBcrit}
\end{equation}
where
\begin{align}
P =  \frac{\Theta _{0}\left( 1+\psi \right) ^{9/2}\left(
eB_{0}\right) ^{2}N_{\mathrm{lin}}}{\pi \left( T_{e}+T_{i}\right)
m_{i}N_{0}}&  \approx 0.4\left(  1+\psi\right)
^{9/2}\nonumber\\
\times\left(  \frac{1000\mathrm{K} }{T_{e}+T_{i}}\right)\!\!  \left(
\frac{B_{0}}{0.5\times10^{-4}\mathrm{T} }\right)^{2} &\!\! \left(
\frac{N_{\mathrm{lin}}}{10^{14}\mathrm{m}^{-1}}\right)\!\!\left(
\frac{10^{11}\mathrm{m}^{-3}}{N_{0}}\right)\!.  %
\label{P}
\end{align}
If this criterion is satisfied when the trail initially forms, then
the instability starts generating plasma irregularities. If the
instability persists for the sufficient time, then turbulence will
develop and partially saturate through nonlinear processes. Because
$\rho(t)$ monotonically increases with time, see
Eq.~(\ref{rhotinterp}), then at some moment, $t=t_{\mathrm{FB}}$,
the two sides of Eq.~(\ref{FBcrit}) become equal. At this moment,
the linear growth of the FB instability starts being replaced by
linear damping and irregularities will diffuse away. At
$t<t_{\mathrm{FB}}$, the linear FB instability sustain plasma
turbulence at a certain level, while at $t>t_{\mathrm{FB}}$ there is
no more free energy to sustain the turbulence, so that
irregularities will quickly disappear.

The instability duration $t_{\mathrm{FB}}$ depends critically upon
the altitudinal parameter $\psi$ and the effective trail-background
interaction cross-section $\sigma_{\mathrm{eff}}\equiv
N_{\mathrm{lin}}/N_0$. Due to this, the nighttime conditions with
low $N_0$ will produce longer lived meteor trail than will the
daytime conditions. Figure \ref{dimantFig13} shows the altitude
dependence of the instability duration $t_{\mathrm{FB}}$ for several
constant $\sigma_{\mathrm{eff}}$ during daytime and nighttime
conditions. Notice clear peaks of $t_{\mathrm{FB}}$ at some
intermediate altitudes which increase with $\sigma_{\mathrm{eff}}$.
The non-specular echo boundary for a given trail, like that in
Fig.~\ref{dimantFig1}, should roughly follow the altitude dependence
of $t_{\mathrm{FB}}$. However, because column plasma density varies
along the meteor trail and due to other inhomogeneities, we expect
more variability than this model predicts. Measurements of the
evolution of the trail echoes, in combination with other
observations, should enable us to retrieve useful information about
meteors, ionosphere and atmosphere. Implementation of this procedure
requires better models of instability generation than that used
above, as well as models of ablation and ionization to give improved
estimates of $N_{\mathrm{lin}}$
\cite{Dyrud:Interpretation02,Dyrud2005JASTP}.

\section{Summary\label{summary}}

In this paper, we have described analytical theory and
finite-element simulations of trail diffusion and fields for the
mutually orthogonal trail axis and magnetic field. Unlike previous
models, this theory includes both the trail and the background
ionospheric plasma. This has two major effects: (1) a natural
restriction on the ambipolar electric field that otherwise would
infinitely grow with distance from the trail and (2) a later-stage
transition from sharply anisotropic (for $\psi\ll 1$) diffusion to
nearly isotropic one. The former is important for plasma
instabilities responsible for non-specular radar echoes, while the
latter is important for interpreting specular radar echoes.

A key element of the present treatment is the introduction of the
residual potential, $\phi_{\mathrm{res}}$, defined by
Eq.~(\ref{phiadd}). Its gradient describes the total force acting on
electrons. Due to high electron mobility along the magnetic field
$\vec{B}_0$, the typical scale of $\phi_{\mathrm{res}}$ spatial
variations in this directions are much larger than that in the
perpendicular direction, while the typical scales of trail density
variations in both directions are comparable. This fact, which had
not been realized in earlier simulations of meteor trail diffusion,
requires setting the simulation box boundary along $\vec{B}_0$ far
from the trail boundary and simultaneously resolving the two
different scales parallel to $\vec{B}_0$. In our simulations, we
have overcome the computational difficulties by employing a
finite-element software FlexPDE with the adaptive cell structure.

Based on the insight from simulations and using the large ratio of
the electron and ion mobilities along the magnetic field, we have
developed an approximate analytical approach. In this approach, the
problems of trail diffusion and of spatial distribution of the
residual potential are treated separately, while the coupling
between the two is made via parameters and approximate solutions. In
particular, Gaussian approximation of the trail peak has allowed us
to to treat the residual potential in terms of one parameter $\rho$.
This parameter is proportional to the square of the trail peak
dispersion, as described by Eq.~(\ref{rho}), and monotonically grows
with time. Due to high electron mobility along the magnetic field,
the total force acting on electrons in this direction is much
smaller than the corresponding components in other directions. For
the particular case of mutually orthogonal trail axis and the
magnetic field, this has allowed us to reduce the original 2D
description to a 1D linear integrodifferential equation given in two
different forms by Eq.~(\ref{integrodiff}). This governing equation
is for a residual potential in the near-trail zone and it has a
unique solution depending upon $\rho$. The approximate solution of
Eq.~(\ref{integrodiff}) depends on the range of $\rho$, as described
in Sect.~\ref{solution of equations}. Using this solution, we have
obtained the approximate expression for $\rho(t)$,
Eq.~(\ref{rhotinterp}), which closes the entire description of trail
diffusion and fields. Note that the near-trail potential, which is
easily spread along the magnetic field, may create significant
disturbances of the background plasma beyond the trail, as described
in Sect.~\ref{density disturbances}.

Comparison of analytical theory with simulations have demonstrated
good agreement between the results with one exception. At a later
stage of trail diffusion, the parameter $\rho$ for the residual
potential should be replaced by an adjusted parameter $\tilde\rho$
as described in Sect.~\ref{comparison of theory}. We have identified
the nature of this deviation analytically in terms of the
disturbances of the background plasma, but to quantitatively relate
$\tilde\rho$ and $\rho$ we have invoked simulations.

The analytical theory and simulations have allowed us to estimate
the spatial distribution of the ambipolar electric field within and
near the plasma trail. These fields are crucial for plasma
instabilities responsible for generation of field-aligned electron
density irregularities observed by high-power large-aperture (HPLA)
radars as non-specular echoes. Measuring the characteristics of
non-specular echoes and some other characteristics should allow one
to retrieve an important information on the meteoroids and the
surrounding atmosphere.

\begin{acknowledgments}
Work was supported by National Science Foundation Grants
No.~ATM-9986976, ATM-0332354, and ATM-0334906. Authors thank L.
Dyrud and T. Lin for their help and fruitful discussions.
\end{acknowledgments}

\appendix

\section{Self-similar solution\label{self-similar solution}}

In this Appendix, we obtain explicit expressions for the
self-similar solution (SSS) proposed in the general form, but not
found explicitly, by Jones \cite{JonesDiffusion1991}. In our
notations, this solution (denoted below by superscript
`$\mathrm{ss}$') follows the ansatz
$n(\vec{r},t)=n^{\mathrm{ss}}(x,y,t)$ and
$\phi_{\mathrm{res}}(\vec{r},t)=\phi_{\mathrm{res}}^{\mathrm{ss}}
(x,y,t)$, where
\begin{equation}
n^{\mathrm{ss}}(x,y,t)=\frac{C}{t}\exp\left(  -\ \frac
{A_{xx}x^{2}+A_{yy}y^{2}+A_{xy}xy}{4Dt}\right),%
\label{nSSanz}
\end{equation}%
\begin{equation}
\phi_{\mathrm{res}}^{\mathrm{ss}}
(x,y,t)=\frac{B_{xx}x^{2}+B_{yy}y^{2}+B_{xy}xy}{4Dt} + \mathrm{const},%
\label{phiSSanz}
\end{equation}
with positive diagonal coefficients $A_{ii}$, $B_{ii}$ and
\begin{equation}
A_{xx}A_{yy}>\frac{A_{xy}^{2}}{4},\qquad B_{xx}B_{yy}>\frac{B_{xy}^{2}}{4}.
\label{ineq2}%
\end{equation}
This is the solution to Eq.~(\ref{eqss}) in an infinite and
homogeneous neutral atmosphere with no background plasma, provided
the diffusion starts from an infinitely thin and dense plasma column
with a given line density. The electron Hall velocities give rise to
the non-diagonal coefficients, $A_{xy}=B_{xy}$. Inequalities given
by Eq.~(\ref{ineq2}) mean that the contours of the constant density
and the residual potential form ellipses in the $xy$-plane, whose
major axes are rotated with respect to the $x$ and $y$ axes through
a common angle $\chi$ determined by
\begin{equation}
\tan2\chi=\frac{A_{xy}}{A_{xx}-A_{yy}}=\frac{B_{xy}}{B_{xx}-B_{yy}}
\label{tanchi}%
\end{equation}
The constant $C$ in (\ref{nSSanz}) is expressed in terms of the
conserved linear density (along the $z$ coordinate),
$N_{\mathrm{lin}}$, as
\begin{equation}
C=\frac{\left(  4A_{xx}A_{yy}-A_{xy}^{2}\right)  ^{1/2}}{8\pi D} \,
\frac{N_{\mathrm{lin}}
}{N_{0}}. %
\label{C}
\end{equation}
For arbitrary electron and ion mobilities it is hard to obtain
explicit analytical expressions for the coefficients $A_{ij}$ and
$B_{ij}$. However, under restrictions described by Eq.~(\ref{Q}), to
leading order accuracy with respect to the
small parameters $Q^{-1}$, $\psi Q^{-1}$, $(\psi Q)^{-1}$, we obtain%
\begin{align}
A_{xx}  &  =1+B_{xx},\ \ \ B_{xx}=\frac{\sin^{2}\theta}{\psi},\nonumber\\
A_{yy}  &  =1+B_{yy},\ \ \ B_{yy}=\frac{1+\cos^{2}\theta}{Q}=\Theta_{0}%
^{2}\left(  \frac{1+\cos^{2}\theta}{\sin^{2}\theta}\right)\!,\label{SSsolution}\\
A_{xy}  &  =B_{xy}=-\ \frac{2\mu}{Q}\ \frac{\sin^{2}\theta}{\psi}%
=-\ \frac{2\Omega_{i}}{\nu_{in}}\cos\theta.\nonumber
\end{align}
The non-diagonal coefficient, $A_{xy}=B_{xy}$, is always small
compared to $A_{xx}$, and $A_{yy}\approx 1$, but not necessarily to
$B_{xx}$ and $B_{yy}\ll 1$. The rotation angle, Eq.~(\ref{tanchi}),
is small $\vert \tan2\chi\vert \approx{2\nu_{en}
\cos\theta}/({\Omega_{e}\sin^{2}\theta})$. Because
\[
\frac{A_{xy}^{2}}{4A_{xx}A_{yy}}\approx\frac{\mu^{2}}{Q^{2}}\ \frac{\sin
^{4}\theta}{\psi\left(  \psi+\sin^{2}\theta\right)  }=\frac{\Theta_{0}^{2}%
\cos^{2}\theta}{\psi+\sin^{2}\theta}\ll1,
\]
we have
\begin{equation}
C\approx\frac{N_{\mathrm{lin}}}{4\pi D
N_0}\left(1+\frac{\sin^{2}\theta}{\psi}\right)^{1/2}.%
\label{CC}
\end{equation}
The residual potential is stretched along the coordinate $y$ in
accord with the qualitative discussion in Sect.~\ref{formulation},
which holds under condition
\begin{equation}
\theta^{2}\gg\frac{\nu_{en}}{\Omega_{e}}=\Theta_{0}\sqrt{\psi}.%
\label{suffic}
\end{equation}
This restriction due to the electron Hall velocity is stronger than
that of Eq.~(\ref{>psi*theta0}). For the trail strictly
perpendicular to $\vec{B}$, $\cos\theta=0$, $\mu=0$, Hall velocity
is directed along $z$ and plays no role in 2D trail diffusion.

To the same accuracy, the particle fluxes are given by
\[
\Gamma_{ex}=\Gamma_{ix}=\frac{x}{2t}\
n^{\mathrm{ss}}(x,y,t),\qquad\!\!\Gamma
_{ey}=\Gamma_{iy}=\frac{y}{2t}\ n^{\mathrm{ss}}(x,y,t),
\]%
\begin{equation}
\Gamma_{iz}=0,\qquad\Gamma_{ez}=-\ \frac{\Omega_{e}}{\nu_{en}}\frac
{x\sin\theta}{2t}\ n^{\mathrm{ss}}(x,y,t) \label{Gammas}%
\end{equation}
In the SSS, the flux components in both $x$ and $y$ directions are
equal for electrons and ions. The only disparity is in the flux
component along the $z$-axis due to electron Hall velocity. The net
electric current is directed along the trail axis with the current
density
\begin{equation}
j_{z}=\frac{e\Omega_{e}}{\nu_{en}}\left(\frac{x\sin\theta}{2t}\right)\
n^{\mathrm{ss}
}(x,y,t). %
\label{currentdens}
\end{equation}

\section{Residual potential in the far zone\label{far zone}}

In this Appendix, we solve for $\phi_{\mathrm{res}}$ in the far zone
$|\zeta|\gg1$, where $\zeta=y/(Dt)^{1/2}$ is a coordinate parallel
to $\vec{B}_0$. This zone is located well beyond the plasma trail.
Neglecting plasma density disturbances allows us to reduce
Eq.~(\ref{Equ2}) to the Laplace equation
\begin{equation}
\partial_{\xi\xi}\phi_{\mathrm{res}}+\partial_{\zeta^{\prime}\zeta^{\prime}
}\phi_{\mathrm{res}}=0,%
\label{Laplace}
\end{equation}
where we have renormalized the variable $\zeta$ as $
\zeta^{\prime}=\alpha\zeta$ with
\begin{equation}
\alpha=\left(  \frac{1+\psi}{Q}\right)  ^{1/2}=\Theta_{0}\left( 1+\psi\right)
^{1/2}\ll1. \label{alpha}
\end{equation}
In terms of $\xi$ and $\zeta^{\prime}$, the near zone described in
the following Appendix, reduces to a thin strip extended along
$\xi$. Assuming that potential disturbances have typical scales of
order unity with respect to both coordinates, we can approximate
this strip by a cut in the $\xi,\zeta^{\prime}$-plane at
$\zeta^{\prime}=\pm0$. Here the signs $\pm$ mean infinitesimal
offsets from zero to either positive or negative directions. Because
of the mirror symmetry with respect to the $\xi$-axis,
Eqs.~(\ref{phi+-}), we will consider only the positive half-space,
$\zeta^{\prime}>0$.

The nearly constant value of the residual potential in the near zone represents
a boundary condition for $\phi_{\mathrm{res}}(\xi,\zeta^{\prime})$ at the cut,
\begin{equation}
\left[  \phi_{\mathrm{res}}(\xi,\zeta^{\prime})\right]  _{\zeta^{\prime}=\pm
0}\approx\phi_{\mathrm{res}}^{0}(\xi). %
\label{granusl}
\end{equation}
Other boundary conditions are given by
$\phi_{\mathrm{res}}(\xi,\zeta^{\prime})\rightarrow0$ as $
\xi,\zeta^{\prime}\rightarrow\pm\infty$. The solution of
Eq.~(\ref{Laplace}) with these boundary conditions (the Dirichlet
problem for the upper half-space, $\zeta'>0$) is given by
\begin{eqnarray}
\phi_{\mathrm{res}}(\xi,\zeta^{\prime})&=&\frac{\zeta^{\prime}}{\pi}
\int_{-\infty}^{\infty}\frac{\phi_{\mathrm{res}}^{0}(\tau)\,d\tau}
{(\tau-\xi)^{2}
+\zeta^{\prime2}}\nonumber\\
&=&\frac{\alpha\zeta}{\pi}\int_{-\infty}^{\infty}\frac{\phi
_{\mathrm{res}}^{0}(\tau)\,d\tau}{(\tau-\xi)^{2}+\alpha^{2}\zeta^{2}}%
\label{solu}
\end{eqnarray}
In particular, for $\xi^{2} +\zeta^{\prime2}\gg 1$, Eq.~(\ref{solu}) reduces to
a 2D quadrupole,
\begin{equation}
\phi_{\mathrm{res}}(\xi,\zeta^{\prime})\approx \frac{C(\xi^{2}
-\zeta^{\prime 2})}{(\xi^{2}+\zeta^{\prime2})^{2}}.
\label{quadr}
\end{equation}
In polar coordinates, $r$ and $\theta$ are defined as
$\xi=r\cos\theta $, $\zeta^{\prime}=r\sin\theta$, the quadrupole
potential is $\phi_{\mathrm{res}}(r,\theta)\approx
(C\cos2\theta)/r^{2}$, while the electric field lines of force are
determined by $(\sin2\theta)/r^{2}=\mathrm{const}$. Here the
constant $C$ is determined by the distribution of
$\phi_{\mathrm{res}}(\xi)$ along the cut $\zeta=0$, as discussed in
Appendix~\ref{integrodifferential equations}.

Now we obtain general relations which follow from Laplace
Eq.~(\ref{Laplace}) and will be used in Appendix~\ref{near-zone
potential} to derive a closed equation for
$\phi_{\mathrm{res}}^0(\xi)$ when combined with the residual
potential in the near zone. Introducing a complex coordinate,
\begin{equation}
Z=\xi+i\zeta^{\prime},%
\label{z}
\end{equation}
we consider $\zeta^{\prime}>0$ as a complex half-plane and introduce
a complex potential,
\begin{equation}
\Psi\left(  Z\right)  =\phi_{\mathrm{res}}(\xi,\zeta^{\prime}) +
iW\left(\xi,\zeta^{\prime}\right), \label{Psi}
\end{equation}
where the function $W(\xi,\zeta)$ also satisfies Laplace
Eq.~(\ref{Laplace}). This is an analytic function of the complex
coordinate $Z$ with $\phi_{\mathrm{res}}$ and $W$ related by the
Cauchy-Riemann equations,
\begin{equation}
\partial_{\xi}\phi_{\mathrm{res}}=\partial_{\zeta^{\prime}}W,\qquad
\partial_{\zeta^{\prime}}\phi_{\mathrm{res}}=-\ \partial_{\xi}W.
\label{CauRiem}
\end{equation}
At the cut, $\zeta^{\prime}=+0$, we denote $ w^{0}(\xi)\equiv
W(\xi,+0)$, so that from Eq.~(\ref{CauRiem}) we have
\begin{equation}
\partial_{\xi}\phi_{\mathrm{res}}^{0}=\left.  \partial_{\zeta^{\prime}}W\right|
_{\zeta^{\prime}=+0},\qquad\left.
\partial_{\zeta}\phi_{\mathrm{res}}\right|_{\zeta =+0}=-\
\partial_{\xi}w^{0}. \label{CaRi}
\end{equation}
According to the mirror symmetry, Eq.~(\ref{phi+-}), the function $\phi
_{\mathrm{res}}^{0}(\xi)$ is even, while $w^{0}(\xi)$ is odd,
\begin{equation}
\phi_{\mathrm{res}}^{0}(\xi)=\phi_{\mathrm{res}}^{0}(-\xi), \qquad
w^{0}(\xi)=-w^{0}(-\xi)\,.%
\label{evenodd}
\end{equation}
At large $Z$, the residual potential
$\phi_{\mathrm{res}}\propto1/|Z|^{2}$ decreases faster than $1/|Z|$
as $|Z|\rightarrow\infty$. According to the Cauchy-Goursat theorem,
$\oint_{C}\Psi(Z)dZ=0$, so that we have
\begin{equation}
\int_{-\infty}^{\infty}\phi_{\mathrm{res}}^{0}(\xi)d\xi=0\,,
\qquad\int_{-\infty}^{\infty}w^{0}(\xi)d\xi=0. %
\label{intphi=0}
\end{equation}
While the integral relation for $w^{0}(\xi)$ is trivial because
$w^{0}$ is odd, the integral relation for the even function
$\phi^{0}_{\mathrm{res}}(\xi)$ represents an important constraint.

Now we proceed with the complex potential $\Psi(Z)$. In the upper
half-plane, we consider the continuous contour $C$, which includes
the axis $\zeta^{\prime}=+0$ with an infinitesimal half-circle
around $Z^{\prime }=\xi+i0$, and the infinite half-circle,
$|Z|\rightarrow\infty$. Because $\Psi$ is an analytic function and
the pole $Z=Z^{\prime}$ is beyond the area closed by this contour,
the residue theorem yields
$\oint_{C}{\Psi(Z)}/(Z-Z^{\prime})\,dZ=0$, or
\begin{equation}
\mathcal{P}\!\!\int_{-\infty}^{\infty}\frac{\Psi\left(\tau\right)}
{\tau-\xi}\,d\tau=i\pi\Psi\left(  \xi\right)  , \label{KraKroComp}
\end{equation}
where $\mathcal{P}$ denotes the principal value of the integral along the real
axis. Separating in Eq.~(\ref{KraKroComp}) the real and imaginary part, we
obtain
\begin{subequations}
\label{KraKro}
\begin{align}
\phi_{\mathrm{res}}^{0}(\xi)  &
=\frac{1}{\pi}\,\mathcal{P}\!\!\int_{-\infty}^{\infty}
\frac{w^{0}\left(  \tau\right)  }{\tau-\xi}\,d\tau,\label{KraKro1}\\
w^{0}(\xi)  &  =-\,\frac{1}{\pi}\,\mathcal{P}\!\!\int_{-\infty}^{\infty}
\frac{\phi_{\mathrm{res}}^{0}\left(  \tau\right)  }{\tau-\xi}\,d\tau.
\label{KraKro2}
\end{align}
\end{subequations}
These equations are equivalent to the well-known Kramers-Kronig
dispersion relations in optics, plasma physics, etc., which are
derived in the same way. Applied to an arbitrary function
$\phi_{\mathrm{res}}^{0}(\xi)$ [or $w^{0}(\xi)$], Eq.~(\ref{KraKro})
is also known as the Hilbert transform \cite{Papoulis, Bracewell}.
Equation (\ref{KraKro}) will be applied in Appendix~\ref{near-zone
potential} and Sect.~\ref{solution of equations}.

\section{Near-zone potential\label{near-zone potential}}

In this Appendix, using the results of Appendix~\ref{far zone}, we
derive the governing equation for the residual potential in the near
zone, $|\zeta|\ll 1/\alpha$, where $\alpha =\Theta
_{0}(1+\psi)^{1/2}\ll 1$ is defined by Eq.~(\ref{alpha}). To obtain
a closed equation for $\phi_{\mathrm{res}}(\xi)$, we write
Eq.~(\ref{Equ}) for $\mu=0$ in a conservative form as
\begin{subequations}
\label{Eqperp}%
\begin{eqnarray}
\partial_{t}\left(t\Delta n\right) & -&\frac{\partial_{\xi}\left(\xi\Delta
n\right)  +\partial_{\zeta}\left(  \zeta\Delta n\right)
}{2}-\vec\nabla^{2}\Delta n\nonumber\\
&-&\vec\nabla\cdot\left[  \left(  1+\Delta n\right)  \vec\nabla\phi_{\mathrm{res}%
}\right] =0,%
\label{Eqperp1}
\end{eqnarray}
\begin{eqnarray}
\partial _{\xi }\left[ \left( 1+\Delta n\right)
\partial _{\xi }\phi _{\mathrm{res}}\right] &+&\frac{1}{\alpha ^{2}}\ \partial _{\zeta
}\left[ \left( 1+\Delta n\right) \partial _{\zeta }\phi
_{\mathrm{res}}\right] \nonumber\\
&+&\frac{\vec\nabla ^{2}\Delta n}{1+\psi }=0, %
\label{Eqperp2}
\end{eqnarray}
\end{subequations}
where $\Delta n\equiv \Delta N/N_0$ is the total density disturbance
which includes both the plasma trail and the background plasma
disturbance. We assume that the major density disturbances are
concentrated within the near zone, $\zeta\ll 1/\alpha$, where the
residual potential is only weakly $\zeta$-dependent. In this zone,
which overlaps with the far zone $\zeta\gg 1$, we represent the
residual potential as
\begin{equation}
\phi _{\mathrm{res}}(\xi,\zeta )=\phi
_{\mathrm{res}}^{\mathrm{far}}(\xi,\zeta) +\delta \phi
(\xi,\zeta),\qquad |\delta \phi| \ll
|\phi_{\mathrm{res}}^{\mathrm{far}}|, %
\label{finot}
\end{equation}
where $\phi _{\mathrm{res}}^{\mathrm{far}}(\xi,\zeta)$ is the
component in the far zone and therefore satisfies the Laplace
equation described in Appendix~\ref{far zone}. Because most of the
potential changes occur in the background plasma (far zone), our
assumption $|\delta \phi| \ll |\phi_{\mathrm{res}}^{\mathrm{far}}|$
is well justified as can be seen in the example solution shown in
Fig.~\ref{PotenDens}. The function $\phi
_{\mathrm{res}}^{\mathrm{far}}(\xi,\zeta)$ varies along $\zeta$ with
a large scale-length typical for the far zone, so that within the
near-zone it varies approximately linearly
\begin{equation}
\phi_{\mathrm{res}}^{\mathrm{far}}(\xi,\zeta)\approx
\phi_{\mathrm{res}}^{\mathrm{far}}(\xi,0) + \zeta\partial_\zeta
\phi_{\mathrm{res}}^{\mathrm{far}}(\xi),%
\label{lineardob}
\end{equation}
where $|\zeta\partial_\zeta \phi_{\mathrm{res}}^{\mathrm{far}}| \ll
|\phi_{\mathrm{res}}^{\mathrm{far}}(\xi,0)|$, with the derivative
$\partial_\zeta \phi_{\mathrm{res}}^{\mathrm{far}}$ remaining nearly
$\zeta$-independent across a transitional zone between the near and
far zones as seen in Fig.~\ref{PotenDens}. The perturbation $\delta
\phi(\xi,\zeta)$, associated with strong density disturbances in the
near zone, is relatively small, but its $\zeta$-derivative is not
small compared to
$\partial_\zeta\phi_{\mathrm{res}}^{\mathrm{far}}$. Furthermore, it
is this perturbation that provides a smooth transition from the zero
$\zeta$ derivative of $\phi_{\mathrm{res}}(\xi,\zeta)$ at the
symmetry center, $\left.\partial_\zeta
\phi_{\mathrm{res}}\right|_{\zeta=0}=0$,
\begin{equation}
\left.\partial_\zeta \delta\phi_{\mathrm{res}}\right|_{\zeta=0}=
-\partial_\zeta
\phi_{\mathrm{res}}^{\mathrm{far}},%
\label{deriv-}
\end{equation}
to the finite derivatives $\partial_\zeta\phi_{\mathrm{res}}
\approx\partial_\zeta\phi_{\mathrm{res}}^{\mathrm{far}}$ in the far
zone where $\delta \phi(\xi,\zeta)$ gradually disappears. Now we
subtract from Eq.~(\ref{Eqperp2}) the Laplace equation for
$\phi_{\mathrm{res}}^{\mathrm{far}}$ and integrate over $\zeta $
from $0$ to $\zeta =\zeta _{1}$, where $\zeta_1\ll \alpha^{-1}$ is
located in the transitional zone where both $\delta \phi$ and
$\partial_\zeta\delta \phi$ are zero. Also, at $\zeta=\zeta_1$ we
neglect the density disturbances $\Delta n$ and the perturbations
$\delta\phi_{\mathrm{res}}$ with their derivatives, except in the
term $\propto 1/\alpha^2$. We will also ignore within the range
$|\zeta|<\zeta_1$ all density disturbances other than the trail
plasma itself. (This introduces some error which we correct in
Sect.~\ref{weakest case} and Appendix~\ref{effective integral}.) The
range $|\zeta|<\zeta_1$ includes practically the entire trail, so
that we can extend the upper limit of all density integrals to
infinity. As a result, using Eq.~(\ref{deriv-}), we obtain
\begin{equation}
\partial _{\xi }\left( \Delta N_\zeta\,\partial _{\xi }\phi
_{\mathrm{res}}^{\mathrm{far}}\right) +\frac{1}{\alpha ^{2}}\,\partial _{\zeta
}\phi _{\mathrm{res}}^{\mathrm{far}} +\left( \frac{1}{1+\psi }\right)
\partial
_{\xi \xi }^{2}\left( \Delta N_\zeta\right) =0.%
\label{prok}
\end{equation}
Here
\begin{equation}
\Delta N_{\zeta_1}(\xi)\equiv\int_{0}^{\zeta _{1}}\Delta n\, d\zeta
\approx
\Delta N_{\mathrm{Trail}}(\xi)\equiv\int_{0}^{\infty }\Delta n_{\mathrm{Trail}}\ d\zeta .%
\label{extend}
\end{equation}
where $\Delta n_{\mathrm{Trail}}$ is the part of the total density
disturbance associated only with the trail plasma.

To good accuracy, the trail plasma distribution is described by a Gaussian
distribution, see Sect.~\ref{numerical analysis} and \ref{trail diffusion},
\begin{equation}
\Delta n_{\mathrm{Trail}}(\xi ,\zeta ,t) =
\frac{n_{0}t_{0}}{t}\sqrt{\frac{q_{0}}{q(t)}}\exp \left[ -\left(
\frac{\xi
^{2}}{4q(t)}+\frac{\zeta ^{2}}{4} \right) \right],%
\label{Gauss}
\end{equation}
where $q_0=\psi/(1+\psi)$ and $q(t)$ gradually varies between $q_0$
and 1. Because $\phi_{\mathrm{res}}^{\mathrm{far}}$ is the far-zone
potential, we apply Eq.~(\ref{CaRi}), $\partial _{\zeta}
\phi_{\mathrm{res}}^{\mathrm{far}} = -\,\alpha\partial _{\xi
}w^{0}(\xi)$. Bearing in mind Eq.~(\ref{lineardob}), in
Eq.~(\ref{prok}) we can eliminate all $\zeta$-dependence, $\phi
_{\mathrm{res}}^{\mathrm{far}}(\xi,\zeta)\approx
\phi_{\mathrm{res}}^{0}(\xi)
\equiv\phi_{\mathrm{res}}^{\mathrm{far}}(\xi,0)$, and obtain
\begin{equation}
\partial _{\xi }\left( \Delta N_{\zeta_1}\,
\partial _{\xi }\phi_{\mathrm{res}}^{0}\right)
- \frac{1}{\alpha}\,\partial_{\xi
}w^{0}+\left(\frac{1}{1+\psi}\right)
\partial _{\xi \xi }^{2}\Delta N_{\zeta_1}=0,%
\label{nepka}
\end{equation}
where
\begin{equation}
\Delta N_{\zeta_1}(\xi)\approx\frac{\sqrt{\pi
}n_{0}t_{0}}{t}\sqrt{\frac{q_{0}}{q(t)}}\exp \left( -\ \frac{\xi
^{2}}{4q(t)}\right).%
\label{G}
\end{equation}
Integrating Eq.~(\ref{nepka}) with the boundary condition $w^{0}\rightarrow 0$
at $\left\vert \xi \right\vert \rightarrow \infty $ and expressing $w^{0}$ in
terms of $\phi_{\mathrm{res}}^{0}$ according to (\ref{KraKro2}), we obtain
\begin{equation}
\partial _{\xi }\phi_{\mathrm{res}}^{0}+\frac{1}{\pi \alpha \Delta N_{\zeta_1}}\ \mathcal{P}
\!\!\int_{-\infty }^{\infty }\frac{\phi_{\mathrm{res}}^{0}(\chi
)}{\chi -\xi }\ d\chi =
\frac{\xi }{2\left( 1+\psi \right) q(t)}\,.%
\label{prom}
\end{equation}
Renormalizing the potential and coordinate as follows,
\begin{equation}
\varphi =2\left( 1+\psi \right) \phi_{\mathrm{res}}^{0},\qquad \xi
=\sqrt{q}\eta ,
\label{reno}
\end{equation}
we arrive at an integrodifferential equation
\begin{equation}
\partial _{\eta }\varphi +\frac{2\rho e^{\eta ^{2}/4}}{\pi ^{3/2}}\mathcal{P}%
\!\!\int_{-\infty }^{\infty }\frac{\varphi (\tau )}{\tau -\eta }\ d\tau =\eta ,
\label{FF2}
\end{equation}%
which depends only upon one dimensionless parameter
\begin{equation}
\rho(t) =\frac{\sqrt{\pi q(t)}}{2\alpha \Delta
N_{\zeta_1}(0)}=\gamma q(t)t,\qquad \gamma
=\frac{1}{2\Theta _{0}n_{0}t_{0}\sqrt{\psi }} =\frac{\rho_{0}}{t_{0}q_{0}}\,.%
\label{rhop}
\end{equation}
Here
\begin{equation}
\rho _{0}\equiv \rho(t_0)=\frac{\sqrt{\psi }}{2\Theta _{0}n_{0}\left( 1+\psi
\right) }.%
\label{rho0}
\end{equation}
Multiplying Eq.~(\ref{FF2}) by $\exp ({-\eta ^{2}/4})$, applying the
Hilbert transform, Eq.~(\ref{KraKro}), and using the identity
\begin{equation*}
\mathcal{P}\!\!\int_{-\infty }^{\infty }\frac{1}{\tau -\eta } \left(
\mathcal{P}\!\!\int_{-\infty }^{\infty }\frac{\varphi \left( {\chi }\right)
}{{\chi }-\tau }\,d{\chi }\right) d\tau =-\pi ^{2}\varphi (\eta ),
\end{equation*}
we arrive at a different form of integrodifferential Eq.~(\ref{FF2}),
\begin{equation}
\rho \varphi (\eta )-\frac{1}{2\sqrt{\pi }}\ \mathcal{P}\!\!\int_{-\infty
}^{\infty }\frac{\partial _{\tau }\varphi (\tau )}{\tau -\eta }\,e^{-\tau
^{2}/4}d\tau =S(\eta ).%
\label{FF1}
\end{equation}
The function $S(\eta )$ in the RHS of Eq.~(\ref{FF1}) stems from the
integration
\begin{equation}
S(\eta) =-\ \frac{1}{2\sqrt{\pi }}\ \mathcal{P} \!\!\int_{-\infty
}^{\infty }\frac{\tau e^{-\tau ^{2}/4}}{\tau -\eta }\ d\tau %
\label{Sin}
\end{equation}
and can be recast as
\begin{eqnarray}
S(\eta) &=&\eta e^{-\,\eta ^{2}/4}\!\!\int_{0}^{\eta /2}\!\! e^{\tau
^{2}}d\tau -1\nonumber\\
&=&-\left[ \frac{i\sqrt{\pi }}{2}\, \eta e^{-\,\eta
^{2}/4}\mathrm{erf}\left(
\frac{i\eta }{2}\right) +1\right],%
\label{S}
\end{eqnarray}
where $\mathrm{erf}(x)$ denotes the standard error-function.

Now we discuss some properties of the function $S(\eta)$, which will
be used in Sect.~\ref{solution of equations} and others. Firstly, we
observe that Eq.~(\ref{Sin}) shows that $S(\eta )$ and $(\sqrt{\pi
}/2) \eta \exp(-\eta ^{2}/4)$ form a Hilbert transform pair,
Eqs.~(\ref{KraKro}a,b), so that
\begin{equation}
\mathcal{P}\!\!\int_{-\infty }^{\infty }\frac{S(\tau )}{\tau -\eta }\,d\tau =
\frac{\pi ^{3/2}}{2}\,\eta \exp \left( -\,\frac{\eta ^{2}}{4}\right) .%
\label{intS}
\end{equation}
Considering the double integral
\begin{equation*}
K=\int_{-\infty }^{\infty }d\eta \ \mathcal{P}\!\!\int_{-\infty }^{\infty }
\frac{F(y)}{y-\eta }\ dy,
\end{equation*}
with an arbitrary integrable function $F(y)$ and changing the order
of integration, we verify that $K=0$. Applying this to
Eq.~(\ref{Sin}), we conclude that the even function $S(\eta )$ obeys
\begin{equation}
\int_{0}^{\infty }S(\eta )\,d\eta =0.%
\label{intS=0}
\end{equation}
The function $S(\eta)$ has important power-series approximations at
sufficiently small and large values of $|\eta|$. The function $S(\eta)$ can be
represented as an infinite Taylor series,
\begin{eqnarray}
S(\eta) & =&-1+\frac{1}{2}\sum_{n=1}^{\infty}\left(  -1\right)
^{n-1}\frac{\eta^{2n}\left(  n-1\right)  !}{\left(  2n-1\right)!}
\nonumber\\
&=&-1+\frac{\eta^{2}}{2}-\frac{\eta^{4}}{2\times6}+\frac{\eta^{6}}
{2\times6\times10}-\ldots %
\label{series}
\end{eqnarray}
which is convergent for all values of $\eta$. A truncated series
with a few first terms approximates the function at $|\eta| <2$
well. At $|\eta| >4$, the function $S(\eta)$ can be approximated
well by an asymptotic series,
\begin{equation}
S(\eta)  \approx 2\sum_{n=1}^{n_{\max}}\frac{\left(2n-1\right)
!}{(n-1) !\eta^{2n}}=\frac{2}{\eta^{2}}+\frac{2\times6}
{\eta^{4}}+\frac{2\times6\times10}{\eta^{6}}+\ldots %
\label{asympt}
\end{equation}
Unlike the Taylor series, this asymptotic series is divergent, so
that the total number of terms $n_{\max}$ should not be too large.

To conclude this Appendix, we calculate the integral
\begin{equation}
J(\eta)=-\,\frac{1}{2\sqrt{\pi}}\ \mathcal{P}\!\!\int_{-\infty}^{\infty}
\frac{\partial_{\tau}S(\tau)}{\tau-\eta}\,e^{-\tau^{2}/4}d\tau,%
\label{J}
\end{equation}
needed to obtain the first-order correction of the later-stage
residual potential in Sect.~\ref{late stage diffusion}. Exact
integration yields
\begin{eqnarray}
J(\eta)&=&\left(  1-\frac{\eta^{2}}{2}\right)  e^{-\eta^{2}/2}\left[
\frac {1}{2}\left(  \int_{0}^{\eta/2}e^{y^{2}}dy\right)
^{2}-\frac{\pi}{8}\right] \nonumber\\
&+&\frac{1}{2}\left(  \eta
e^{-\eta^{2}/4}\int_{0}^{\eta/2}e^{y^{2}}dy-\frac
{1}{2}\right).  %
\label{Jexact}
\end{eqnarray}
The function $J(\eta)$ looks qualitatively as $S(\eta)$ and to a good accuracy
can be approximated by a simpler expression
\begin{equation}
J(\eta)\approx\tilde{J}(\eta)=lS(p\eta),%
\label{Jprime}
\end{equation}
where the constants $l$ and $p$ are given by%
\begin{subequations}
\label{lp}%
\begin{align}
l  & =\frac{\pi+2}{8}\approx 0.643,\label{l}\\
p  & =\left(  \frac{3+\pi}{1+\pi/2}\right)  ^{1/2}\approx
1.546.\label{p}%
\end{align}
\end{subequations}
We have chosen these constants to provide the best parabolic fit
between $J(\eta)$ and $\tilde{J}(\eta)$ at $\eta=0$. The biggest
difference between the exact and approximate expressions for
$J(\eta)$ is near the maxima of $J(\eta)$ ($|\eta|\approx 2$), where
it reaches about 0.016 (less than $10\%$, see
Fig.~\ref{comparison1}). In all other locations, the functions
$J(\eta)$ and $\tilde{J}(\eta)$ are much closer to each other.

\section{Properties of residual potential \label{integrodifferential
equations}}

In this Appendix, we discuss general properties of solutions of
Eq.~(\ref{integrodiff}). The solution of $\varphi(x)$ has the
following important property,
\begin{equation}
\int_{0}^{\infty}\varphi(x)dx=0,%
\label{int=0}
\end{equation}
which follows from general properties of analytic functions [see
Eq.~(\ref{intphi=0})] and symmetry (\ref{phi+-}), and is necessary
for the consistency of Cauchy type integrodifferential
Eq.~(\ref{integrodiff}). We may rewrite this relation in terms of
the $x\ $component of the electric field
$E_{x}=-\partial_{x}\varphi$. Expressing the residual potential as
$\varphi(x)=\int_{x}^{\infty}E_{\tilde{x}}d\tilde{x}$, substituting
this in Eq.~(\ref{int=0}), and changing the order of integration, we
obtain
\begin{equation}
\int_{0}^{\infty}E_{x}(x)xdx=0.%
\label{Exdx=0}
\end{equation}
In addition, Eqs.~(\ref{integrodiff}a,b) impose restrictions on
possible asymptotic behavior of the residual potential at large
distances. For $\left\vert \eta\right\vert \gg1$, we expect the
function $\varphi(\eta)$ to behave as
\begin{equation}
\varphi(\eta)\simeq\sum_{k=1}^{k_{\max}}\frac{C_{k}}{\eta^{2k}},%
\label{asympto}
\end{equation}
which corresponds to the expansion of the electric potential in
multipoles. Note that the lowest-order coefficient $C_1$, in accord
with Eq.~(\ref{reno}), relates to the quadrupole coefficient $C$ in
Eq.~(\ref{quadr}) as $C_1=2(1+\psi)C$. The maximum value of $k$,
$k_{\max}$, is determined by $\eta$ (the asymptotic series may
diverge as $k\rightarrow\infty$). The electric field proportional to
$\partial_{\eta}\varphi$ tends to zero as $\left\vert
\eta\right\vert \rightarrow\infty$. This means that the RHS\ of Eq.
(\ref{Form2}), i.e., the term $\eta$, should asymptotically match
the integral term in the LHS\ of Eq. (\ref{Form2}). The
exponentially growing factor in front of the principal value
integral requires the integral to decrease as
$|\eta|\rightarrow\infty$ faster than $|\eta|^{-n}$ with positive
$n$. Analyzing the asymptotic behavior of the integral, one can
obtain the following restriction,
\begin{equation}
\int_{-\infty}^{\infty}\left(  \varphi\left(  \eta\right)  -\sum_{k=1}^{m}%
\frac{C_{k}}{\eta^{2k}}\right)  \eta^{2m}d\eta=0,%
\label{hypot}
\end{equation}
valid for any positive integer $m$ [in the case of $m=0$, we obtain
Eq.~(\ref{int=0})]. From Eq.~(\ref{hypot}) it follows that the
residual potential cannot be an exponentially decreasing function of
$\eta$ as $|\eta|\rightarrow\infty$, but must have a power-law
asymptotic behavior (as required by the multipole expansion).
Indeed, all coefficients $C_{n}$ cannot equal zero because no
non-zero $\varphi(x) $ could satisfy Eq.~(\ref{hypot}) for all
positive integer $m$.

The self-similar solution (SSS) for the residual potential, defined
in Appendix~\ref{self-similar solution} (to the accuracy of an
arbitrary constant), in variables defined by Eq.~(\ref{reno}) can be
written as $\varphi(0)+\eta^{2}/2$. Introducing the difference
between the actual potential $\varphi(\eta)$ and the SSS,
\begin{equation}
\delta\varphi(\eta)\equiv\varphi(\eta)-\varphi(0)-\frac{\eta^{2}}{2},%
\label{deltaPhi}
\end{equation}
we rewrite Eqs.~(\ref{integrodiff}) as a Hilbert transform pair:
\begin{subequations}
\label{F1F2}
\begin{align}
&  \partial_{\eta}\delta\varphi(\eta)e^{-\ \frac{\eta^{2}}{4}}=-\ \frac{2\rho
}{\pi^{3/2}}\ \mathcal{P}\!\!\int_{-\infty}^{\infty}\frac{\varphi(\tau)}%
{\tau-\eta}\,d\tau.%
\label{F2}\\
&  \rho\varphi(\eta)=\frac{1}{2\sqrt{\pi}}\ \mathcal{P}\!\!\int_{-\infty}%
^{\infty}\frac{\partial_{\tau}\delta\varphi(\tau)}{\tau-\eta}\,e^{-\ \frac
{\tau^{2}}{4}}d\tau,%
\label{F1}
\end{align}
\end{subequations}
We expect $ \delta\varphi(\eta)\approx -\lambda\eta^{2}$ as
$|\eta|\ll 1$ with a positive constant $\lambda$ and
$\delta\varphi(\eta)\approx -\eta^{2}/2$ as $|\eta
|\gg1,\label{nerdelPhi2}$. Considering the limits of small and large
values of $|\eta|$, we obtain integral relations
\begin{eqnarray}
\varphi(0) & =&\frac{1}{\sqrt{\pi}\rho}\int_{0}^{\infty}
\frac{\partial_{\tau
}\delta\varphi(\tau)}{\tau}\,e^{-\ \frac{\tau^{2}}{4}}d\tau,\nonumber\\
\left.  \frac{d\varphi(\eta)}{d(\eta^{2})}\right|_{\eta=0} &
=&\frac{1} {\sqrt{\pi}\rho}\int_{0}^{\infty}\frac{1}{\tau}\
\partial_{\tau}\left(
\frac{\partial_{\tau}\delta\varphi(\tau)}{\tau}\,e^{-\
\frac{\tau^{2}}{4}
}\right)  d\tau,\nonumber\\
\lim_{|\eta| \rightarrow \infty}[\varphi(\eta)\eta^{2}] & \equiv &
2\left(  1+\psi\right)  C_{1}\label{constr}\\
&=&-\ \frac{1}{\sqrt{\pi}\rho}
\int_{0}^{\infty}\partial_{\tau}\delta\varphi(\tau)\ \tau e^{-\
\frac{\tau^{2} }{4}}d\tau, \nonumber
\end{eqnarray}
which describe the parabolic behavior of the residual potential near
its minimum and express the highest-order multipole coefficient in
terms of deviation from the self-similar potential.
Equations~(\ref{int=0}), (\ref{hypot}), (\ref{constr}), as well as
the exact relations obtained in the following Appendix, can be used
for checking the solution.

\section{Density disturbance integrals\label{DDI}}

In this Appendix, we derive an important exact property of density
disturbance integrals (DDIs), which have been applied for monitoring
the simulation accuracy, see Sect.~\ref{numerical analysis} and
\ref{comparison of theory}. We define the two DDIs by integrating
parallel and perpendicular to $\vec{B}_0$,
\begin{subequations}
\begin{eqnarray}
\Delta N_{\zeta}(\xi,t)&=&\int_{-\infty}^{\infty}\Delta
n(\xi,\zeta,t)d\zeta,\label{DelNzeta}\\
\Delta N_{\xi}(\zeta,t)&=&\int_{-\infty}^{\infty}\Delta
n(\xi,\zeta,t)d\xi ,%
\label{DelNxi}
\end{eqnarray}
\end{subequations}
where $\Delta n$ include both the plasma trail density and
disturbances of the background plasma. If we solve Eq.~(\ref{Equ})
with the self-similar solution (SSS) as the initial condition then
the DDIs multiplied by $t$, for all coordinates $\zeta$ or $\xi$,
are strictly conserved and equal to those for the SSS, even though
the SSS is, in general, invalid.

To prove this and obtain the explicit analytical expressions for the
DDIs, we start from Eqs.~(\ref{Eqperp}a,b). Adding them, we
eliminate $\xi$-derivatives of $\phi_{\mathrm{res}}$,
\begin{eqnarray}
\partial_{t}\left(  t\Delta n\right)  &-&\frac{\partial_{\xi}\left(  \xi\Delta
n\right)  +\partial_{\zeta}\left(\zeta \Delta n\right)  }{2}
-\frac{\psi\,\vec\nabla^{2}\Delta n}{1+\psi}\nonumber\\ &+&
\frac{1-\alpha^2}{\alpha^2}\,\partial_{\zeta
}\left[  \left(  1+\Delta n\right)  \partial_{\zeta}\phi_{\mathrm{res}%
}\right]  =0\,. %
\label{sub111}
\end{eqnarray}
Because both the density disturbance and the potential go to zero at
infinity, we integrate this equation along the coordinate $\zeta$
from $-\infty$ to $\infty$ and obtain
\begin{equation}
\partial_{t}\left(  t\Delta N_{\zeta}\right)  -\partial_{\xi}\left[  \left(
\frac{\xi}{2}+\frac{\psi}{1+\psi}\,\partial_{\xi}\right)  \Delta N_{\zeta
}\right]  =0. \label{intzeta}%
\end{equation}
Similarly, multiplying Eq.~(\ref{Eqperp2}) by $\alpha^2$ and adding
with (\ref{Eqperp1}), we eliminate $\zeta$-derivatives of
$\phi_{\mathrm{res}}$. After the integration over $\xi$ we obtain
\begin{equation}
\partial_{t}\left(  t\Delta N_{\xi}\right)  =\partial_{\zeta}\left[  \left(
\frac{\zeta}{2}+\,\partial_{\zeta}\right)  \Delta N_{\xi}\right].
\label{intxi}%
\end{equation}
If we use the SSS as the initial condition at $t=t_{0}$ in
Eqs.~(\ref{intzeta}) and (\ref{intxi}) then $\left[
\partial_{t}\left(  t\Delta N_{\xi}\right)  \right]  _{t=t_{0}}=0$
and$\ \left[  \partial_{t}\left(  t\Delta N_{\zeta}\right)  \right]
_{t=t_{0}}=0$. The unique solutions of Eqs.~(\ref{intzeta}) and
(\ref{intxi}) will keep these relations in their self-similar form,
\begin{subequations}
\label{zexi}
\begin{align}
\Delta N_{\zeta}(\xi,t)  &  =\frac{2C}{t}\,\sqrt{\pi}\,\exp\left[  -\left(
\frac{1+\psi}{\psi}\right)  \frac{\xi^{2}}{4}\right]  ,\label{zeta1}\\
\Delta N_{\xi}(\zeta,t)  &  =\frac{2C}{t}\,\sqrt{\frac{\pi\psi}{1+\psi}}
\,\exp\left(  -\ \frac{\zeta^{2}}{4}\right)  , \label{xi1}
\end{align}
\end{subequations}
for all $t$. This completes the proof.

\section{Effective integral of density disturbances\label{effective integral}}

In this Appendix, we calculate the effective integral over near
zone, as discussed in Sect.~\ref{weakest case}. If the trail density
is the major contribution to the integral over the effective region
then $\Delta N_{\zeta_1}\approx\Delta N_{\mathrm{Trail}}(\xi)$,
where $\Delta N_{\mathrm{Trail}}(\xi)=\sqrt{\pi
q_0/q(t)}\,(n_0t_0/t)\exp[-\xi^2/4q(t)]$ and $q_0=\psi/(1+\psi)$
[see Eqs.~(\ref{extend}) and (\ref{G}) in Appendix~\ref{near-zone
potential}]. If, however, we extend the integral over $\zeta$ to the
entire half-axis $0<\zeta<\infty$, then we will include the entire
background plasma disturbances along this half-axis and obtain
$\Delta N_\infty =\Delta
N_{\zeta}/2(\xi)=\sqrt\pi\,(n_0t_0/t)\exp(-\xi^2/4q_0)$, where
$\Delta N_{\zeta}$ is the density disturbance integral (DDI) over
the entire $\zeta$ axis, Eq.~(\ref{zeta1}). For $q(t)\simeq 1$, the
difference between $\Delta N_{\mathrm{Trail}}$ and $\Delta
N_{\infty}$ is significant. The two integrals, however, are both
Gaussian functions of $\xi$ with peaks at $\xi=0$. Their integrals
over the entire $\xi$-axis, i.e., the 2D integrals of the trail
density and the total density disturbances over the entire
half-space $\zeta>0$, are equal, $\int_{-\infty}^{\infty}\Delta
N_{\mathrm{Trail}}d\xi = \int_{-\infty}^{\infty}\Delta N_{\infty}
d\xi =2\pi\sqrt{q_0}(n_0t_0/t)$. The true function $\Delta
N_{\zeta_1}(\xi)$, taken over the effective region within the near
zone, $0<\zeta<\zeta_1$, should combine the entire contribution from
the trail density with a part of the background density
disturbances. It is natural to assume that $\Delta N_{\zeta_1}(\xi)$
varies between the two Gaussian functions, $\Delta
N_{\mathrm{Trail}}(\xi)$ and $\Delta N_{\infty}(\xi)$, and can be
approximated by another Gaussian function with the peak at $\xi=0$.
This function should have the same integral over $\xi$, so that it
is determined by one parameter $\tilde q$,
\begin{equation}
\Delta N_{\zeta_1}(\xi)=\left(\frac{\pi q_0}{\tilde
q}\right)^{1/2}\frac{n_0t_0}{t}
\exp\left(-\,\frac{\xi^2}{4\tilde q}\right).%
\label{tildeq}
\end{equation}
The reason why the integral over $\xi$ should be the same is that at
any $\zeta$ beyond the trail the corresponding integral of the
background density disturbances along $\xi$, Eq.~(\ref{xi1}), is
exponentially small, as confirmed by our simulations.

The parameter $\tilde q(t)$ is similar to $q$ in
$N_{\mathrm{Trail}}$ and is uniquely related to the peak value
$\Delta N_{\zeta_1}(0)$. From the derivation of
Appendix~\ref{near-zone potential}, it is clear that the adjusted
parameter $\tilde\rho$ is related to $\tilde q$ by the same relation
as $\rho$ to $q$, Eq.~(\ref{rho}), i.e., $\tilde\rho=\gamma \tilde q
t$. Thus to find $\tilde\rho$, we need to estimate $\Delta
N_{\zeta_1}(0)$,
\begin{equation}
\Delta N_{\zeta_1}(0)=\left(\frac{\pi q_0}{\tilde
q}\right)^{1/2}\frac{n_0t_0}{t}=n_0\left(\frac{\pi \rho_0
t_0}{\tilde
\rho t}\right)^{1/2}.%
\label{N1(0)}
\end{equation}
The part of the corresponding integral stemming from the trail
density is determined by $q$ or by $\rho=\gamma qt$. To find the
additional part stemming from the background density disturbances,
we use Eq.~(\ref{dnsmall}), where we should replace $\rho$ by
$\tilde\rho$ because these disturbances are determined by the
residual potential. While the constant value of $\delta
n=1/1(1+\psi)\tilde\rho$ is easy to integrate, it is not so easy to
determine the upper limit of integration, i.e., the exact value of
$\zeta_1$.

It is clear, however, that $\zeta_1$ should lie within the near
zone, Eq.~(\ref{nearzone}), so that $\zeta_1=\beta /\alpha$, where
$\alpha=(1+\psi)^{1/2}\Theta_0$, Eq.~(\ref{alpha}), determines the
typical scale ($\sim\alpha^{-1}$) of the residual potential
variation along $\zeta$ and $\beta$ is a numerical factor of order
unity or less. Adding the two contributions and using
Eq.~(\ref{N1(0)}), we obtain
\begin{equation}
n_0\left(\frac{\pi t_0\rho_0}{t\tilde\rho}\right)^{1/2} =
n_0\left(\frac{\pi t_0\rho_0}{t\rho}\right)^{1/2}+
\frac{\beta}{2(1+\psi)^{3/2}\Theta_0\tilde\rho}\,.
\label{propro}
\end{equation}
Manipulation with Eqs.~(\ref{gamma}) and (\ref{rho00}) yields
$2(1+\psi)^{3/2}\Theta_0n_0(t_0\rho_0)^{1/2}=(1+\psi)/{\gamma^{1/2}}$,
so that we can recast Eq.~(\ref{propro}) as a quadratic equation for
$\sqrt{\tilde\rho/\rho}$,
\begin{equation}
\frac{\tilde{\rho}}{\rho}-\sqrt{\frac{\tilde{\rho}}{\rho}}+\frac{\tilde\beta}{2}=0,
\label{DelNbeta}
\end{equation}
where
\begin{equation}
\tilde\beta = \frac{2\beta}{1+\psi}\,\sqrt{\frac{\gamma t}{\pi
\rho}}%
\,.%
\label{B}
\end{equation}
Solving Eq.~(\ref{DelNbeta}), we obtain
\begin{equation}
\tilde\rho(t)=\frac{\rho(t)}{2}\left(1-\tilde\beta \pm
\sqrt{1-2\tilde\beta}\right).
\label{tilderho}
\end{equation}
Equation~(\ref{tilderho}) contains parameters $\beta$ or
$\tilde\beta$ which are still unknown functions of $\rho$ and
$\psi$. From Eq.~(\ref{tilderho}) it is clear that there exists an
upper restriction on them ($\rho(t)/\gamma t\rightarrow 1$ as
$\rho\rightarrow\infty$),
\begin{equation}
\tilde\beta\leq \tilde\beta_{\max}=\frac12\,, \qquad \beta\leq
\beta_{\max}=\frac{\sqrt\pi(1+\psi)}{4}.%
\label{Bmax}
\end{equation}
If we start from small values of $\rho$, as assumed by our theory,
then we have to choose the solution with the `plus' sign in front of
the square root corresponding to $\tilde\rho\rightarrow\rho$ as
$\tilde\beta\rightarrow 0$. In this case, the ratio
$\tilde\rho/\rho$ decreases with increasing $\tilde\beta$, so that
$\tilde\rho$ cannot be less than $\rho/4$. In our simulations,
however, we found cases when, at least for some time, $\tilde\rho$
followed Eq.~(\ref{tilderho}) with the `minus' sign in front of the
square root, so that $\tilde\rho$ was less than $\rho/4$. These
special cases ($\psi=0.05$, $\Delta n_0 \lesssim 50$) usually start
from sufficiently large values of $\rho$ which are beyond the
assumptions of our analytic theory.

While the strongest discrepancy between $\tilde\rho$ and $\rho$
takes place for $\rho\gg 1$, the deviation between the two
parameters starts developing at $\rho\sim 1$. Although the above
heuristic derivation for $\tilde\rho$ is only valid for $\rho\gg 1$,
we can formally extend Eqs.~(\ref{DelNbeta})--(\ref{tilderho}) to
$\rho\sim 1$ and determine the function $\beta(\rho)$ numerically
using comparison with simulations, see Sect.~\ref{comparison of
theory}. For $\psi\geq 0.05$, to good accuracy, the numerically
found coupled functions $\beta(\rho)$ and $\tilde\beta(\rho)$ can be
approximated by
\begin{equation}
\beta(\rho)\approx\frac{0.46\rho}{2.2+\rho}\,,\qquad
\tilde\beta(\rho)\approx
\frac{0.92\rho}{(1+\psi)(2.2+\rho)}\sqrt\frac{\gamma t}{\pi\rho}\,.%
\label{betaemp}
\end{equation}
Note that for $\psi=0.05$ at large time, $\tilde\beta$
asymptotically reached the maximum value of
$\tilde\beta_{\max}=1/2$. We extrapolate this to smaller $\psi$
keeping the same $\rho$-dependence as in Eq.~(\ref{betaemp}),
\begin{equation}
\beta(\rho)\approx\frac{\sqrt\pi(1+\psi)\rho}{4(2.2+\rho)} \,,\qquad
\tilde\beta(\rho)\approx
\frac{\rho}{2(2.2+\rho)}\,\sqrt\frac{\gamma t}{\rho}\,.%
\label{betaemp1}
\end{equation}

\bibliography{geophys2005}

\end{document}